\documentclass[aps,prd,preprint,superscriptaddress,nofootinbib,onecolumn,11pt,notitlepage]{revtex4-2}

\usepackage{silence}
\usepackage[utf8]{inputenc}
\usepackage{amsmath}
\usepackage{amssymb}
\usepackage{graphicx}

\usepackage{underscore}
\usepackage{setspace}
\usepackage{etoolbox}
\makeatletter
\pretocmd{\frontmatter@abstract@produce}
  {\vspace{2.5em}} 
  {}{}
\patchcmd{\frontmatter@abstract@produce}
  {\vskip200\p@\@plus1fil \penalty-200\relax \vskip-200\p@\@plus-1fil}
  {}
  {}{}
\apptocmd{\frontmatter@abstract@produce}
  {\clearpage}     
  {}{}
\makeatother
\usepackage{titlesec}
\titlespacing*{\subsection}{0pt}{2.5ex plus 0.5ex minus 0.2ex}{1ex plus 0.2ex}

\usepackage[dvipsnames, table]{xcolor}
\usepackage{booktabs}
\usepackage[normalem]{ulem}
\usepackage{csquotes}
\usepackage{xstring}
\usepackage{multirow}
\usepackage{subcaption}
\usepackage{hyperref}
\usepackage{cleveref}

\newcommand{\internalcite}[2][]{%
  \IfSubStr{#2}{,}{Refs.}{Ref.}~\IfStrEq{#1}{}{\cite{#2}}{\cite[#1]{#2}}%
}
\let\citealp\internalcite 
\newcommand{\switch}{\ensuremath{\Lambda_s}CDM}
\newcommand{\nosroll}{Agnos. Reion.}
\newcommand{\me}{Varying $m_\mathrm{e}$}
\newcommand{\modrec}{4param. Mod. Rec.}
\newcommand{\wzdr}{WZDR}
\newcommand{\emg}{EMG}
\newcommand{\rnr}{RnR}
\newcommand{\thawing}{Thaw. Grav.}
\newcommand{\ede}{EDE}
\newcommand{\neff}{\ensuremath{\Delta N_\mathrm{eff}}}
\newcommand{\sidr}{SIDR}
\newcommand{\lcdm}{\ensuremath{\Lambda}CDM}
\newcommand{\idedm}{iDM-DE}
\newcommand{\coldnede}{NEDE}
\newcommand{\hotnede}{DRMD}

\expandafter\def\csname AppendDictw0wa\endcsname{, when a CPL dark energy characterized by parameters $w_0$\,, $w_a$ is assumed. The datasets are CMB, CMB+BAO, CMB+BAO+SN, and CMB+BAO+SN+$M_B$ prior as shown in the legend.}

\newcommand{\permodelplot}[2]{%
    \begin{figure}[htbp]
        \centering
        \includegraphics[width=0.99\linewidth]{plots/permodel/appendix_#1\ifblank{#2}{}{_#2}.pdf}
        \caption{68\% and 95\% CL contours of the \csname ModelDict#1\endcsname{} model\csname AppendDict#2\endcsname{}%
        \ifcsname specialtext#1\endcsname
          \csname specialtext#1\endcsname
        \fi
        }
        \label{app:fig:#1#2}
    \end{figure}%
}

\begin{document}

\title{The \texorpdfstring{$H_0$}{H0} world cup II: A comprehensive competition between proposed Hubble tension solutions}

\author{Nils Sch\"oneberg}
\affiliation{University Observatory, Faculty of Physics, Ludwig-Maximilians-Universit\"at,
Scheinerstr. 1, 81677 M\"unchen, Germany}
\affiliation{Excellence Cluster ORIGINS, Boltzmannstrasse 2, 85748 Garching, Germany}
\author{Vivian Poulin}
\affiliation{Laboratoire Univers et Particules de Montpellier (LUPM), Universit\'e de
Montpellier \& CNRS, Place Eug\`ene Bataillon, 34095 Montpellier Cedex 05, France}
\author{Angelo G. Ferrari}
\affiliation{INFN--Bologna, Via C. Berti Pichat 6/2, 40127 Bologna, Italy}
\author{Fabio Finelli}
\affiliation{INAF--Osservatorio di Astrofisica e Scienza dello Spazio di Bologna,
Via Piero Gobetti 101, 40129 Bologna, Italy}
\affiliation{INFN--Bologna, Via C. Berti Pichat 6/2, 40127 Bologna, Italy}
\author{Julien Lesgourgues}
\affiliation{Institute for Theoretical Particle Physics and Cosmology (TTK), RWTH Aachen
University, D-52056 Aachen, Germany}
\author{Luca Morelli}
\affiliation{Dipartimento di Fisica e Astronomia “Augusto Righi”, Università di Bologna, Via Piero Gobetti 93/2, I-40129 Bologna, Italy}
\affiliation{INAF--Osservatorio di Astrofisica e Scienza dello Spazio di Bologna,
Via Piero Gobetti 101, 40129 Bologna, Italy}
\affiliation{INFN--Bologna, Via C. Berti Pichat 6/2, 40127 Bologna, Italy}
\author{Markus R. Mosbech}
\affiliation{Institute for Theoretical Particle Physics and Cosmology (TTK), RWTH Aachen
University, D-52056 Aachen, Germany}
\affiliation{Institute for Theoretical Particle Physics (TTP), Karlsruhe Institute of
Technology (KIT), 76128 Karlsruhe, Germany}
\author{Ravi Kumar Sharma}
\affiliation{Institute for Theoretical Particle Physics and Cosmology (TTK), RWTH Aachen
University, D-52056 Aachen, Germany}
\author{Th\'eo Simon}
\affiliation{Laboratoire Univers et Particules de Montpellier (LUPM), Universit\'e de
Montpellier \& CNRS, Place Eug\`ene Bataillon, 34095 Montpellier Cedex 05, France}
\affiliation{Laboratoire de Physique Nucléaire et de Hautes Energies (LPNHE), CNRS/IN2P3 \& Sorbonne Université, 4 place Jussieu, 75005 Paris, France}

\begin{abstract}
Cosmology stands at a crossroads. The Hubble tension has reached a nominal significance above $7\sigma$, while analyses combining DESI BAO and Type Ia supernova data show emerging hints of departures from $\Lambda$CDM. Meanwhile, high-precision CMB measurements from ACT and SPT enable a timely and more stringent reassessment of proposed solutions to the tension. In this paper, we revisit the $H_0$ Olympics, a systematic contest comparing proposed alternatives to $\Lambda$CDM using common datasets, likelihoods, and statistical criteria. In this updated edition, the \enquote{$H_0$ World Cup,} we subject fourteen representative solutions to a common analysis of current CMB, BAO, and SN data. The contenders span four broad mechanisms: late-time modifications of the expansion history, modified recombination, additional pre-recombination radiation, and early non-radiative energy injection. Relative to the original analysis, the present competition includes models and mechanisms proposed in the intervening years and evaluates all contenders using both Bayesian and Frequentist tests of tension and model performance, letting the neutrino mass sum vary. We further test if late-time extensions through curvature or the Chevallier-Polarski-Lindner (CPL) dark energy parametrization can aid the success of the models. Finally, we subject the leading contenders to dedicated robustness tests involving alternative CMB likelihoods and multipole cuts, supernova samples, large-scale-structure information, and big-bang nucleosynthesis constraints. This framework assesses both the ability of each mechanism to ease the Hubble tension and the robustness of our conclusions to datasets and analysis choices.
\end{abstract}

\begin{flushright}
        {\large \tt TTK-26-23}\\
        {\large \tt TTP26-28}
\end{flushright}
\maketitle

\section{Introduction}
Over the past decade an inconsistency has emerged between local measurements of the current expansion rate of the Universe -- known as the Hubble constant, $H_0$ -- and the indirect inference using the $\Lambda$CDM model calibrated on pre-recombination data. Since about 2013 with the release of the Planck satellite CMB data, this \enquote{Hubble tension} has continued to grow in significance. More recent observations of the cosmic microwave background (CMB) from the Planck satellite \cite{Planck:2018nkj}, the Atacama cosmology telescope (ACT) \cite{ACT:2025blo}, and the South Pole Telescope (SPT) \cite{SPT-3G:2025bzu} have been used to infer a value of the Hubble constant in the $\Lambda$CDM cosmological model, yielding a value of $H_0=(67.24 \pm 0.35)$km/s/Mpc \cite{SPT-3G:2025bzu}. On the other hand, very recently, experts in different techniques for local measurements of the Hubble constant have participated in a workshop to determine a consensus value of the Hubble constant from local observations, analyzing their respective data in a joint pipeline that yielded $H_0 = (73.50\pm 0.81)$km/s/Mpc \cite{H0DN:2025lyy}. Taken at face value, these measurements now yield a Hubble tension of $7.1\sigma$, which corresponds to a random chance of disagreement of roughly 1 in 770 billion. This leads to the inevitable conclusion that either the data model is incorrect or the $\Lambda$CDM model does not provide a sufficiently accurate description of our Universe.

A biased data analysis pipeline could be caused, for example, by an incomplete assessment of systematic uncertainties plaguing a given measurement, or otherwise potentially wrong assumptions about the underlying data (for example by assuming the cosmological principle). While such a possibility cannot be disproven because of \enquote{unknown unknowns} that may affect a given measurement \cite{61d4609e-39c8-35e6-875d-0c1dc5ef1fae}, there are good indications that the origin of the tension may not reside in the data alone. First, there is a large body of literature that investigates different possible systematics both on the side of the local distance ladder and of the CMB \cite{Riess:2023bfx,Tristram:2025you,H0DN:2025lyy,CosmoVerseNetwork:2025alb,2026arXiv260617008A,ACT:2025blo,SPT-3G:2025bzu}, and still did not identify a convincing way to reconcile the measurements. Second, and perhaps more importantly, a large variety of fundamentally different techniques now exists to determine the Hubble constant from both early and late Universe data.\footnote{We use the term \enquote{early-Universe} data for observables that primarily constrain cosmological conditions before the onset of structure formation, such as the CMB or BBN, even when their extraction requires modeling or subtracting later astrophysical effects. In these cases, the information content is dominated by physical processes that decoupled from the thermal bath before or around recombination.}
On the one hand, beyond the use of different CMB datasets---such as WMAP \cite{2013ApJS..208...20B}, Planck, ACT, or SPT---high-precision early-Universe constraints on $H_0$ can also be obtained by combining big-bang-nucleosynthesis (BBN) measurements of the baryon abundance with large-scale-structure (LSS) observations. 
These lead to values of the Hubble constant below $70$km/s/Mpc in $\Lambda$CDM, and are summarized for example in \citealp{Verde:2023lmm,CosmoVerseNetwork:2025alb}. On the other hand, the consensus report on the Hubble constant \cite{H0DN:2025lyy} shows that the different model-independent local measurements agree largely on the fact that the value lies above $70$km/s/Mpc. 
This is also supported by other local observations like the time delay of strongly lensed quasars or the astrophysical modeling of supernovae of type II using the expanding photosphere method, see \citealp{Verde:2023lmm,CosmoVerseNetwork:2025alb} and references therein. 
Interestingly, observations of cosmic chronometers and inferences from the ages of the oldest astrophysical objects similarly point towards low values of the Hubble constant around $68 \pm 4$km/s/Mpc, see also \citealp{Verde:2023lmm,CosmoVerseNetwork:2025alb}, though here the uncertainty is a little too large to make definite conclusions. 
Many other local measurement methods (such as gravitational waves as bright standard sirens or the use of fast radio bursts) currently do not have enough precision to weigh in on the Hubble tension significantly, but are expected to improve in the future \cite{Verde:2023lmm,CosmoVerseNetwork:2025alb}. 

Given that at this point there are multiple \enquote{families} of measurements pointing towards a Hubble tension \cite{Verde:2023lmm}, it is vital that all of these methods are thoroughly investigated, with a special focus on the ones that give the tightest constraints and are thus driving the Hubble tension. 

Given the compelling convergence of evidence for a robust tension from different datasets and independent analyses, it is crucial that the community identifies the most promising alternatives to the $\Lambda$CDM standard model of cosmology and builds on these to construct first-principle models in which all cosmological datasets can be brought into concordance. Some attempts at extensive categorization of alternatives have been performed in \citealp{DiValentino:2021izs,CosmoVerseNetwork:2025alb}. However, to outline which of these models are currently the most successful at reducing the Hubble tension and to compare their goodness of fit, the models need to be subjected to the same data likelihoods in the same analysis pipeline. Such comparisons have already been carried out, for example, in
\citealp{Schoneberg:2021qvd,Escudero:2022rbq,Khalife:2023qbu,ACT:2025tim},
including by some of us in the previous {\it Hubble Olympics}
\citealp{Schoneberg:2021qvd}, where a variety of models were tested under
common conditions and ranked relative to one another. A comparison of proposed solutions is also important for identifying predictions beyond the value of $H_0$ itself. Once one or more viable solutions are identified (if any), their additional signatures can be worked out and tested against existing or future independent observations.

Since the original Hubble Olympics paper \cite{Schoneberg:2021qvd}, both new models and new data have become available, in particular from ACT, SPT, and DESI, together with new SNIa and $H_0$ measurements. These data have important implications for the status of $\Lambda$CDM, including a preference for dynamical dark energy at the $\sim 3\sigma$ level \cite{DESI:2025fii}, and significantly affect many models proposed to resolve the $H_0$ tension (e.g., \cite{Poulin:2025nfb,Sharma:2026ngx}). This motivates a timely reassessment of the previous results. In this work, we therefore extend the previous literature by investigating a broad range of models with current cosmological datasets, evaluating several performance metrics, and constructing a comparative ranking of model performance. In particular, as part of the baseline analysis, we consider a free neutrino mass sum, and extend the analysis through variations where we consider a curved background geometry, a Chevallier–Polarski–Linder (CPL) dark energy equation of state, or a running of the primordial power spectrum.
We then subject the best performing models to further tests designed to probe the robustness of the results.

Our results are presented in two companion papers. Paper I \cite{PaperI} gives a concise overview of the baseline competition and its main implications. This paper, Paper II, presents the {\it $H_0$ world cup} competition in full details: in \cref{sec:models}, we introduce the contending models, their additional parameters and the priors we use; in \cref{sec:metrics}, we define the metrics used to evaluate them, fixed before inspecting the results; and in \cref{sec:competition}, we present the results of the {\it group stage} competition and select most successful models to advance to a {\it knockout-round}. We subject these {\it finalist models} to further robustness tests and cross-checks in \cref{sec:golden}, before concluding in \cref{sec:conclusions}.

\section{Models}\label{sec:models}

In this section we summarize and describe all the models that enter our competition. Given the large number of possible models and the limited time and computational power available, we could only let a small selection enter the competition.

We have focused on recent proposals that show promise in resolving the Hubble tension when faced with CMB, BAO and SNIa data. A key requirement was that models were implemented in an up-to-date version of the Boltzmann code \texttt{CLASS} \cite{Lesgourgues:2011re} (v$3.3$). This ensures that all relevant cosmological observables (including those based on the evolution of perturbations like the CMB anisotropy power spectrum) can be computed efficiently and with the required accuracy. Notable models not included involve self-interacting neutrinos \cite{Lancaster:2017ksf,Kreisch:2019yzn} or majorons \cite{Escudero:2019gvw,Escudero:2021rfi}, which were strongly constrained by Planck polarization and large-scale structure data \cite{Camarena:2023cku,Camarena:2024daj,Poudou:2025qcx}, as well as the \enquote{mirror world} of \citealp{Cyr-Racine:2021oal,Ge:2022qws,Greene:2023cro,Greene:2024qis}, which by construction leaves the CMB invariant under a change in $H_0$\,, but is potentially constrained by Big Bang Nucleosynthesis. We expect that a larger selection of models will be available in upcoming model comparison efforts, such as the one initiated by the CosmoVerse initiative \cite{CosmoVerseNetwork:CosmologyCompilation}. 

Many of the investigated models lack a deep theoretical foundation (in terms of a UV-complete extension of the standard model of particle physics and a corresponding Lagrangian formulation in quantum field theory), and are to be understood as phenomenological toy models exhibiting certain mechanisms that are useful for solving the Hubble tension. The question in this manuscript is not \enquote{What is the correct first-principles theory to describe nature, assuming the Hubble tension is real?} but instead \enquote{Which of the proposed mechanisms to ease the Hubble tension are most successful from a phenomenological perspective?} As such, we remain agnostic about the ways in which such models can be realized in a quantum field theory.

For each of the models listed below, we briefly summarize the theoretical context and motivations (where does this idea/model come from), the properties that are particularly relevant for our analysis, the numerical code that we use to predict observables, the list of additional free parameters, and the prior assumed on each of them. In section titles, the name of the model is followed by a subtitle indicating which general mechanism the model uses to solve the Hubble tension.
We have grouped these into four competing groups:
\begin{enumerate}
    \item Group L: Competitors that incorporate \textit{late} time (post-recombination) modifications.
    \item Group M: Competitors that are based on \textit{modifications} of the process of recombination.
    \item Group R: Competitors that involve an increased amount of \textit{radiation}.
    \item Group E: Competitors that change the inferred value of $H_0$ primarily through modifications of the \textit{early} (pre-recombination) Universe, but are not part of groups~M or~R.
\end{enumerate}
All these models are extensions of the $\Lambda$CDM paradigm, which includes here the effects of massive neutrinos, treating the summed neutrino mass as a free parameter. Thus, the base model contains the seven parameters listed in \cref{tab:parameters}, for which we adopt wide flat priors. We note that the $H_0$ prior is somewhat informative in some cases -- but typically we are not interested in extreme values of $H_0$ below 60 km/s/Mpc or above 80 km/s/Mpc. The massive neutrinos are modeled as three degenerate species with the same mass -- this has been shown in \citealp{Lesgourgues:2004ps,Archidiacono:2020dvx} to be sufficiently accurate in representing both inverted and normal hierarchy scenarios.
\begin{table}[t]
    \centering
    \begin{tabular}{c|c c}
        Parameter & Lower bound & Upper bound \\ \hline
        \rule{0pt}{1em}$\Omega_\mathrm{b} h^2$ & 0.017 & 0.027 \\
        $\Omega_\mathrm{cdm} h^2$ & 0.09 & 0.15 \\
        $\ln 10^{10}A_\mathrm{s}$ & 2.6 & 3.5 \\
        $H_0$ [km/s/Mpc] & 60 & 80 \\
        $n_\mathrm{s}$ & 0.9 & 1.1 \\
        $\tau_\mathrm{reio}$ & 0.004 & 0.1 \\
        $\sum m_\nu$ [meV]& 0 & 30\,000
    \end{tabular}
    \caption{Bounds on the parameters considered for the standard $\Lambda$CDM cosmological model.}
    \label{tab:parameters}
\end{table}

Unlike \citealp{Schoneberg:2021qvd,Khalife:2023qbu}, we systematically investigate how additional freedom in the form of either dynamical dark energy, spatial curvature, or a deviation from a power-law primordial power spectrum affects our results. We therefore also test whether a combination of early- and late-time ingredients could help resolve the tension, as well as the implication of early-time modifications for the preference towards dynamical dark energy. Since we always treat the summed neutrino mass as a free parameter, we explore at the same time the impact of these physical ingredients on the tension between cosmology and laboratory bounds on the neutrino mass scale.

In order to efficiently explore the model parameter spaces (which can have complicated shapes, see \cref{app:permodel}) we use the Online Learning Emulator (OL\'E) of \citealp{Gunther:2025xrq}.\footnote{The OL\'E code is available at \url{https://github.com/svenguenther/OLE}.} This tool, fully integrated within MontePython \cite{Brinckmann:2018cvx,Audren_2013}, builds an accurate emulator of all the observables computed by the Boltzmann code  while sampling the parameter space, in order to accelerate future evaluations. OL\'E tests the emulator accuracy during this process to ensure that the final posteriors are not biased by emulation errors. 
The resulting speedup enables us to explore the many combinations of the different datasets and models in this work efficiently.

\subsection{Axion-like Early Dark Energy [Group E]}

\textbf{Motivation:} 
The possibility that a non-negligible dark energy component may be present prior to last scattering has been explored for over a decade \cite{Doran:2000jt,Wetterich:2004pv,Doran:2006kp}. However, these scenarios attracted renewed interest in the context of the Hubble tension, as it was shown in \citealp{Karwal:2016vyq,Poulin:2018cxd} that they may provide a compelling solution.
Models of Early dark energy (EDE) typically invoke a scalar field energy density which remains constant at epochs prior to recombination, behaving as a dark energy component during this period, until it starts diluting typically at a redshift close to matter-radiation equality. 
This effectively provides a time-localized boost to the expansion history that allows these models to reduce the sound horizon $r_s$\,. Due to the degeneracies inherent in the angular power spectra of the CMB, the sound horizon and the Hubble constant are anti-correlated, leading to a larger inferred value of the Hubble constant.
Since the original proposal of \citealp{Karwal:2016vyq,Poulin:2018cxd}, a wide range of realizations and extensions of the underlying scalar-field framework have been proposed and studied extensively in the literature \cite{Smith:2019ihp,Agrawal:2019lmo,Lin:2019qug,Alexander:2019rsc,Sakstein:2019fmf,Das:2020wfe,Niedermann:2019olb,Niedermann:2020dwg,Niedermann:2021vgd,Ye:2020btb,Simon:2023hlp,Berghaus:2019cls,Freese:2021rjq,Braglia:2020bym,Sabla:2021nfy,Sabla:2022xzj,Gomez-Valent:2021cbe,Moss:2021obd,Guendelman:2022cop,Karwal:2021vpk,McDonough:2021pdg,Wang:2022nap,Alexander:2022own,McDonough:2022pku,Nakagawa:2022knn,Gomez-Valent:2022bku,MohseniSadjadi:2022pfz,Kojima:2022fgo,Rudelius:2022gyu,Oikonomou:2020qah,Tian:2021omz,Maziashvili:2021mbm}. 
 
In this work, we consider a realization of EDE through a modified axion potential introduced in \citealp{Kamionkowski:2014zda,Karwal:2016vyq,Poulin:2018dzj,Poulin:2018cxd,Smith:2019ihp},
\begin{equation}\label{eq:potential}
V_n(\Theta) = m^2 f^2[1-\cos (\Theta)]^n,
\end{equation}
where $m$ is the axion mass, $f$ is the axion decay constant, and $\Theta \equiv \phi/f$ is a dimensionless (re-normalized) field variable defined on the interval $-\pi \leq \Theta \leq \pi$. This potential provides a phenomenological generalization of the well-motivated axion-like potential, which is recovered for $n=1$, and arises in string-theoretic constructions \cite{Svrcek:2006yi,Douglas:2006es,Arvanitaki:2009fg,Marsh:2015xka, McDonough:2022pku,Cicoli:2023qri}.

We assume that the field initially satisfies slow-roll conditions, enforced by the large Hubble friction at early times, and without loss of generality restrict the initial field value to $0 \leq \Theta_i \leq \pi$. The model thus contains four fundamental parameters, ${m, f, n, \Theta_i}$. Since the background and perturbation dynamics are only weakly dependent on $n$ within the range $2 \lesssim n \lesssim 4.5$ \cite{Agrawal:2019lmo,Smith:2019ihp}, we follow Refs.~\cite{Poulin:2018cxd,Smith:2019ihp} and fix $n=3$.

As is customary, rather than working directly with $m$ and $f$, we describe the model in terms of the fractional energy density $f_{\rm EDE}(z_c)$—approximately the maximum contribution of EDE to the total energy budget—and the critical redshift $z_c$ at which the field becomes dynamical. Note that this translates into non-trivial priors on the fundamental parameters \cite{Hill:2020osr}.
The remaining degree of freedom enters through the dynamics of linear perturbations, fully characterized by the effective sound speed $c_s^2$. Physically, this quantity is related to the curvature of the potential near the initial field value and, once the other parameters are fixed, is uniquely determined by $\Theta_i$ \cite{Poulin:2018dzj,Smith:2019ihp}.

The qualitative evolution of the model proceeds as follows (see Refs.~\cite{Poulin:2018dzj,Smith:2019ihp} for details). At early times, the scalar field is frozen by Hubble friction and behaves as a dark energy component. When the Hubble rate falls below the axion mass, the field becomes dynamical and undergoes damped oscillations about the minimum of the potential. The subsequent dilution of the energy density is governed by the period-averaged equation of state, which is approximately given by $w(n) = (n-1)/(n+1)$.

The model has been confronted to a wide variety of datasets, including recent CMB and LSS data, with varying degree of success (or constraints) in alleviating the tension \cite{Hill:2020osr,Ivanov:2020ril,DAmico:2020ods,Murgia:2020ryi,Smith:2020rxx,Simon:2022adh,Simon:2023hlp,ACT:2025tim,Poulin:2025nfb,chaussidon:2025,SPT-3G:2025vyw}. For reviews on the topic, we refer to \citealp{Kamionkowski:2022pkx,Poulin:2023lkg,McDonough:2023qcu}.
Part of our goal in this work is precisely to clarify the observational status of axion-like EDE in light of up-to-date data.

\vspace*{0.5\baselineskip}
\noindent \textbf{Analysis:} To perform our analyses, we use
{\sc AxiCLASS} \citealp{Smith:2019ihp}, which we updated to version \texttt{v3.3} of the parent code {\sc class} \cite{Lesgourgues:2011re}. 
The code is publicly available at \url{https://github.com/PoulinV/AxiCLASS}. 
\vspace*{0.5\baselineskip}

\noindent \textbf{Parameters and priors:} The EDE model considered here extends $\Lambda$CDM by three additional parameters: (1) the critical redshift $z_c$\,, or equivalently the critical scale factor $a_c = 1/(1+z_c)$, at which the scalar field becomes dynamical, (2) the fractional energy density contributed by the field at that epoch, $f_{\rm EDE}(a_c)$, and (3) the initial dimensionless field value $\Theta_i$\,.  We adopt the following flat priors: $f_{\rm EDE}(z_c) \in [0.0001, 0.3]$, $\log_{10}(z_c) \in [3, 4]$, and $\Theta_i \in [0.01, 3.1]$.

\subsection{Cold New Early Dark Energy [Group E]}
\textbf{Motivation:} The axion-like EDE has sparked interest for the development of models that better address the underlying theoretical motivation and/or improve the inherent fine-tuning problem related to the question: \enquote{Why is the axion mass such that the axion-like EDE decays exactly before recombination?} In addition, the original axion-like EDE model slightly increases the value of $\sigma_8$ due to an unavoidable increase in the CDM density that leads to earlier matter-radiation equality. Recent cosmic shear and galaxy clustering analyses find a value of $S_8\equiv \sigma_8 \sqrt{\Omega_\mathrm{m}/0.3}$ compatible with other observations within the $\Lambda$CDM framework, but still constrain higher values of $S_8$ predicted in models such as EDE.

To circumvent theoretical and observational issues, one particularly promising proposal is Cold New Early Dark Energy (cold NEDE)\cite{Niedermann:2020dwg,Niedermann:2023ssr,Cruz:2023lmn,Chatrchyan:2024xjj}. This model belongs to the class of EDE scenarios in which a scalar field undergoes a first-order phase transition. In short, the NEDE scalar field $\psi$ is initially trapped in a false vacuum of its potential. The height of the potential barrier is controlled by a second (trigger) scalar field $\phi$, which has an ultra-light mass $m$. As $\phi$ slowly rolls down its potential, the barrier separating the false and true minima decreases, eventually allowing $\psi$ to quantum tunnel through it.

This mechanism can be described by a two-field potential in an eV-scale realization inspired by models of first-order inflation:
\begin{equation}
V(\psi, \phi) = \frac{\lambda}{4}\psi^4 + \frac{\beta}{2} M^2 \psi^2 - \frac{\alpha}{3} M \psi^3 + \frac{1}{2} m^2 \phi^2 + \frac{1}{2} \tilde{\lambda} \phi^2 \psi^2.
\end{equation}

Here, the quadratic mass term $\beta$ and cubic coupling $\alpha$ are assumed to be of order unity, while the quartic self-coupling $\lambda$ is smaller than one. Additional higher-order terms (not shown) are expected within an effective field theory framework. The mechanism relies on a hierarchy of scales: the tunneling field $\psi$ is governed by eV-scale physics (i.e., $M \sim \mathrm{eV}$), whereas the trigger field $\phi$ is extremely light, with mass $m \sim 10^{-27}\,\mathrm{eV}$, so that its evolution becomes relevant only around matter--radiation equality. The heavy field therefore plays the role of EDE while the lighter field helps reduce $S_8$ (similarly to ultra-light axions).

The NEDE model is characterized by four  parameters: the fractional energy density at the transition, $f_{\rm NEDE}$, the transition redshift $z_c$ (or equivalently the transition time $\tau_*$), the post-transition equation-of-state parameter $w_{\rm NEDE}$, and the present-day density parameter of the trigger field, $\Omega_\phi$. These quantities are related to the underlying model parameters $(\alpha,\beta,\lambda,M,\phi_{\rm ini})$ through
\begin{equation}
f_{\rm NEDE}
=
\frac{\bar{\rho}_{\rm NEDE}(\tau_*)}{\rho_{\rm tot}(\tau_*)}
=
c_\delta
\frac{\alpha^4}{36\lambda^3}
\frac{M^4}{M_{\rm Pl}^2 H_*^2},
\end{equation}
where
\begin{equation}
c_\delta
=
\frac{1}{216}
\left(3+\sqrt{9-4\delta}\right)^2
\left(3-2\delta+\sqrt{9-4\delta}\right),
\qquad \delta=\frac{9\lambda\beta}{\alpha^2}.
\end{equation}
The trigger field $\phi$ subsequently behaves as a fuzzy dark matter component, whose present-day abundance is determined by the initial field displacement,
\begin{equation}
\Omega_\phi
\simeq
0.4
\left(\frac{1+z_c}{5000}\right)
\left(\frac{\phi_{\rm ini}}{M_{\rm Pl}}\right)^2
(1-f_{\rm NEDE}).
\end{equation}

\noindent \textbf{Analysis:} We use the publicly available {\sc class} implementation of NEDE, that can be found at \url{https://github.com/NEDE-Cosmo/TriggerCLASS} and is upgraded to version \texttt{v3.3} of {\sc class}.

\vspace*{0.5\baselineskip}
\noindent \textbf{Parameters and Priors:} This model has four additional free parameters compared to $\Lambda$CDM, which are varied in following range: $f_{\rm NEDE} \in [0,0.3]$, $\log_{10}z_{c}\in [2.8,3.7]$, $3w \in [1,3]$, $\Omega_{\rm \phi}\in [0,0.3]$.

\subsection{Early Modified Gravity and Rock'n'Roll [Group E]}\label{ssec:emg}

\textbf{Motivation:} Early modifications of gravity can successfully address the $H_0$ tension 
\cite{Rossi:2019lgt,Braglia:2020iik,Zumalacarregui:2020cjh,Ballardini:2020iws,Braglia:2020bym,FrancoAbellan:2023gec}.
The coupling $\xi$ between the Ricci scalar, or equivalently the total trace of the energy-momentum tensor, and a non-minimally scalar field $\sigma$ can drive the scalar field which controls the gravitational constant. Since the trace is effectively zero during the relativistic regime, $\sigma$ can be set at rest at early times. The scalar field then moves after matter-radiation equality and alters the expansion history around and after recombination.
The overall effect of having a different gravitational constant in the relativistic regime and a non-trivial dynamics of the field after matter-radiation equality can induce a positive degeneracy between the coupling $\xi$ and $H_0$\,, allowing larger values for $H_0$ than in $\Lambda$CDM when compared with observations, even in the simplest models \cite{Umilta:2015cta,Ballardini:2016cvy}. 

Here we consider the Early Modified Gravity (EMG) model introduced in Ref. \cite{Braglia:2020auw}, which is described by the following action, 
\begin{equation}
S = \int d^4x \sqrt{-g} \left[ (M_{\rm pl}^2+\xi \sigma^2)\frac{R}{2} -\frac{g^{\mu \nu}}{2} \partial_\mu \sigma\partial_\nu \sigma 
-\Lambda -\frac{\lambda\sigma^4}{4} \right] +S_{\rm m} \,,
\label{EMG}
\end{equation}
where $R$ is the Ricci scalar, $S_{\rm m}$ is the action for matter fields, 
and $\xi, \ \lambda$ are dimensionless constants. In this model, 
we have a weaker gravitational strength at early times, and the quartic potential leads to a fast rolling of $\sigma$ towards the minimum, in such a way that the effective Newtonian constant is automatically consistent with the tight constraints from laboratory experiments and on post-Newtonian parameters from Solar System measurements. 

For $\lambda = 0$, the model reduces to the case of a non-minimally coupled massless scalar field considered in \citealp{Braglia:2020iik}, 
whereas for $\xi = 0$ it reduces to the Rock 'n' Roll (henceforth RnR) model of \citealp{Agrawal:2019lmo}, which is a particular EDE model. The RnR model is interesting both as the $\xi \rightarrow 0$ limit of EMG and as an EDE model with a different potential and one parameter less than in \cref{eq:potential}.

\vspace*{0.5\baselineskip}
\noindent \textbf{Analysis:} We use a modified version of {\sc class} \texttt{v3.3} dedicated to scalar-tensor theory in the Jordan frame, called {\sc ClassIG}, originally presented in \citealp{Umilta:2015cta} and compared to other codes in \citealp{Bellini:2017avd}. In the analysis, a shooting algorithm determines the value of $\Lambda$ required to obtain a chosen value of $H_0$, which is used as the primary parameter with a flat prior. We use a primordial Helium-4 mass fraction from a corresponding shift in $N_\mathrm{eff}$ that would result in the same altered Hubble rate at early times.

\vspace*{0.5\baselineskip}
\noindent \textbf{Parameters and Priors:} EMG has three additional free parameters compared to 
$\Lambda$CDM,
which are varied as in \cite{FrancoAbellan:2023gec} in the following range: $\xi \sigma_i^2/M_{\rm pl}^2 \in [0,0.2]$, $\sigma_i/M_{\rm pl} \in [0.03,0.9]$, $\alpha_\mathrm{EMG}  \equiv \log_{10} (3.516 \cdot 10^{109} \lambda)/2 \in [0.6,3.5]$. 
In the RnR case, we fix $\xi=0$ and we vary $\sigma_i/M_{\rm pl} \in [0,0.9]$, $\alpha \in [0.6,3.5]$.

\subsection{Thawing gravity [Group E+L]}\label{ssec:thawing}

\textbf{Motivation:} The {\em thawing gravity} (TG) model is another example of a non-minimally coupled scalar field as in \cref{ssec:emg}, but with an exponential potential rather than a quartic potential plus a cosmological constant:
\begin{equation}
S = \int d^4x \sqrt{-g} \left[ (M_{\rm pl}^2+\xi \sigma^2)\frac{R}{2} -\frac{g^{\mu \nu}}{2} \partial_\mu \sigma\partial_\nu \sigma - V_0 e^{- \lambda_\mathrm{TG} \cdot \sigma/M_{\rm pl}} \right] +S_{\rm m} \,,
\label{eq:TG}
\end{equation}
This model has been introduced in \cite{Ye:2024ywg} in the context of dynamical dark energy as indicated since DESI 2024 BAO data. Indeed, phantom crossing without quantum instabilities can occur in scalar-tensor theories of gravity with a single scalar-field \cite{Gannouji:2006jm}. This model's background cosmology has been previously studied in \citealp{Uzan:1999ch}. Differently from EMG, in which the amplitude of the scalar field is rapidly damped during the coherent oscillations at the bottom of the potential, in TG the scalar field dynamics is slowly varying, and is potentially not negligible at low redshift as in \cite{Braglia:2020iik,Zumalacarregui:2020cjh,Ferrari:2023qnh,Ferrari:2025egk}. It thus can induce late time modifications of gravity, which are constrained by Solar System and other local measurements if screening mechanisms are not invoked. For this reason, TG cannot be considered purely as an {\em early model} like EMG above and we consider it as representative of a new \enquote{Early plus Late} class of models. In particular, the model has the potential to affect both early and late time physics, depending on the considered parameter regime.  
The value of the coupling $\xi$ plays a major role in how strong the changes in Planck mass are in the early universe, while the potential slope $\lambda_\mathrm{TG}$ is particularly important for the model's late time evolution.

\enlargethispage*{2em}
\vspace*{0.5\baselineskip}
\noindent \textbf{Analysis:} We use the same code as for EMG to numerically evolve the background and perturbations, replacing the potential by the exponential form in \cref{eq:TG}, following \citealp{Ye:2024zpk}. In the analysis, an implicit algorithm determines the value of $V_0$ required to obtain a chosen value of $H_0$, which is used as the primary parameter with a flat prior. Thus, $V_0$ is not varied independently.

\vspace*{0.5\baselineskip}
\noindent \textbf{Parameters and Priors:} Thawing Gravity has three additional free parameters, on which we impose flat priors $\xi \sigma_i^2/M_{\rm pl}^2 \in [-10^{-0.5},-10^{-3}]$, $\sigma_i/M_{\rm pl} \in [-20,-0.03]$, $\lambda_\mathrm{TG} \in [0,5]$. The priors are motivated to cover the region where \citealp{Ye:2024zpk} finds the model to resolve the Hubble tension (at $\xi \sigma_i^2/M_{\rm pl}^2 \to -10^{-1.5}$ and $\sigma_i/M_{\rm pl} \to -3.2$) as well as the $\Lambda$CDM limit ($\xi \to 0$, $\sigma_i/M_{\rm pl} \to 0$), noting that we use the opposite definition of $\xi$ here.

\subsection{Varying effective electron mass [Group M]}\label{ssec:me}

\textbf{Motivation:}
The sound horizon $r_s$ can be reduced by increasing the redshift of recombination. Thanks to the degeneracies inherent to CMB physics, this leads to a larger inferred Hubble constant. The recombination redshift is generally regarded as well determined on the basis of established atomic physics. However, if the fundamental atomic constants at the time of recombination are different from those measured today in the laboratory, the recombination redshift may be shifted. In this work, we use a simple phenomenological model in which we let the electron mass take a different value at and before recombination, $m_\mathrm{e}^\mathrm{early}$, and in the present universe, $m_\mathrm{e}^\mathrm{late}$. This scenario has been investigated thoroughly in \citealp{Hart:2019dxi,Sekiguchi:2020teg,Hart:2021kad,Seto:2022xgx,Chluba:2023xqj,Seto:2024cgo,Baryakhtar:2024rky,Schoneberg:2024ynd,Baryakhtar:2025uxs,Smith:2025grk,Garramone:2026evc}. In particular, \citealp{Baryakhtar:2024rky} discusses a possible motivation for this model and how it may avoid many of the late universe constraints. However, this discussion was carried at the classical level. It appears difficult to build a fully consistent model at the quantum level in which the effective electron mass drifts between recombination and today. If such a drift is realized by coupling the electrons to a light degree of freedom, it is very difficult to avoid constraints on the resulting fifth force. Therefore, instead of regarding this scenario as a first-principle extension of the standard model, it should be seen as a toy model for shifting recombination while keeping the diffusion damping scale unchanged.

It has been shown in \citealp{Smith:2025arq} that this toy model is almost unconstrained by CMB data alone due to an underlying physical mechanism that leaves the CMB power spectrum almost entirely invariant. Indeed, as discussed in \citealp{Sekiguchi:2020teg, Schoneberg:2024ynd, Baryakhtar:2024rky}, the relevant scales (e.g. the sound horizon, the equality scale, or the Silk damping horizon) imprinted in the CMB are all shifted proportionally to $m_\mathrm{e}^\mathrm{early}/m_\mathrm{e}^\mathrm{late}$. The CMB angular power spectrum is then left invariant when the angular diameter distance to the CMB is shifted by the same amount. In a $\Lambda$CDM background, this typically requires $h \propto m_\mathrm{e}^{3.18}$ and $\Omega_\mathrm{m} \propto h^{-1.7}$, leading to low values of $\Omega_\mathrm{m}$ when attempting to solve the Hubble tension, see also \cite[Fig.~3]{Poulin:2024ken}. This provides a clear signature, which even favors this model when using DESI BAO data \cite{Schoneberg:2024ynd}. 

However, the lower value of $\Omega_\mathrm{m}$ is not a generic prediction of this mechanism. It can be avoided by opening up further freedom in the late-universe background evolution, for instance, with spatial curvature or dynamical dark energy. Then, the angular diameter distance to the CMB can be increased by another parameter than $h$, allowing more freedom to compensate for the required shift. As a by-product, the parameter constraints on $\Omega_\mathrm{m}$ from low-redshift probes become much wider in this case. \citealp{Schoneberg:2021qvd} showed that this mechanism results in a slight preference for non-zero spatial curvature in presence of a varying electron mass. In this work, we wish to cross-check whether such a preference still exists. 

\vspace*{0.5\baselineskip}
\noindent \textbf{Analysis:} We employ a simplified toy model with an instantaneous transition of the electron mass from its early to its late time value, see \cite[Secs.~2,3]{Schoneberg:2024ynd} for justifications on this approximation. To perform our analyses, we use a modified version of {\sc class} presented in \citealp{Schoneberg:2024ynd}, updated to match version \texttt{v3.3}. We also include in our {\sc class} version an adequate BBN interpolation table for this model, obtained by running the code described in \citealp{Escudero:2025kej}, which is available at \url{https://github.com/schoeneberg/class_public_versions}. We also employ the modified BBN code of \citealp{Escudero:2025kej} to compute the variation in $N_\mathrm{eff}$ caused by the electron mass shift. This variation is typically very small and does not significantly affect our results. The impact of BBN constraints on this model are potentially more important (see \citealp{Garramone:2026evc}) and further examined in \cref{ssec:bbn}.
 
\vspace*{0.5\baselineskip}
\noindent \textbf{Parameters and priors:} The varying electron mass model considered here extends $\Lambda$CDM by a single additional parameter $m_\mathrm{e}^\mathrm{early}/m_\mathrm{e}^\mathrm{late}$, for which we adopt the following flat prior: $m_\mathrm{e}^\mathrm{early}/m_\mathrm{e}^\mathrm{late} \in [0.8, 1.2]$.

\subsection{Four-parameter modified recombination history [Group M]}
\textbf{Motivation:}
Another way to reduce the comoving sound horizon $r_s$ is to shift the redshift of recombination through a modified recombination history. For example, \citealp{Lynch:2024gmp,Mirpoorian:2024fka} consider a free-form reconstruction of the ionization rate and find that this may reconcile CMB data with large values of $H_0$\,. A physically motivated mechanism to realize a modified ionization history has been proposed through inhomogeneous recombination, for example in the presence of large primordial magnetic fields (PMF) on small scales \cite{Jedamzik:2020krr,Mirpoorian:2024fka}. 
\citealp{Jedamzik:2023rfd} performed numerical simulations of primordial magnetic fields potentially leading to inhomogeneous recombination and found that the typical outcome is an overall shift of the recombination redshift to larger values along with a bump in the ionization fraction, $x_e(z)$, occurring in the range $1000 < z < 1200$. Subsequently, \citealp{Mirpoorian:2024fka} proposed to approximate a range of modified ionization histories through a 4-parameter phenomenological parametrization, motivated from simulations with PMF,
\begin{equation}
    x_e^\mathrm{modified}(z) = x_e(z-\Delta z) \,\cdot\, \left[1+A_\mathrm{gauss} \exp\left(-\frac{1}{2}\frac{(z-z_\mathrm{gauss})^2}{\sigma_\mathrm{gauss}^2}\right)\right] \, ,
    \label{eq:mod_rec}
\end{equation}
\enlargethispage*{1em}
where $\Delta z$ accounts for the overall shift and the last term for a Gaussian enhancement of the ionization history. According to \citealp{Mirpoorian:2024fka}, this parametrization is sufficiently representative of such inhomogeneous reionization scenarios and leads to similar CMB constraints as the full model. In this work, we adopt this simplified model -- a free-form reconstruction would be difficult due to the large and arbitrary number of free parameters. One should keep in mind that a primordial magnetic field model can still perform significantly better than this simple approximation. See also \citealp{Cyr-Racine:2021oal,Ge:2022qws,Greene:2023cro,Greene:2024qis} for an alternative proposal to shift recombination that is expected to better fit the data.

\vspace*{0.5\baselineskip}
\noindent \textbf{Analysis:} We implemented the 4-parameter model according to \cref{eq:mod_rec} in a modified version of {\sc class} \texttt{v3.3}.
 
\vspace*{0.5\baselineskip}
\noindent \textbf{Parameters and priors:} We assume flat priors on the four parameters in the ranges $\Delta z \ \in [-100, 100]$, $A_\mathrm{gauss} \in [0.01, 0.5]$, $z_\mathrm{gauss} \in [500, 1500]$, and $\sigma_\mathrm{gauss} \in [10,500]$.

\subsection{Free-streaming dark radiation [Group R]} 

\textbf{Motivation:} It is well known that the presence of additional mas   sless free-streaming relic particles enhances the Hubble rate in the early Universe, leading to a reduced sound horizon and, through degeneracies of the CMB, to an increase in the inferred Hubble constant~\cite{Planck:2018jri,Allali:2024cji,Gariazzo:2023hch,Saravanan:2025cyi,Bernal_2016}. The cosmological evolution of such relics is similar to that of massless free-streaming neutrinos. Their abundance can be quantified by the effective neutrino number, $N_{\mathrm{eff}}$, which is defined as the ratio of the energy density of all relativistic species but photons to that of a single neutrino species (computed in the instantaneous decoupling limit). Due to non-instantaneous decoupling and further quantum electrodynamics effects, the contribution from the three active neutrino species is close to  $N_{\mathrm{eff}}^{\mathrm{std}} = 3.044$ \cite{Froustey:2020mcq,Drewes:2024wbw} in the standard model, slightly higher than the na\"ive expectation of 3. If additional free-streaming massless relics are present (such as axions or sterile neutrinos), their energy density contributes to an excess 
$\Delta N_{\mathrm{eff}} \equiv N_{\mathrm{eff}} - N_{\mathrm{eff}}^{\mathrm{std}}$. We remain agnostic as to the specific species contributing to $N_\mathrm{eff}$.

\vspace*{0.5\baselineskip}
\noindent \textbf{Analysis:} The presence of free-streaming dark radiation (DR) is already implemented in the public {\sc class} code, which we employ here in its version \texttt{v3.3}.

\enlargethispage*{1em}
\vspace*{0.5\baselineskip}
\noindent \textbf{Parameters and Priors:} This model has only one extra parameter $N_{\rm ur}\equiv\Delta N_{\mathrm{eff}} = N_{\mathrm{eff}} - N_{\mathrm{eff}}^{\mathrm{std}} $. We impose a flat prior that ensures a positive contribution to the energy density, $N_{\rm ur}\in [0,1]$.

\subsection{Self-interacting  dark radiation [Group R]} 

\textbf{Motivation:} Two of the most impactful cosmological constraints on free-streaming DR come from Silk damping and the neutrino drag effect \cite{Hou:2011ec}. An enhanced radiation density increases the Hubble rate, which in turn decreases the sound horizon and allows for a larger $H_0$. This also impacts photon diffusion close to recombination and enhances Silk damping. Then, the relation between the angular sound horizon and the diffusion damping horizon shifts from its $\Lambda$CDM value, which leaves a measurable imprint on the CMB. However, if DR is strongly self-coupled, the anisotropic stress term in the Boltzmann hierarchy effectively vanishes, and DR behaves as a perfect relativistic fluid. It clusters more efficiently on small scales compared to free-streaming DR, thereby counteracting the increased amount of Silk damping. Furthermore, the sound speed of self-interacting DR remains close to that of the photon-baryon plasma. As a consequence, the neutrino drag effect becomes negligible, and the characteristic phase shift in the acoustic peaks is absent. Since these observational signatures are suppressed, cosmological data can typically accommodate a larger value of $N_{\rm eff}$ than in the standard free-streaming scenario.  Self-interacting DR is thus better able to reduce the Hubble tension.

\vspace*{0.5\baselineskip}
\noindent \textbf{Analysis:} The presence of self-interacting DR is already implemented in the public {\sc class} code, which we employ here in its \texttt{v3.3} version. The original implementation was performed in \texttt{v2.9} and follows \citealp{Lesgourgues:2015wza,Cyr-Racine:2015ihg,Archidiacono:2017slj,Archidiacono:2019wdp}.

\vspace*{0.5\baselineskip}
\noindent \textbf{Parameters and Priors:} This model has one extra parameter named $N^\mathrm{SIDR}_{\rm eff}$. We impose a flat prior that ensures positivity, $N^\mathrm{SIDR}_{\rm eff} \in [0,1]$.

\subsection{Wess-Zumino Dark Radiation [Group R]}

\textbf{Motivation:}
While self-interacting DR alleviates some of the observational consequences of increased Silk damping and neutrino drag, it still distorts the shape of the CMB spectrum compared to the $N_\mathrm{eff}=3.044$ limit, especially at large $\ell$. However, an increase in the DR abundance precisely as wavenumbers visible in the CMB enter the Hubble horizon can suppress the radiation impact on the high-$\ell$ angular power spectrum relative to the low-$\ell$ regime. For example, if the DR increases when scales corresponding to $\ell \sim 800$ enter the Hubble horizon ($z \sim 10^{4.5}$), then the Silk damping and neutrino drag at $\ell \gtrsim 800$ are not strongly enhanced, whereas the reduction of the sound horizon can still be significant. Additionally, this model has been shown to explain successfully a feature in the CMB temperature spectrum that may be interpreted as a slight tension between high and low multipoles, see \cite[Fig.~21]{Planck:2018vyg}. One way to motivate such an increase in the DR abundance at later times is through a dark sector particle becoming non-relativistic and decaying into DR. This is exactly the phenomenology that naturally arises in the supersymmetric Wess-Zumino model \cite{Wess:1974}, hence the name `Wess-Zumino Dark Radiation' (WZDR) adopted here.

\enlargethispage*{2em}
Since WZDR can decrease the sound horizon like any other dark radiation model while avoiding a few unwanted effects of dark radiation on the shape of the high-$\ell$ CMB spectrum, it has been proposed as a better solution to the Hubble tension, see \citealp{Aloni:2021eaq,Schoneberg:2021qvd,Schoneberg:2022grr,Joseph:2022jsf,Cvetko:2025kda,Meiers:2023gft,Sobotka:2023bzr,Joseph:2022yys,Bagherian:2024obh,Zhou:2024igb,Smith:2025zsg}. Additionally, \citealp{Aloni:2023tff,Allali:2021azp} proposed some mechanisms to generate the DR excess only after BBN. The WZDR model is parametrized through the transition redshift $z_t$ at which $N_\mathrm{eff}(z)$ experiences a step-like increase, the amount of DR after the step $N_\mathrm{wzdr}$, and a few other parameters that control the size of the step. Since \citealp{Schoneberg:2022grr} showed that varying the latter parameters does not make a significant difference in reducing the Hubble tension, we fix them like in the original WZDR proposal and assume a fermionic/bosonic pair with $r_g=\frac{7}{4}$ and $g_2=2$. 

\vspace*{0.5\baselineskip}
\noindent \textbf{Analysis:} We employ the publicly available implementation of the model available at \url{https://github.com/schoeneberg/class_public_versions} based on version \texttt{v3.3} of {\sc class}.

\pagebreak[10]
\vspace*{0.5\baselineskip}
\noindent \textbf{Parameters and priors:} Our realization of the WZDR model extends the $\Lambda$CDM model by two parameters, for which we assume flat priors in the ranges $N_\mathrm{wzdr} \in [0.01, 5]$ and $\log_{10}(z_t) \in [4.0, 4.6]$.

\subsection{Dark Radiation Matter Decoupling [Group R]}

\textbf{Motivation:} 
\citealp{Garny:2025kqj} recently discussed another model that features self-interacting dark radiation and an increase in the effective number of neutrino species prior to recombination, which is caused by a first-order phase transition. This model also features a dark matter component interacting with the SIDR at early times. This models builds upon a paradigm dubbed Hot New Early Dark Energy \cite{Niedermann:2021ijp,Niedermann:2021vgd,Garny:2024ums}.

In this model, the Dark sector includes two representations of a dark gauge group SU$(N)$, a dark Higgs field $\Psi$ (spin-$0$ $N$-multiplet) and a massive dark fermion $\chi$ (spin-$\frac{1}{2}$ $N$-multiplet). The dark sector Lagrangian reads
\begin{equation}
    \mathcal{L} = \bar \chi (i \gamma^\mu D_\mu - m_\chi ) \chi + \lvert D \Psi \rvert^2 - V_{\rm cl} (\Psi) - \frac{1}{4} F_{\mu\nu}^a F^{\mu \nu}_a \, ,
\end{equation}
where $F_{\mu\nu}^a$ is the field strength tensor of the dark gauge bosons, $D_{\mu}$ the covariant derivative, and $V_{\rm cl} (\Psi) = - \mu^2 \lvert \Psi \rvert^2 + \lambda \lvert \Psi \rvert^4 + V_0$ the classical Coleman-Weinberg potential of the dark Higgs.

At early times, including the BBN epoch, the SU$(N)$ symmetry is preserved. The dark sector consists, firstly, of a strongly-interacting dark radiation (SIDR) component composed of gauge and Higgs bosons, which remains sub-dominant compared to neutrinos, $\Delta N_\mathrm{eff} \ll 1$; and secondly, of an interacting dark matter (IDM) component $\chi$ which scatters over dark radiation and accounts for a small, constant fraction $f_\mathrm{idm}\ll 1$ of total dark matter. The dominant contribution of dark matter in the Universe, with fraction $(1-f_\mathrm{idm})$, is assumed to come as usual from a decoupled, cold dark matter component.

At some time between BBN and photon decoupling, the Higgs vacuum expectation value breaks the symmetry spontaneously into SU$(N-1)$ according to the Brout-Englert-Higgs mechanism \cite{Higgs:1964,Englert:641592,Kibble:1964}. This phase transition increases the temperature and entropy of the dark radiation component, which brings $\Delta N_\mathrm{eff}$ to sizeable values. Later on, the interaction between SIDR and IDM becomes inefficient, leading to a Dark Radiation Matter Decoupling (DRMD). After that time, SIDR subsists in the form of the massless gauge bosons of SU$(N-1)$ with a constant contribution $\Delta N_\mathrm{eff}^\mathrm{DMDR}$. The components of $\chi$ still contribute to dark matter, but act essentially as decoupled, cold species.

The combined effects of the increase in $\Delta N_\mathrm{eff}$ at the phase transition and SIDR-IDM interactions before dark decoupling have been shown in \citealp{Garny:2025kqj} to be highly promising for reducing the Hubble tension \cite{Garny:2025kqj}.

\noindent \textbf{Analysis:} To perform our analyses, we use the modified version of {\sc class} presented in \citealp{Garny:2025kqj}, which is publicly available at \url{https://github.com/NEDE-Cosmo/DRMD-CLASS}, and which we have updated to the version \texttt{v3.3} of {\sc class}.

 \enlargethispage*{2em}
\vspace*{0.5\baselineskip}
\noindent \textbf{Parameters and Priors:}
The \hotnede{} model has four additional free parameters compared to $\Lambda$CDM, namely the interacting dark matter fraction, $f_{\rm idm} \in [0,1]$, the effective number of SIDR species after BBN, $\Delta N^\mathrm{DRMD}_{\rm eff} \in [0, 3]$, the redshift at which the SIDR-IDM interaction rate becomes exponentially suppressed, $\log_{10}(z_{\rm stop}) \in [2, 5]$, and the initial value of the ratio between the interaction rate and the Hubble rate, $\log_{10} (\mathcal{G}/\mathcal{H})_{\rm ini} \in [2,14]$.
However, \citealp{Garny:2025kqj} noticed that this last parameter is unconstrained, which means that fixing its value does not affect the posteriors of other parameters.
As shown in \cref{fig:DRMD_free_G} of 
\cref{app:permodel},
we reach the same conclusion using the latest data, and, as a consequence, we consider only the first three additional parameters in this work.

\subsection{Sign-Switching cosmological constant [Group L]}

\textbf{Motivation:} One example of a post-recombination change that has been proposed to 
alleviate the Hubble tension is to reverse the sign of the cosmological constant at a fixed redshift, as
investigated for example in \citealp{Akarsu:2021fol,Ibarra-Uriondo:2026zbp,Bouhmadi-Lopez:2026ckz,Khandelwal:2026btl,Akarsu:2022typ,Akarsu:2023mfb,Paraskevas:2024ytz,Soriano:2025gxd,Bouhmadi-Lopez:2025ggl}. These analyses found that values of $H_0$ as large as $\sim 74\mathrm{km/s/Mpc}$ were compatible with some combinations of cosmological data.

\vspace*{0.5\baselineskip}
\noindent \textbf{Analysis:} We implemented in \text{CLASS} \texttt{v3.3} the simple prescription that $\Lambda$ changes of sign at redhsift $z_t$,
\begin{equation}
    \Lambda(z) = \Lambda \,\,\mathrm{sign}(z-z_t)~.
\end{equation}
We note that the precise modeling of the transition did not result in large differences in past analyses \cite{Ibarra-Uriondo:2026zbp,Bouhmadi-Lopez:2026ckz}. We plan to make this modification publicly available at \url{https://github.com/schoeneberg/class_public_versions}. 

\vspace*{0.5\baselineskip}
\noindent \textbf{Parameters and priors:}
The model is characterized by a single additional parameter compared to the $\Lambda$CDM case, $z_t$. We assume a flat prior on $z_t \in [1,10]$.

\subsection{Reionization agnostic [Group L]}\label{ssec:nosroll}

\textbf{Motivation:} 
Several authors have recently raised the possibility that the constraints on the reionization optical depth derived from Planck’s measurements of the reionization bump may not be reliable. In particular, \citealp{Allali:2025wwi,Jhaveri:2025neg,Sailer:2025lxj,Loverde:2024nfi} have examined the implications of this hypothesis for constraints on cosmological parameters and tensions emerging in $\Lambda$CDM, finding a decent reduction in multiple tensions simultaneously. Therefore, we consider that the use of the \texttt{SROLL2} lowl-$\ell$ EE polarization likelihood of \cref{ssec:data} is invalidated either by some unknown systematics, astrophysical foregrounds, or a more complex model of reionization. Strictly speaking, this case is not a `model', but rather a way to reanalyze the $\Lambda$CDM model with a reduced data set. However, to ensure a seamless discussion, we treat it on equal footing with our extended models. 

\vspace*{0.5\baselineskip}
\noindent \textbf{Analysis:} We use the public \texttt{v3.3} version of {\sc class} without further modifications. Instead, we remove the \texttt{SROLL2} likelihood of \cref{ssec:data}.

\vspace*{0.5\baselineskip}
\noindent \textbf{Parameters and priors:} This case does not include any additional parameters.

\subsection{Dark Matter-Dark Energy Interactions [Group L]}\label{ssec:idmde}

\textbf{Motivation:} Phenomenological models in which energy and momentum are transferred between dark matter and dark energy, as described, e.g., in \citealp{Gavela:2009cy,DiValentino:2017iww,DiValentino:2017oaw,DiValentino:2019ffd}, have been proposed to ease the Hubble tension. The energy-momentum transfer can be parameterized as 
\begin{eqnarray}
\nabla_\mu T^\mu_{{\rm (idm)}\nu} &=&Q \,u_{\nu}^{\rm (idm)}/a~, \\
\nabla_\mu T^\mu_{{\rm (ide)}\nu} &=&-Q \,u_{\nu}^{\rm (idm)}/a~,
\end{eqnarray}
where $T^\mu_{{\rm (idm)}\nu}$ and $T^\mu_{{\rm (ide)}\nu}$ are the dark matter dark energy energy-momentum tensors, $Q$ is the interaction rate, and $u_{\nu}^{\rm (idm)}$ is the four-velocity of the dark matter fluid. We assume that  the interaction rate is proportional to the energy density of dark energy, $\rho_\mathrm{ide}$, and parametrize it as
\begin{equation}
Q=\xi_{\rm ide} \mathcal{H} \rho_{\rm ide}~,
\end{equation}
where $\xi_{ide}$ is a dimensionless coupling constant. The background evolution is then given by
\begin{eqnarray}
\dot{{\rho}}_{\rm idm}+3{\mathcal H}{\rho}_{\rm idm}= \xi{\mathcal H}{\rho}_{\rm ide}~, \\
\dot{{\rho}}_{\rm ide}+3{\mathcal H}(1+w_{\rm ide}){\rho}_{\rm ide}= -\xi{\mathcal
  H}{\rho}_{\rm ide}~,
\end{eqnarray}  
where $w_{\rm ide}$ is the equation-of-state parameter of the interacting dark energy fluid. We refer to, e.g., \citealp{Gavela:2009cy,DiValentino:2017iww,DiValentino:2017oaw,DiValentino:2019ffd} for a complete description of the models and the perturbation equations.

\vspace*{0.5\baselineskip}
\noindent \textbf{Analysis:}
We modify {\sc class} \texttt{v3.3} to implement the background and perturbation equations of \citealp{DiValentino:2019jae}, following the description in their Section II. We plan to make this modification publicly available at \url{https://github.com/schoeneberg/class_public_versions}.

\vspace*{0.5\baselineskip}
\noindent \textbf{Parameters and Priors:} We fix the DE equation-of-state parameter to $w_{\rm ide}=-0.999$ to avoid the gravitational instability that appears for $w_{\rm ide}=-1$. This value was already assumed in \citealp{DiValentino:2019jae} when showing that this model may help solving the Hubble tension. Once the DE equation of state is fixed, the model features only one additional parameter compared to $\Lambda$CDM, the interaction rate amplitude $\xi_{\rm ide}$. We assume a flat prior with boundaries $\xi_{\rm ide}\in [-2,0]$. Since the CMB is most sensitive to the dark matter density around/before recombination (as opposed to today), we introduce the effective DM density 
$\Omega_\text{idm,eff}\equiv\Omega_\text{idm}+\xi_\text{ide}\Omega_\text{ide}/(3w_\text{ide}+\xi_\text{ide})$, and we use in our MCMC analysis the parameter $\Omega_\mathrm{idm,eff}h^2$ in replacement of $\Omega_\mathrm{cdm} h^2$.
A dynamical system analysis of the evolution equations has shown that  certain regions of the parameter space can lead to a pathological behavior with an exponential growth of density perturbations. We do not use here any full-shape LSS data, which would forbid this behavior. Instead, we introduce a sharp cut-off in our prior, $\xi_\text{ide} < 3 \Omega_\text{idm} /\Omega_\text{ide}$, to exclude this nonphysical region.

\enlargethispage*{1em}
\section{Tension and model preference metrics}\label{sec:metrics}
As outlined in \citealp{Schoneberg:2021qvd,Khalife:2023qbu}, two questions must be prioritized to qualify the success of a model:
\begin{enumerate}
    \item How well does the model resolve the tension?
    \item How good is the fit of the model to the combined datasets compared to $\Lambda$CDM?
\end{enumerate}
While it might seem like both questions are naturally related (a model capable of easing the tension is expected to fit the combined datasets very well), there are many cases in which this relation is broken. For example, fixing the CDM abundance to $\Omega_\mathrm{cdm} h^2=0.11$ in a $\Lambda$CDM model gives a model in which the tension gets eased, but the fit to the combined dataset is extremely bad \cite{Schoneberg:2021qvd}. In principle, a model could also give a much better fit to some observables (like the CMB) without improving the tension by much, though in practice this is not often the case. In this work, we employ multiple different metrics described in the following to answer each of the two questions. These metrics provide largely consistent conclusions in most cases, except for a few exceptions which we highlight in the main text.

Beyond choosing some relevant questions to quantify the tension, one has to pick up a quantitative way to answer them. In statistics, these issues are tackled differently by two major schools of thought, the Frequentist and Bayesian schools. Bayesian statistics is ideally designed to deal with cosmological observations stemming from a single realization of an underlying stochastic theory. However, it is also subject to somewhat arbitrary choices of priors. To remain agnostic, general and as informative as possible, we will address each of the two questions with both Bayesian and Frequentist statistical tools.

In the following description of tension metrics, the letter $A$ refers to a baseline dataset and $B$ to a dataset whose level of tension with $A$ is to be evaluated. In our case, $A$ typically corresponds to a combination of CMB, SN, and BAO data, while $B$ corresponds to measurements of the Hubble constant $H_0$ or the supernova calibration parameter $M_B$\,. In \cref{sec:tension_metrics,sec:model_fitting_metrics}, we review the metrics that address questions 1 and 2, respectively.

\subsection{Tension metrics} \label{sec:tension_metrics}
\subsubsection{The Gaussian tension}\label{ssec:gaussian}
The Gaussian tension ($Q_\mathrm{DM}$ in the notation of \cite{Raveri:2021wfz}) is computed simply from the parameter difference weighted by the combined covariance matrix
\begin{equation}
    \Delta_\mathrm{Gauss} = \sqrt{(\mu_A-\mu_B)^T (C_A + C_B)^{-1} (\mu_A-\mu_B)}
\end{equation}
with mean $\mu$ and covariance $C$. When applied to a single parameter (like $H_0$ or $M_\mathrm{B}$) this metric reduces to $|\mu_A-\mu_B|/\sqrt{\sigma_A^2+\sigma_B^2}$. This tension metric typically fails to capture cases with non-Gaussian posteriors, as it completely relies on the first two moments of a posterior distribution to characterize it fully. We keep it for reference.

\subsubsection{Frequentist non-Gaussian tension metric: The maximum a posteriori distance}\label{ssec:qdmap}
In a Frequentist analysis, this limitation can be resolved by estimating the difference of the maximum a posteriori (DMAP) of the parameter values ($Q_\mathrm{DMAP}$ in \citealp{Raveri:2021wfz}). The DMAP evaluates the square root of the difference in maximum likelihoods for the data $A$ alone and for the combination of $A+B$:
\begin{equation}
    \Delta_\mathrm{DMAP} = \sqrt{\chi^2_{{\rm min},A+B} - \chi^2_{{\rm min},A}-\chi^2_{{\rm min},B}}~.
\end{equation}
In the case of purely Gaussian posteriors this criterion reduces to the $\Delta_\mathrm{Gauss}$ criterion, but for other distributions it provides a suitable generalization that relies only on the underlying likelihoods and not the prior bounds.

\subsubsection{Bayesian non-Gaussian tension metric: The parameter shift}\label{ssec:kde}
As an alternative to the Frequentist $\Delta_\mathrm{DMAP}$ metric, we also compute the Bayesian {\it parameter shift} statistic, which generalizes the Gaussian tension metric to non-Gaussian posteriors (used also in \citealp{Raveri:2021wfz,Leizerovich:2023qqt,Khalife:2023qbu}, and dubbed $Q_{\rm MPCL}$ in \citealp{Khalife:2023qbu}). 
If one draws two independent realizations $(\theta, \theta')$ of the probability $p_A(\theta)$ of the dataset $A$ and of the probability $p_B(\theta')$ of the dataset $B$, the probability distribution of the parameter difference $\Delta \theta = \theta-\theta'$ is given by:
\begin{equation}
p(\Delta\theta)=\int p_A(\theta)\, p_B(\theta-\Delta\theta)\,\mathrm{d}\theta\,.
\end{equation}
Then, the probability that there is a tension between the data sets (given the model) can be reformulated as the probability that the model provides a better description of the combined data set when different parameters are used ($\Delta \theta \neq 0$) rather than the same parameters ($\Delta \theta = 0$). 

\noindent This is obtained simply by integrating $p(\Delta \theta)$ over the region where it is larger than $p(\Delta \theta= 0)$:
\begin{equation} \label{eq:parameter_shift_prob}
    p(\mathrm{shift~is~required)} = \int_{p(\Delta \theta) > p(\Delta \theta=0)} p(\Delta \theta) \mathrm{d}\Delta\theta~.
\end{equation}
To facilitate comparison with other tension metrics, we convert the above probability into a Gaussian significance,
\begin{equation}
    \Delta_\mathrm{shift} = \sqrt{2} \,\,\mathrm{erf}^{-1}
    \left(
    p(\mathrm{shift~is~required)}
    \right)~,
\end{equation}
with $\mathrm{erf}$ denoting the error function. For Gaussian posteriors, $\Delta_\mathrm{shift}$ also reduces to the $\Delta_\mathrm{Gauss}$ criterion. 
To compute the parameter shift metric, we use the \texttt{tensiometer} python package\footnote{\url{https://github.com/mraveri/tensiometer}} \cite{Raveri:2019}.

Unlike the DMAP criterion, which only uses the underlying likelihoods for the comparison, the parameter shift criterion relies on integrated posterior volumes. Consequently, it can be affected by posterior-volume  and projection effects that are absent in the likelihood-based metric.
However, it has the advantage of naturally penalizing more complex models in a Bayesian manner.

\subsection{Model comparison metrics} \label{sec:model_fitting_metrics}
\subsubsection{Frequentist approach: Akaike information criterion}\label{ssec:aic}

The Akaike information criterion (AIC) provides a simple measure of model performance that balances goodness of fit against model complexity. Inspired by information-theoretic considerations and the principle of Occam's razor, the AIC favors models that achieve a better fit to the data while remaining as simple as possible. It is defined as
\begin{equation}
    \mathrm{AIC} = \chi^2_{{\rm min},A+B} + 2 \,N_\mathrm{param}~,
\end{equation}
where $N_\mathrm{param}$ is the number of free parameters for a given model and $\chi^2_{\mathrm{min},A+B}$ denotes the minimum $\chi^2$ obtained from the combined $A+B$ dataset. The second term penalizes models with larger numbers of free parameters, thereby discouraging unnecessary complexity.

To assess whether an extended model is preferred over the standard $\Lambda$CDM cosmology, we compute the difference
\begin{equation}
    \Delta \mathrm{AIC}
    =
    \mathrm{AIC}_{\mathrm{model}}
    -
    \mathrm{AIC}_{\Lambda\mathrm{CDM}} \, .
\end{equation}
By evaluating the AIC criterion on the combined $A+B$ dataset, we essentially ask if the model is a better description of the joint data than the $\Lambda$CDM standard cosmological model. 
The value of $\Delta \mathrm{AIC}$ should at least be negative (indicating a better fit) and optimally be below some threshold. The rule of thumb derived from information theory is that $\Delta \mathrm{AIC} < -10$ can be interpreted as strong support for the model with the lower AIC. 

\subsubsection{Bayesian approach: the Bayes Factor}\label{ssec:BF}
The Bayes factor provides a Bayesian measure of the relative support that the data gives to two competing models.
It is computed from the ratio of their Bayesian evidences (or difference of log-evidences), which we estimate using the \texttt{MCEvidence} code\footnote{\url{https://github.com/yabebalFantaye/MCEvidence}} \cite{Heavens:2017afc}. Assuming equal prior probabilities for the models, the Bayes factor is equal to the posterior odds ratio and is commonly interpreted using Jeffreys' scale \cite{Trotta:2008qt},
\begin{equation}
    {\rm BF} = \frac{p(D|M_1)}{p(D|M_2)} \qquad \Rightarrow \qquad \ln {\rm BF} = \ln Z_{M_1} - \ln Z_{M_2} \equiv \ln p(D|M_1) - \ln p(D|M_2)~,
\end{equation}
where $p(D|M)\equiv Z_M$ denotes the Bayesian evidence for model $M$ given the dataset $D$. 
The evidence is obtained by marginalizing the likelihood over the model parameter space,
\begin{equation}
    Z_M
    =
    \int
    \mathcal{L}(D|\theta,M)\,
    \pi(\theta|M)\,
    \mathrm{d}\theta
    \, ,
\end{equation}
where $\mathcal{L}$ is the likelihood and $\pi$ the prior distribution.

Like the AIC, the Bayes factor rewards models that provide a good description of the data while disfavoring unnecessarily complex models. However, unlike the AIC, this Occam penalty arises naturally from the evidence integral through the averaging of the likelihood over the prior volume. Consequently, the Bayes factor depends explicitly on the adopted parameter priors. We outline the priors used in this work in \cref{sec:models}. Since some of the considered extensions lack a compelling theoretical motivation for a unique prior choice, the resulting Bayes factors should be interpreted in the context of the prior assumptions adopted in this analysis.

\subsection{Thresholds}\label{ssec:thresholds}

To determine which models will advance to the knockout round, we set thresholds on each metric.
These thresholds are motivated by requiring a significant improvement in the combined fit (with the $M_B$ prior), either by reaching $-\Delta \mathrm{AIC} > 10$ or $\ln {\rm BF} > 3$.

\section{The group stage}\label{sec:competition}

\begin{figure}[tp]
    \centering
    \includegraphics[width=0.95\linewidth]{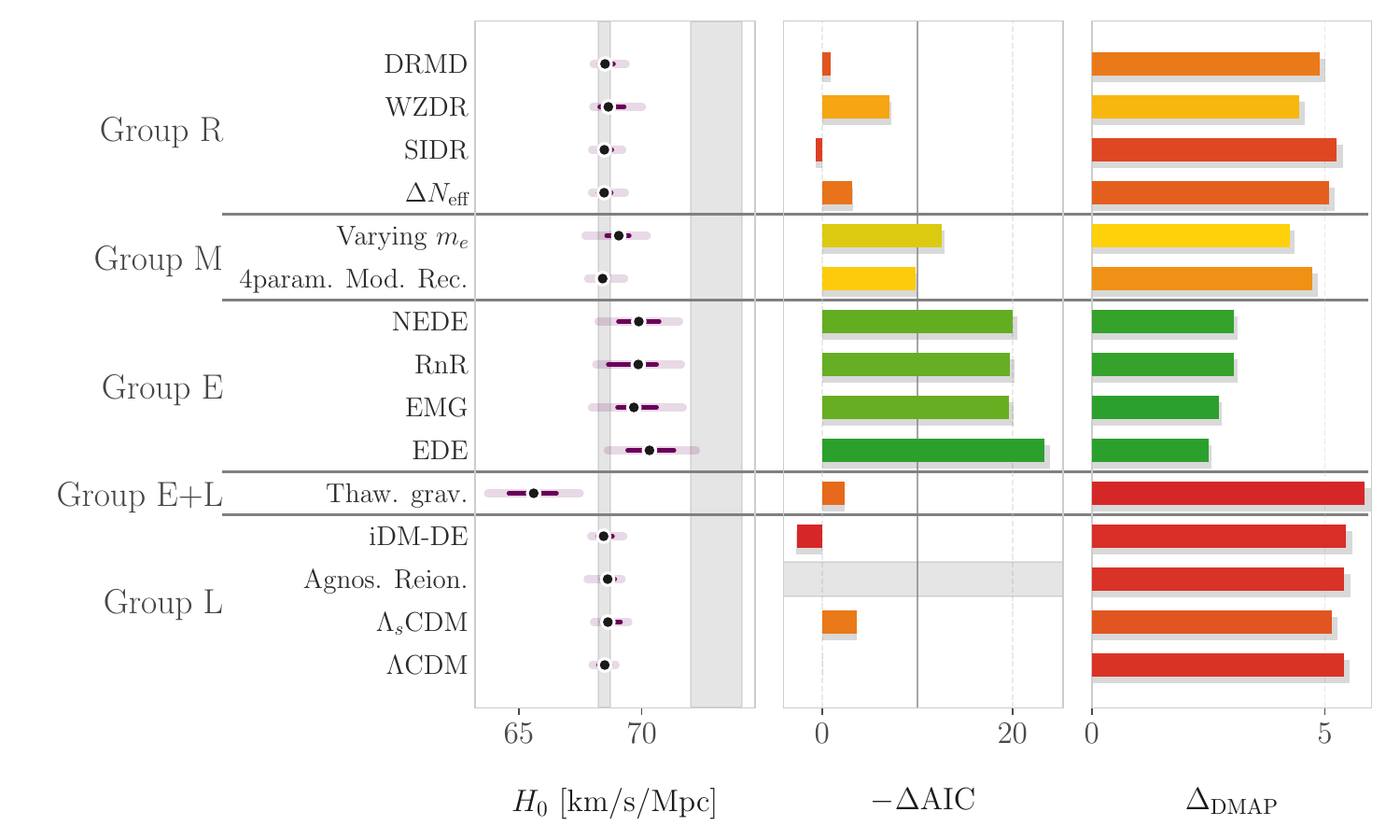}
    \caption{\label{fig:summary_frequentist} Frequentist summary of the group-stage performance for the baseline dataset. The left panel shows the Frequentist $1\sigma$ and $2\sigma$ confidence intervals on $H_0$ based on the profile likelihood ratio method \cite{Herold:2024enb}, compared to the $\Lambda$CDM CMB+BAO+SN and local measurement bands (gray). The middle panel shows the AIC improvement relative to $\Lambda$CDM, plotted as $-\Delta\mathrm{AIC}$ with $\Delta\mathrm{AIC}\equiv \mathrm{AIC}_{\rm model}-\mathrm{AIC}_{\Lambda{\rm CDM}}$, so that larger positive bars indicate a stronger preference over $\Lambda$CDM. The right panel shows the non-Gaussian DMAP tension with the local calibration dataset, with smaller values indicating better agreement. The best-performing models are early dark energy and modified gravity models (group E), while models with enhanced radiation and late-time modifications (groups R, L, E+L) remain close to the $\Lambda$CDM tension.}
\end{figure}

\begin{figure}[!t]
    \centering
    \includegraphics[width=0.95\linewidth]{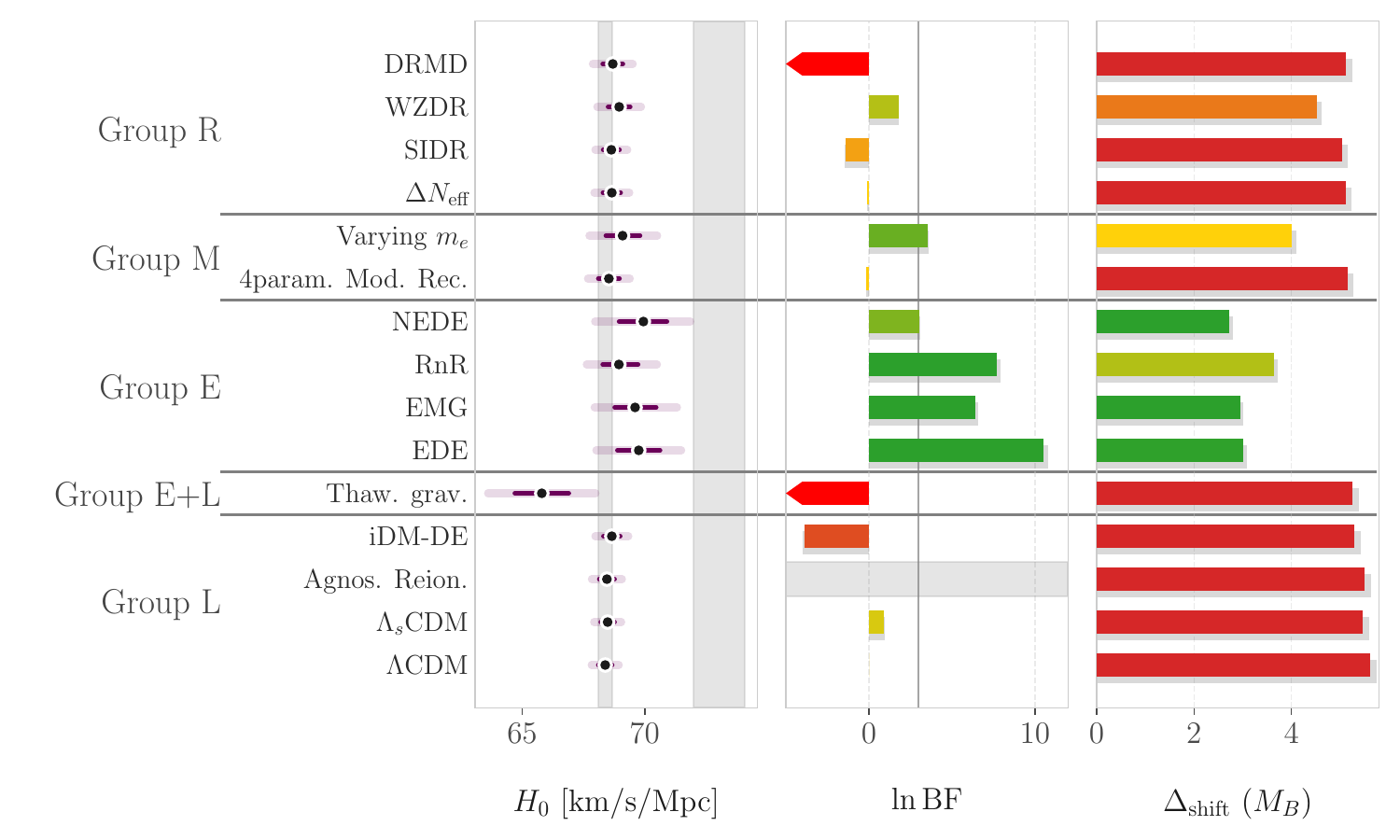}
    \caption{Bayesian summary of the group-stage performance for the baseline dataset. The left panel shows the marginalized $H_0$ credible intervals from CMB+BAO+SN, compared to the $\Lambda$CDM CMB+BAO+SN and local measurement bands (gray). The middle and right panels show the tension with the calibrated supernova absolute magnitude $M_B$ estimated with a parameter-shift statistic and with a Gaussian approximation. The same hierarchy as in \cref{fig:summary_frequentist} is visible: early dark energy and modified gravity models (group E) move $H_0$ furthest toward the local value and yield the smallest $M_B$ tension, whereas models with enhanced radiation, late-time modifications, and modified recombination (groups R, L, M) generally leave a larger residual discrepancy.}
    \label{fig:summary_bayesian}
\end{figure}

We first confront each model from \cref{sec:models} to the data outlined below in \cref{ssec:data}. We grade the model performance using the metrics outlined in \cref{sec:metrics}. Other data combinations will be considered separately in \cref{sec:golden,ssec:noACT}. Further figures including the results for considered each model and data combination can be found at \citealp{ZENODO}.\footnote{The direct link is \url{https://doi.org/10.5281/zenodo.21277319}.}

\subsection{The data}\label{ssec:data}

\begin{itemize}
    \item \textbf{CMB:} For the Planck primary anisotropies, we consider the Planck Public Release 4 (PR4) analysed with the \texttt{NPIPE} processing pipeline~\cite{Planck:2020olo} and described by the \texttt{CamSpec v12.7} likelihood at $\ell>30$~\cite{Efstathiou:2019mdh,Rosenberg:2022sdy}, together with the low-$\ell$ \texttt{plik} TT likelihood (\texttt{Commander})~\cite{Planck:2019nip} and the \texttt{SROLL2} EE likelihood~\cite{Delouis:2019bub,Pagano:2019tci} at $\ell<30$. For ACT DR6 ~\cite{ACT:2025blo} and SPT-3G D1~\cite{SPT-3G:2025bzu} primary anisotropies, we use the corresponding `lite' likelihoods (marginalized internally over nuisance parameters) as implemented in the \texttt{candl}\footnote{\url{https://github.com/Lbalkenhol/candl}} framework \cite{Balkenhol:2024sbv}. We cut the \texttt{CamSpec} likelihood to $\ell<1000$ in TT,TE and $\ell<600$ in EE (following the prescription adopted in \citealp{ACT:2025tim} to avoid double-counting).\footnote{Given that the Planck likelihoods are used in a multipole-cut form, it is interesting to check whether some of the nuisance parameters become unconstrained. We analyze the extent of and correlation with the cosmological parameters in \cref{app:nuisance_corr} and observe that there are no strong correlations or unreasonable nuisance parameter values.} For CMB lensing, we use the SPT-3G 2yr lensing likelihood~\cite{SPT-3G:2024atg} and the ACT DR6 lensing likelihood~\cite{ACT:2023kun} in its combined ACT+Planck variant.
    \item \textbf{BAO:} We use DESI DR2 BAO data~\cite{DESI:2025zgx,DESI:2025zpo} from bright galaxies ($0.1<z<0.4$), luminous red galaxies ($0.4<z<1.1$), emission line galaxies ($0.8<z<1.6$), quasars ($0.8<z<2.1$), and Lyman$-\alpha$ forest quasars ($1.8<z<4.2$). 
    \item \textbf{SN:} We also involve (uncalibrated) luminosity distance measurements from the Type Ia supernovae (SNe Ia) of the Pantheon+ catalog in the range ${0.01<z<2.3}$ \cite{Brout:2022vxf}.
    \item \textbf{$M_B$ prior:} We adopt a Gaussian prior $M_B =  -19.253 \pm 0.027$ on the standardized SNIa absolute magnitude $M_B$ given in \cite{Riess:2021jrx}. Using a prior on this variable is motivated in \citealp{Camarena:2021jlr} as a preferable alternative to a direct prior on $H_0$\,. In the SH0ES collaboration results \cite{Riess:2021jrx} this prior gives $H_0 = (73.04 \pm 1.04) \mathrm{km/s/Mpc}$, but in our case we find $H_0 = (73.26 \pm 0.98) \mathrm{km/s/Mpc}$ when combining with the full Pantheon+ dataset described above.\footnote{The slight deviation in value and uncertainty can be explained by two main differences compared to the analysis of \citealp{Riess:2021jrx}; they limit the supernova range to $z<0.15$ and fix $q_0 = -0.55$.}
\end{itemize}
In our baseline analysis, we only employ the combination CMB+BAO+SN. Other data combinations will be considered later but only for finalist models, including in particular BBN constraints related to the discussion in \citealp{Giovanetti:2026aku}.

\begin{table}[tp]
\centering
\begin{tabular}{cccc}
\toprule
Model & $\Delta_\mathrm{DMAP}$ & $\Delta_\mathrm{DMAP} (+\Omega_k)$ & $\Delta_\mathrm{DMAP} (+w_0 , w_a)$ \\
\midrule \arrayrulecolor[HTML]{CCCCCC} 
\lcdm & $5.41 \sigma$ & $5.26 \sigma$ & $5.48 \sigma$ \\
\switch & $5.15 \sigma$ &  $5.15 \sigma$ & --- \\
\nosroll & $5.41 \sigma$ & $5.23 \sigma$ & $5.67 \sigma$  \\
\idedm & $5.45 \sigma$ & $5.23 \sigma$ & --- \\ \midrule
\thawing & $5.85 \sigma$ & $5.88 \sigma$ & --- \\ \midrule
\ede & {\cellcolor[HTML]{FFB3B3}} $2.51 \sigma$ & {\cellcolor[HTML]{FFB3B3}} $2.52 \sigma$ & {\cellcolor[HTML]{FFB3B3}} $3.24 \sigma$ \\
\emg & {\cellcolor[HTML]{FFB3B3}} $2.72 \sigma$ & {\cellcolor[HTML]{FFB3B3}} $2.89 \sigma$ & --- \\
\rnr & {\cellcolor[HTML]{FFB3B3}} $3.05 \sigma$ & {\cellcolor[HTML]{FFB3B3}} $2.63 \sigma$ & --- \\
\coldnede & {\cellcolor[HTML]{FFB3B3}} $3.06 \sigma$ & {\cellcolor[HTML]{FFB3B3}} $2.87 \sigma$ & {\cellcolor[HTML]{FFB3B3}} $3.45 \sigma$ \\ \midrule
\modrec & $4.72 \sigma$ & $4.48 \sigma$ & $5.43 \sigma$ \\
\me & $4.25 \sigma$ & $3.54 \sigma$ & $4.74 \sigma$ \\ \midrule
\neff & $5.09 \sigma$ & $5.04 \sigma$ & $5.29 \sigma$ \\
\sidr & $5.26 \sigma$ & $5.19 \sigma$ & $5.43 \sigma$ \\
\wzdr & $4.45 \sigma$ & $4.46 \sigma$ & $4.90 \sigma$ \\
\hotnede & $4.90 \sigma$ & $4.87 \sigma$ & $4.90 \sigma$ \\\arrayrulecolor{black}
\bottomrule
\end{tabular}
\caption{\label{tab:baseline_QDMAP} For each model in the competition (including their extensions through curvature, $+\Omega_k$, or CPL dark energy, $+w_0, w_a$), Frequentist non-Gaussian tension metric $\Delta_\mathrm{DMAP}$ as defined in \cref{ssec:qdmap}. Cells with a $\Delta_\mathrm{DMAP}<3.5\sigma$ are highlighted in red.}
\end{table}

\begin{table}[tp]
\resizebox{\textwidth}{!}{
\begin{tabular}{ccccccc}
\toprule
Model & $\Delta_\mathrm{shift}$ & $\Delta_\mathrm{Gauss}$ & $\Delta_\mathrm{shift}$ ($+\Omega_k$) & $\Delta_\mathrm{Gauss}$ ($+\Omega_k$) & $\Delta_\mathrm{shift}$ ($+w_0, w_a$) & $\Delta_\mathrm{Gauss}$ ($+w_0, w_a$) \\
\midrule \arrayrulecolor[HTML]{CCCCCC}
\lcdm & $4.3 \sigma$ & $4.3 \sigma$ & $4.1 \sigma$ & $4.1 \sigma$ & $4.6 \sigma$ & $4.6 \sigma$ \\
\switch & $4.2 \sigma$ & $4.2 \sigma$ & $4.1 \sigma$ & $4.1 \sigma$ & --- & --- \\
\nosroll & $4.2 \sigma$ & $4.2 \sigma$ & $4.2 \sigma$ & $4.2 \sigma$ & $4.8 \sigma$ & $4.6 \sigma$ \\
\idedm & $3.9 \sigma$ & $4.0 \sigma$ & $3.8 \sigma$ & $3.8 \sigma$ & --- & --- \\ \midrule
\thawing & $5.0 \sigma$ & $4.9 \sigma$ & $5.0 \sigma$ & $4.9 \sigma$ & --- & --- \\ \midrule
\ede & {\cellcolor[HTML]{FFB3B3}} $2.6 \sigma$ & {\cellcolor[HTML]{FFB3B3}} $2.7 \sigma$ & {\cellcolor[HTML]{FFB3B3}} $2.4 \sigma$ & {\cellcolor[HTML]{FFB3B3}} $2.5 \sigma$ & {\cellcolor[HTML]{FFB3B3}} $3.4 \sigma$ & {\cellcolor[HTML]{FFB3B3}} $3.4 \sigma$ \\
\emg & {\cellcolor[HTML]{FFB3B3}} $2.4 \sigma$ & {\cellcolor[HTML]{FFB3B3}} $2.6 \sigma$ & {\cellcolor[HTML]{FFB3B3}} $2.7 \sigma$ & {\cellcolor[HTML]{FFB3B3}} $2.7 \sigma$ & --- & --- \\
\rnr & {\cellcolor[HTML]{FFB3B3}} $3.0 \sigma$ & {\cellcolor[HTML]{FFB3B3}} $3.2 \sigma$ & {\cellcolor[HTML]{FFB3B3}} $2.6 \sigma$ & {\cellcolor[HTML]{FFB3B3}} $2.7 \sigma$ & --- & --- \\
\coldnede & {\cellcolor[HTML]{FFB3B3}} $2.2 \sigma$ & {\cellcolor[HTML]{FFB3B3}} $2.3 \sigma$ & {\cellcolor[HTML]{FFB3B3}} $2.2 \sigma$ & {\cellcolor[HTML]{FFB3B3}} $2.2 \sigma$ & {\cellcolor[HTML]{FFB3B3}} $2.9 \sigma$ & {\cellcolor[HTML]{FFB3B3}} $3.1 \sigma$ \\ \midrule
\modrec & $4.0 \sigma$ & $4.0 \sigma$ & $4.0 \sigma$ & $3.9 \sigma$ & $4.7 \sigma$ & $4.6 \sigma$ \\
\me & {\cellcolor[HTML]{FFB3B3}} $3.1 \sigma$ & {\cellcolor[HTML]{FFB3B3}} $3.2 \sigma$ & {\cellcolor[HTML]{FFB3B3}} $3.1 \sigma$ & {\cellcolor[HTML]{FFB3B3}} $3.3 \sigma$ & $3.9 \sigma$ & $4.2 \sigma$ \\ \midrule
\neff & $3.9 \sigma$ & $4.0 \sigma$ & $3.9 \sigma$ & $3.8 \sigma$ & $4.3 \sigma$ & $4.4 \sigma$ \\
\sidr & $4.0 \sigma$ & $4.0 \sigma$ & $3.8 \sigma$ & $3.9 \sigma$ & $4.4 \sigma$ & $4.5 \sigma$ \\
\wzdr & $3.5 \sigma$ & $3.6 \sigma$ & $3.5 \sigma$ & $3.6 \sigma$ & $4.0 \sigma$ & $4.1 \sigma$ \\
\hotnede & $3.8 \sigma$ & $3.9 \sigma$ & $3.8 \sigma$ & $3.8 \sigma$ & $4.3 \sigma$ & $4.3 \sigma$ \\
\arrayrulecolor{black}\bottomrule
\end{tabular}
}
\caption{\label{tab:baseline_tension} For the same models as in~\cref{tab:baseline_QDMAP}, tension in the Hubble constant $H_0$ as assessed through the Bayesian metrics of \cref{ssec:gaussian,ssec:kde}. Cells with $\Delta_\mathrm{shift}<3.5\sigma$ or $\Delta_\mathrm{Gauss}<3.5\sigma$ are highlighted.}
\end{table}

\begin{table}[tp]
\resizebox{\textwidth}{!}{
\begin{tabular}{ccccccc}
\toprule
Model & $\Delta_\mathrm{shift}$ & $\Delta_\mathrm{Gauss}$ & $\Delta_\mathrm{shift}$ ($+\Omega_k$) & $\Delta_\mathrm{Gauss}$ ($+\Omega_k$) & $\Delta_\mathrm{shift}$ ($+w_0, w_a$) & $\Delta_\mathrm{Gauss}$ ($+w_0, w_a$) \\
\midrule  \arrayrulecolor[HTML]{CCCCCC}
\lcdm & $5.6 \sigma$ & $5.6 \sigma$ & $5.3 \sigma$ & $5.3 \sigma$ & $5.4 \sigma$ & $5.4 \sigma$ \\
\switch & $5.5 \sigma$ & $5.5 \sigma$ & $5.3 \sigma$ & $5.3 \sigma$ & --- & --- \\
\nosroll & $5.5 \sigma$ & $5.5 \sigma$ & $5.4 \sigma$ & $5.4 \sigma$ & $5.5 \sigma$ & $5.5 \sigma$ \\
\idedm & $5.4 \sigma$ & $5.3 \sigma$ & $5.1 \sigma$ & $5.1 \sigma$ & --- & --- \\ \midrule
\thawing & $5.2 \sigma$ & $5.9 \sigma$ & $5.8 \sigma$ & $5.8 \sigma$ & --- & --- \\ \midrule
\ede & {\cellcolor[HTML]{FFB3B3}} $3.0 \sigma$ & {\cellcolor[HTML]{FFB3B3}} $3.1 \sigma$ & {\cellcolor[HTML]{FFB3B3}} $2.8 \sigma$ & {\cellcolor[HTML]{FFB3B3}} $2.9 \sigma$ & {\cellcolor[HTML]{FFB3B3}} $3.2 \sigma$ & {\cellcolor[HTML]{FFB3B3}} $3.4 \sigma$ \\
\emg & {\cellcolor[HTML]{FFB3B3}} $2.9 \sigma$ & {\cellcolor[HTML]{FFB3B3}} $3.2 \sigma$ & {\cellcolor[HTML]{FFB3B3}} $3.4 \sigma$ & {\cellcolor[HTML]{FFB3B3}} $3.4 \sigma$ & --- & --- \\
\rnr & $3.6 \sigma$ & $4.0 \sigma$ & {\cellcolor[HTML]{FFB3B3}} $3.1 \sigma$ & {\cellcolor[HTML]{FFB3B3}} $3.3 \sigma$ & --- & --- \\
\coldnede & {\cellcolor[HTML]{FFB3B3}} $2.7 \sigma$ & {\cellcolor[HTML]{FFB3B3}} $2.8 \sigma$ & {\cellcolor[HTML]{FFB3B3}} $2.7 \sigma$ & {\cellcolor[HTML]{FFB3B3}} $2.7 \sigma$ & {\cellcolor[HTML]{FFB3B3}} $3.1 \sigma$ & {\cellcolor[HTML]{FFB3B3}} $3.3 \sigma$ \\ \midrule
\modrec & $5.2 \sigma$ & $5.1 \sigma$ & $5.0 \sigma$ & $5.0 \sigma$ & $5.4 \sigma$ & $5.3 \sigma$ \\
\me & $4.0 \sigma$ & $4.0 \sigma$ & $3.7 \sigma$ & $4.1 \sigma$ & $4.5 \sigma$ & $4.6 \sigma$ \\ \midrule
\neff & $5.1 \sigma$ & $5.1 \sigma$ & $4.9 \sigma$ & $4.9 \sigma$ & $5.4 \sigma$ & $5.1 \sigma$ \\
\sidr & $5.0 \sigma$ & $5.2 \sigma$ & $5.0 \sigma$ & $5.0 \sigma$ & $5.6 \sigma$ & $5.2 \sigma$ \\
\wzdr & $4.5 \sigma$ & $4.6 \sigma$ & $4.4 \sigma$ & $4.6 \sigma$ & $4.7 \sigma$ & $4.7 \sigma$ \\
\hotnede & $5.1 \sigma$ & $5.0 \sigma$ & $4.8 \sigma$ & $4.8 \sigma$ & $5.3 \sigma$ & $5.0 \sigma$ \\
\arrayrulecolor{black}\bottomrule
\end{tabular}
}
\caption{\label{tab:baseline_tension_2} Same as \cref{tab:baseline_tension} but showing the results for the standardized SNIa absolute magnitude $M_B$ instead of $H_0$. This tension also incorporates effects beyond the pure Hubble tension, such as the calibration mismatch between BAO and SNeIa data.}
\end{table}

\begin{table}[tp]
\resizebox{\textwidth}{!}{
\begin{tabular}{cccccccccc}
\toprule
Model Name & \multicolumn{3}{c}{Total $\chi^2$} & \multicolumn{3}{c}{$\Delta \chi^2$} & \multicolumn{3}{c}{-$\Delta \mathrm{AIC}$} \\
 &  & +$\Omega_k$ & +$w_0$, $w_a$ &  & +$\Omega_k$ & +$w_0$, $w_a$ &  & +$\Omega_k$ & +$w_0$, $w_a$ \\
\midrule \arrayrulecolor[HTML]{CCCCCC} 
\lcdm & 5970.34 & 5966.50 & 5961.64 & 0.00 & 0.00 & 0.00 & 0.00 & 0.00 & 0.00 \\
\switch & 5964.74 & 5963.42 & --- & -5.60 & -3.08 & --- & 3.60 & 1.08 & --- \\
\nosroll &5575.66 & 5573.22 & 5570.98 & -394.68 & -393.28 & -390.66 & --- & --- & ---  \\
\idedm & 5969.01 & 5966.40 & --- & -1.33 & -0.10 & --- & -2.67 & -3.90 & --- \\ \midrule
\thawing & 5964.00 & 5964.44 & --- & -6.34 & -2.06 & --- & 2.34 & -1.94 & --- \\ \midrule
\ede & 5940.94 & 5940.76 & 5938.60 & -29.40 & -25.74 & -23.04 & {\cellcolor[HTML]{FFB3B3}} 23.40 & {\cellcolor[HTML]{FFB3B3}} 19.74 & {\cellcolor[HTML]{FFB3B3}} 17.04 \\
\emg & 5944.70 & 5944.87 & --- & -25.64 & -21.63 & --- & {\cellcolor[HTML]{FFB3B3}} 19.64 & {\cellcolor[HTML]{FFB3B3}} 15.63 & --- \\
\rnr & 5946.60 & 5944.87 & --- & -23.74 & -21.63 & --- & {\cellcolor[HTML]{FFB3B3}} 19.74 & {\cellcolor[HTML]{FFB3B3}} 17.63 & --- \\
\coldnede & 5944.32 & 5942.90 & 5939.86 & -26.02 & -23.60 & -21.78 & {\cellcolor[HTML]{FFB3B3}} 20.02 & {\cellcolor[HTML]{FFB3B3}} 17.60 & {\cellcolor[HTML]{FFB3B3}} 15.78 \\ \midrule
\modrec & 5952.56 & 5949.08 & 5951.46 & -17.78 & -17.42 & -10.18 & 9.78 & 9.42 & 2.18 \\
\me & 5955.76 & 5950.02 & 5953.20 & -14.58 & -16.48 & -8.44 & {\cellcolor[HTML]{FFB3B3}} 12.58 & {\cellcolor[HTML]{FFB3B3}} 14.48 & 6.44 \\ \midrule
\neff & 5965.16 & 5964.04 & 5959.51 & -5.18 & -2.46 & -2.13 & 3.18 & 0.46 & 0.13 \\
\sidr & 5967.03 & 5965.74 & 5961.01 & -3.31 & -0.76 & -0.63 & -0.69 & -3.24 & -3.37 \\
\wzdr & 5959.28 & 5959.26 & 5955.78 & -11.06 & -7.24 & -5.86 & 7.06 & 3.24 & 1.86 \\
\hotnede & 5963.48 & 5962.46 & 5955.56 & -6.86 & -4.04 & -6.08 & 0.86 & -1.96 & 0.08 \\
\arrayrulecolor{black}\bottomrule
\end{tabular}
}
\caption{\label{tab:baseline_chi2} For the same model as in table \ref{tab:baseline_QDMAP}, minimum $\chi^2 = -2\ln \mathcal{L}$ values, difference $\Delta \chi^2$ compared to the $\Lambda$CDM model, and $-\Delta$AIC compared to $\Lambda$CDM (as defined in \cref{ssec:aic}) in a fit to CMB+BAO+SN+$M_B$ data. Cells with $-\Delta{\rm AIC} > 10$ correspond to strong evidence and are marked in red.}
\end{table}

\begin{table}[tp]
\resizebox{\textwidth}{!}{
\begin{tabular}{ccccccc}
\toprule
Model & $\ln Z$ & $\ln Z (+\Omega_k)$ & $\ln Z (+w_0, w_a)$ & $\ln$ BF & $\ln$ BF $(+\Omega_k)$ & $\ln$ BF $(+w_0, w_a)$ \\
\midrule \arrayrulecolor[HTML]{CCCCCC}
\lcdm & $-3018.58\pm0.01$ & $-3022.93\pm0.01$ & $-3020.38\pm0.02$ & $0.00\pm0.02$ & $0.00\pm0.01$ & $0.00\pm0.03$ \\
\switch & $-3017.66\pm0.09$ & $-3023.11\pm0.17$ & --- & $0.92\pm0.09$ & $-0.18\pm0.17$ & --- \\
\idedm & $-3022.48\pm0.04$ & $-3026.48\pm0.05$ & --- & $-3.90\pm0.04$ & $-3.55\pm0.05$ & --- \\ \midrule
\thawing & $-3024.70\pm0.12$ & $-3029.04\pm0.08$ & --- & $-6.12\pm0.12$ & $-6.11\pm0.08$ & --- \\ \midrule
\ede & $-3008.07\pm0.07$ & $-3019.09\pm0.00$ & $-3014.54\pm0.00$ & {\cellcolor[HTML]{FFB3B3}} $10.51\pm0.07$ & {\cellcolor[HTML]{FFB3B3}} $3.84\pm0.01$ & {\cellcolor[HTML]{FFB3B3}} $5.84\pm0.02$ \\
\emg & $-3012.17\pm0.06$ & $-3015.55\pm0.07$ & --- & {\cellcolor[HTML]{FFB3B3}} $6.41\pm0.06$ & {\cellcolor[HTML]{FFB3B3}} $7.38\pm0.07$ & --- \\
\rnr & $-3010.87\pm0.04$ & $-3012.14\pm0.05$ & --- & {\cellcolor[HTML]{FFB3B3}} $7.71\pm0.04$ & {\cellcolor[HTML]{FFB3B3}} $10.79\pm0.05$ & --- \\
\coldnede & $-3015.55\pm0.04$ & $-3020.46\pm0.16$ & $-3020.66\pm0.14$ & {\cellcolor[HTML]{FFB3B3}} $3.03\pm0.05$ & $2.47\pm0.16$ & $-0.28\pm0.14$ \\ \midrule
\modrec & $-3018.75\pm0.24$ & $-3023.64\pm0.51$ & $-3026.09\pm0.21$ & $-0.17\pm0.24$ & $-0.71\pm0.51$ & $-5.71\pm0.21$ \\
\me & $-3015.05\pm0.12$ & $-3018.02\pm0.20$ & $-3020.33\pm0.10$ & {\cellcolor[HTML]{FFB3B3}} $3.53\pm0.12$ & {\cellcolor[HTML]{FFB3B3}} $4.91\pm0.20$ & $0.05\pm0.10$ \\ \midrule
\neff & $-3018.68\pm0.01$ & $-3023.70\pm0.02$ & $-3022.18\pm0.02$ & $-0.09\pm0.02$ & $-0.77\pm0.02$ & $-1.80\pm0.03$ \\
\sidr & $-3020.00\pm0.04$ & $-3024.93\pm0.02$ & $-3022.64\pm0.03$ & $-1.42\pm0.04$ & $-2.00\pm0.02$ & $-2.26\pm0.04$ \\
\wzdr & $-3016.82\pm0.04$ & $-3023.73\pm0.10$ & $-3021.57\pm0.24$ & $1.77\pm0.04$ & $-0.80\pm0.10$ & $-1.19\pm0.24$ \\
\hotnede & $-3025.02\pm0.07$ & $-3029.50\pm0.12$ & $-3027.78\pm0.10$ & $-6.44\pm0.07$ & $-6.57\pm0.12$ & $-7.40\pm0.10$ \\
\arrayrulecolor{black}\bottomrule
\end{tabular}
}
\caption{\label{tab:baseline_evidence} For the same models as in~\cref{tab:baseline_QDMAP}, log-evidence $\ln Z$ and log-Bayes-Factor $\ln \mathrm{BF}$ as defined in \cref{ssec:BF}. Cells with $\ln \mathrm{BF} > 3$ indicating a model strongly favored over $\Lambda$CDM are highlighted. The uncertainties are the ones reported by MCEvidence, but additional independent tests (e.g. thinning the chains) suggest that the uncertainty is likely closer to $\pm 0.5$.}
\end{table}

\subsection{Baseline}\label{ssec:baseline} 

\begin{figure}[tp]
    \centering
    \includegraphics[width=0.48\linewidth]{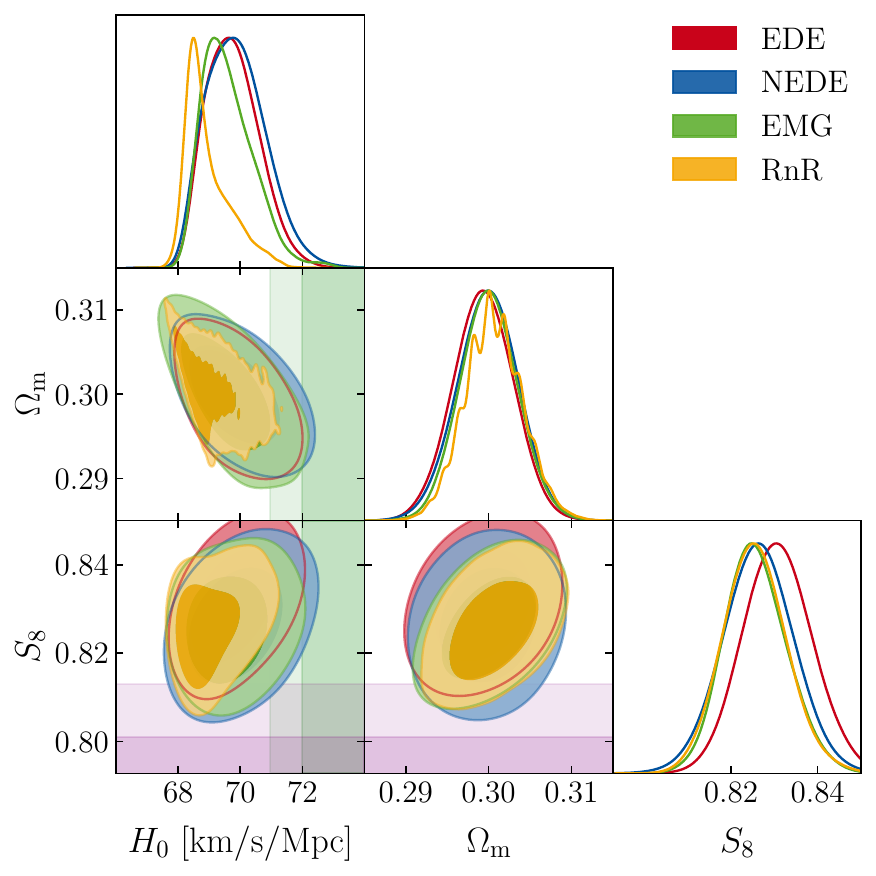}
    \includegraphics[width=0.48\linewidth]{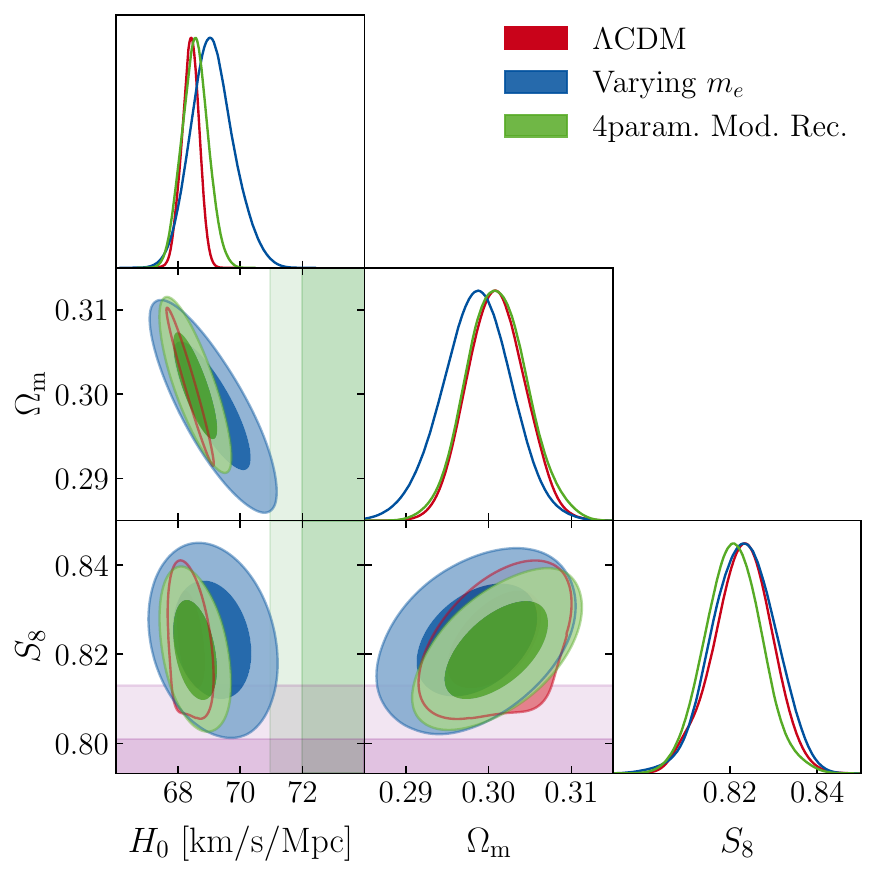}
    \includegraphics[width=0.48\linewidth]{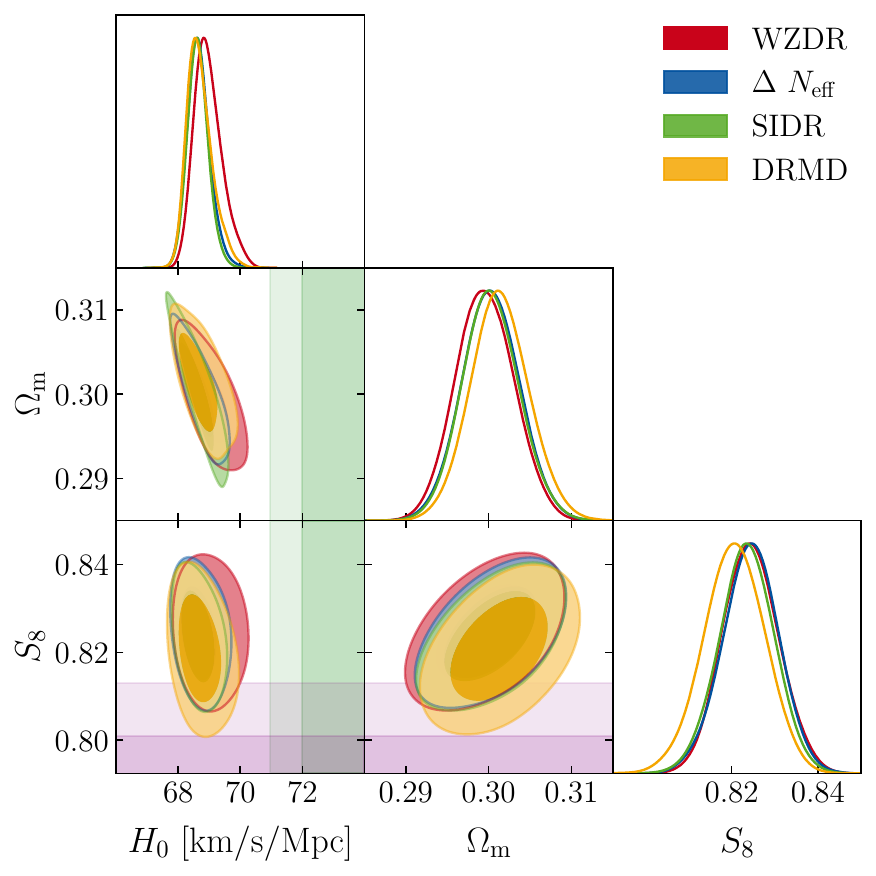}
    \includegraphics[width=0.48\linewidth]{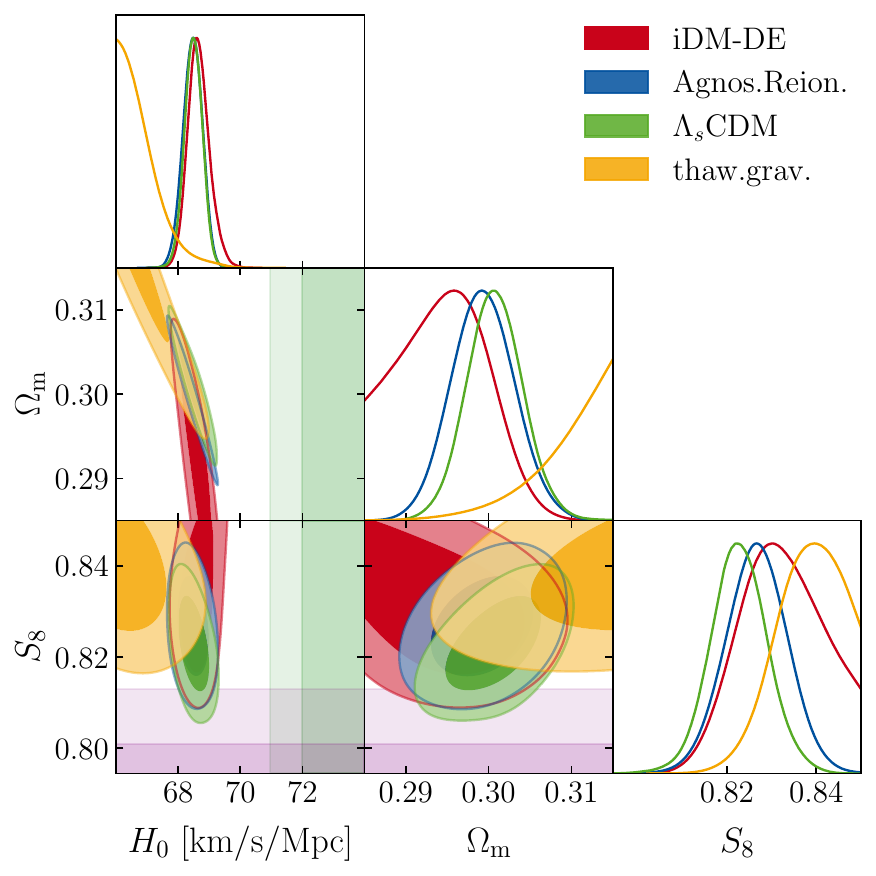}
    \caption{68\% and 95\% credible regions for the models competing in the $H_0$ world cup (group stage) fit to CMB+BAO+SN data. {\it Top left:} Group E. {\it Top right:} Group M plus $\Lambda$CDM. {\it Bottom left:} Group R. {\it Bottom right:} Groups L and E+L. The green bands show the $H_0$ values preferred by SH0ES \cite{Riess:2021jrx}, while the purple bands stands for $S_8$ values indicated by the DES Y6 3x2pt photometric analysis in $\Lambda$CDM \cite{DES:2026fyc}. For posteriors with other datasets, see \cref{app:noSBS,app:permodel}.}
    \label{fig:overviews}
\end{figure}

In our baseline analysis, we employ $\Lambda$CDM with a free neutrino mass sum as the reference model for comparison. The models in \cref{sec:models} extend it with the additional parameters described in that section. Their performance according to our various metrics is reported in \cref{tab:baseline_chi2,tab:baseline_tension,tab:baseline_tension_2,tab:baseline_QDMAP,tab:baseline_evidence}, and summarized in \cref{fig:summary_bayesian,fig:summary_frequentist}. Models that satisfy the threshold criteria of \cref{ssec:thresholds} are highlighted in red. \Cref{fig:overviews} shows the posteriors for the parameters $(H_0, \Omega_{\rm m}, S_8)$ in each model.\\

\enlargethispage*{2em}
\noindent Overall, the rankings are relatively consistent across the various metrics.
\begin{enumerate}
    \item Group E shows the best overall performance, reducing the tension to around the $3\sigma$ level while achieving $\Delta$AIC values of order 20 or larger. The log-Bayes factor is also typically above 5, corresponding to decisive evidence, except for the cold NEDE model, for which it is around 3, corresponding to strong evidence. Within this group, the axion EDE model performs best according to Frequentist metrics, with $-\Delta\mathrm{AIC}=23.4$ and $\Delta_\mathrm{DMAP}=2.5\sigma$, and reaches the largest Bayes factor in the competition, $\ln \mathrm{BF}\simeq 10.5$. However, it is slightly outperformed in the Bayesian tension metrics by the cold NEDE model, for which $\Delta_\mathrm{shift}$ and $\Delta_\mathrm{Gauss}$ are both around $2.8\sigma$.

    \item Group M shows intermediate performance. The varying-electron-mass model reaches a residual tension of around $4\sigma$, with $-\Delta\mathrm{AIC}=12.6$ and a log-Bayes factor of about 3.5, corresponding to very strong evidence. The 4-parameter modified recombination model achieves a good fit improvement, with $-\Delta\mathrm{AIC}=9.8$, but does not substantially reduce the tension metrics $\Delta_\mathrm{shift}$, $\Delta_\mathrm{Gauss}$, or $\Delta_\mathrm{DMAP}$, which remain above $5\sigma$, similarly to $\Lambda$CDM. Its log-Bayes factor is also close to zero, indicating no meaningful Bayesian preference over $\Lambda$CDM.

    \item Group R performs only marginally better than $\Lambda$CDM. The most successful model in this group is WZDR, which reduces the tension to roughly $4.5\sigma$ and reaches $-\Delta\mathrm{AIC}\simeq 8$, with a log-Bayes factor of about 1.8, corresponding to substantial evidence. The other group R models perform worse, with negative log-Bayes factors and residual tensions around $5\sigma$.

    \item Groups L and E+L perform worst overall. These models generally do not significantly reduce the $H_0$ tension and do not achieve competitive AIC improvements. The best-performing model in this group is the sign-switching cosmological constant model, which reaches a modest improvement, $\Delta\mathrm{AIC}\simeq -3.6$, and a log-Bayes factor of about 0.92, corresponding to anecdotal evidence. However, the residual tension remains above $5\sigma$ for this model, as for the other contenders in groups L and E+L.
\end{enumerate}

The weaker performance of group M, group E+L, and especially group R compared to previous literature \citep{Schoneberg:2021qvd,Schoneberg:2022grr} can be attributed to the inclusion of ACT data, which prefer a higher amplitude of the angular power spectrum in the CMB damping tail, c.f. \citealp[Figs.~41,42]{ACT:2025blo} and \citealp[Fig.~3]{Poulin:2025nfb}. We discuss this point further in \cref{ssec:noACT}, where we check how these results shift if ACT data are removed.

\begin{figure}[tp]
    \centering\makebox[\textwidth][c]{%
    \includegraphics[width=1.1\linewidth]{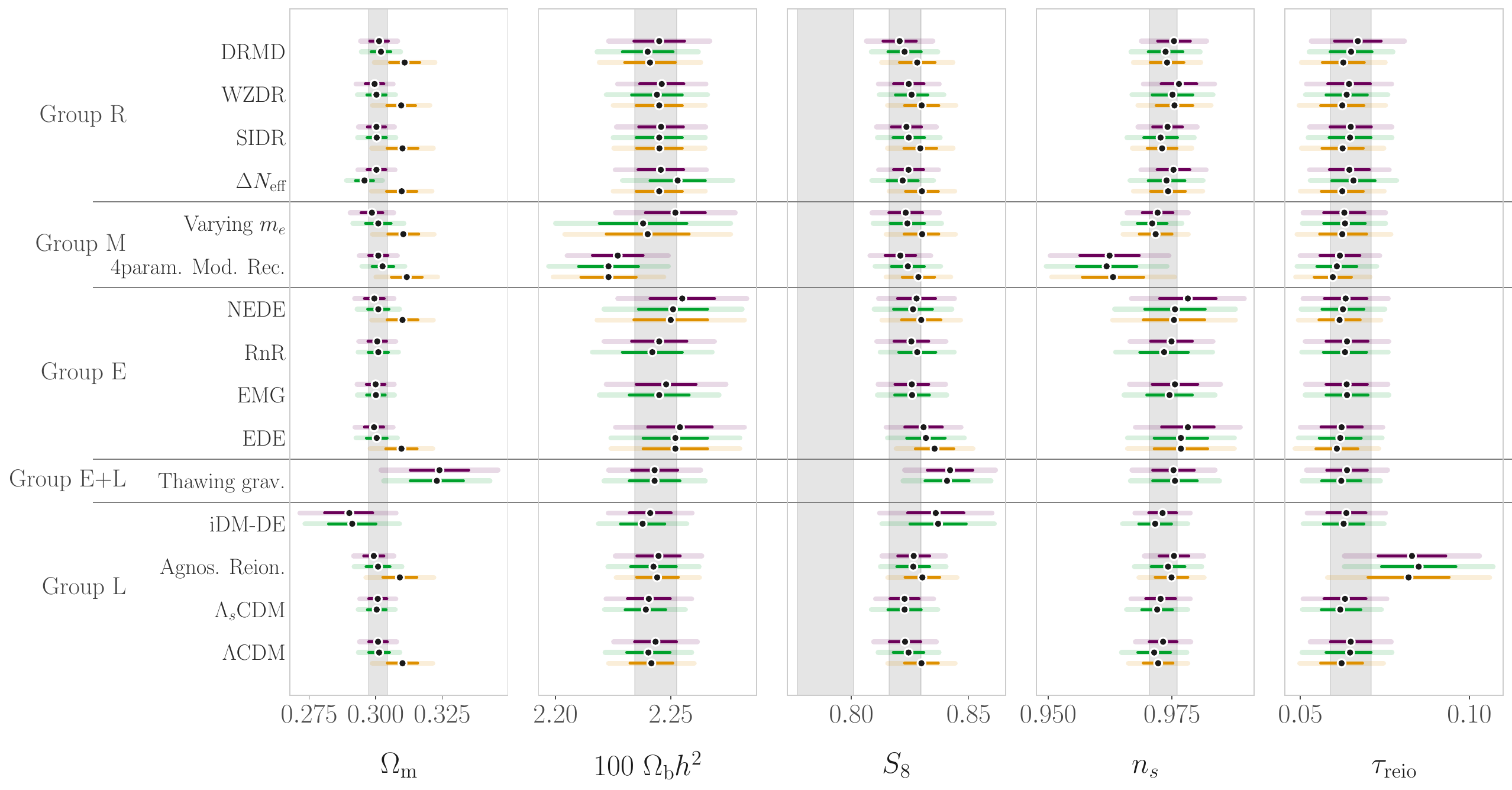}
    }
    
    \caption{Whisker plot of constraints for the  CMB+BAO+SN data, as summarized also in \cref{tab:constr_BS,tab:constr_BS_Ok,tab:constr_BS_w0wa}. The purple error bars correspond to the baseline case, green error bars correspond to a curved geometry, and orange error bars correspond to an extension of the model through CPL dark energy. For posteriors with other datasets, see \cref{app:noSBS,app:permodel}.}
    \label{fig:whisker}
\end{figure}
\begin{figure}[tp]
    \centering
    \includegraphics[width=0.49\linewidth]{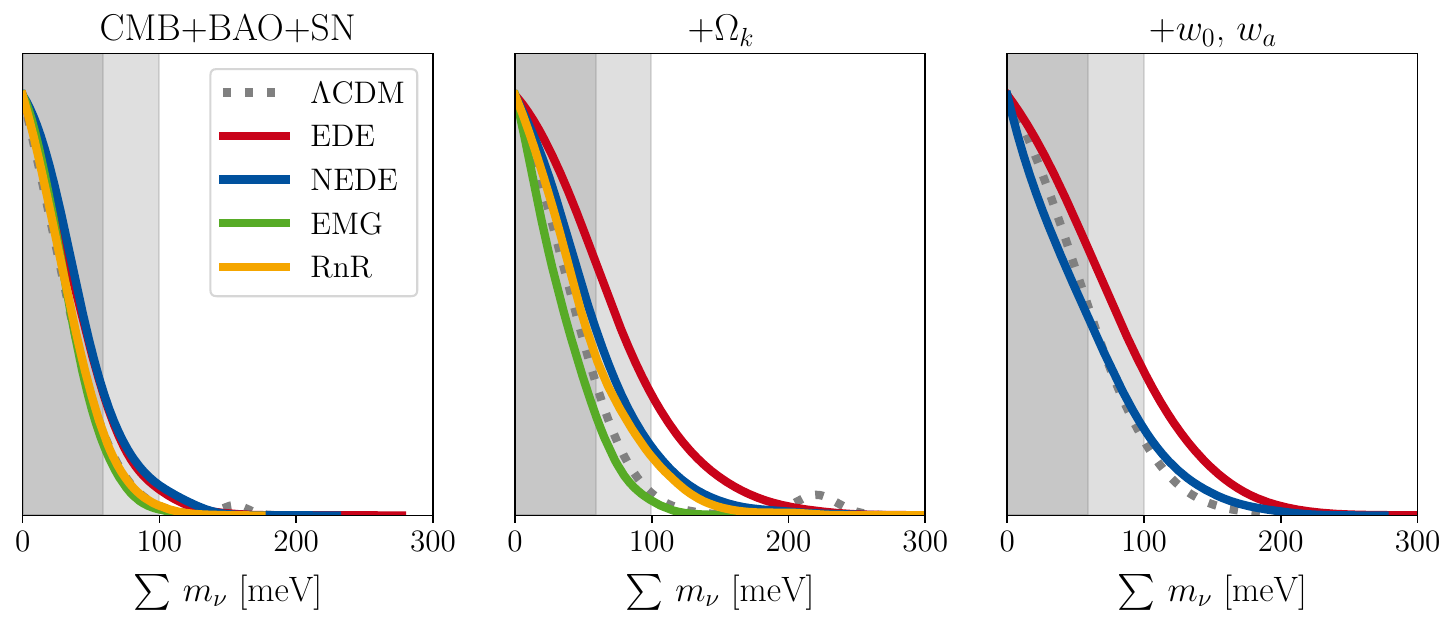}
    \includegraphics[width=0.49\linewidth]{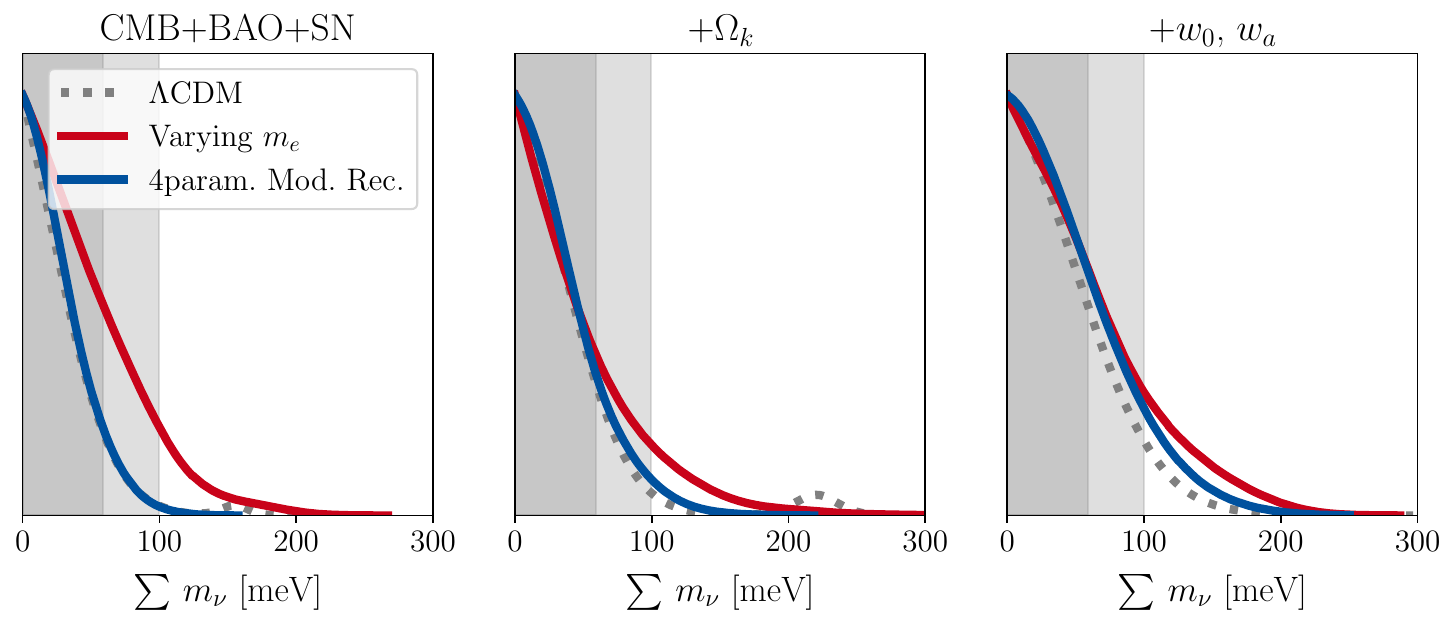}
    \includegraphics[width=0.49\linewidth]{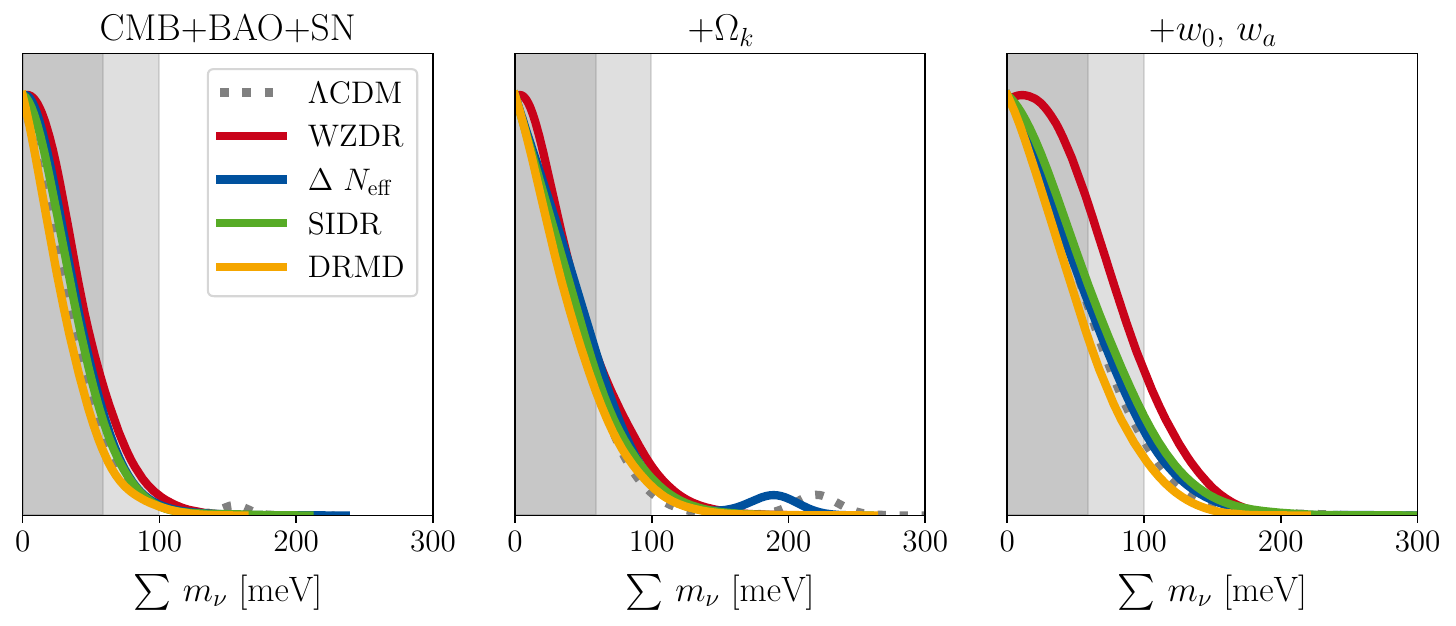}
    \includegraphics[width=0.49\linewidth]{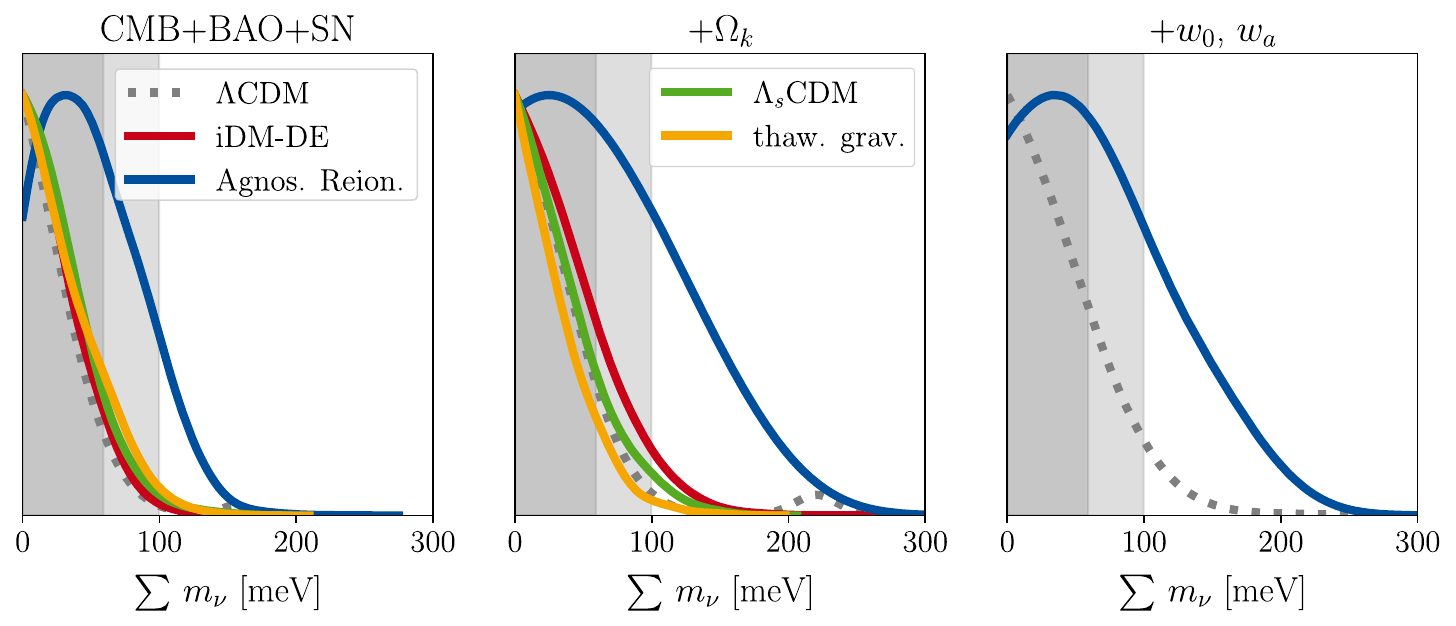}
    \caption{One dimensional posteriors for the neutrino mass sum parameter for the various contenders of this competition, including their extensions through curvature ($+\Omega_k$) or CPL dark energy ($+w_0, w_a$). For comparison, we also show the corresponding $\Lambda$CDM constraints as a dashed gray line in each case, and the neutrino oscillation lower mass sum bounds as vertical bands in dark gray (normal hierarchy, $\sum m_\nu > 59\mathrm{meV}$) and light gray (inverted hierarchy, $\sum m_\nu > 100\mathrm{meV}$).}
    \label{fig:mnu_1d}
\end{figure}
\begin{figure}[tp]
    \centering
    \centering\makebox[\textwidth][c]{%
    \begin{minipage}{1.2\textwidth}
    \includegraphics[width=0.99\linewidth]{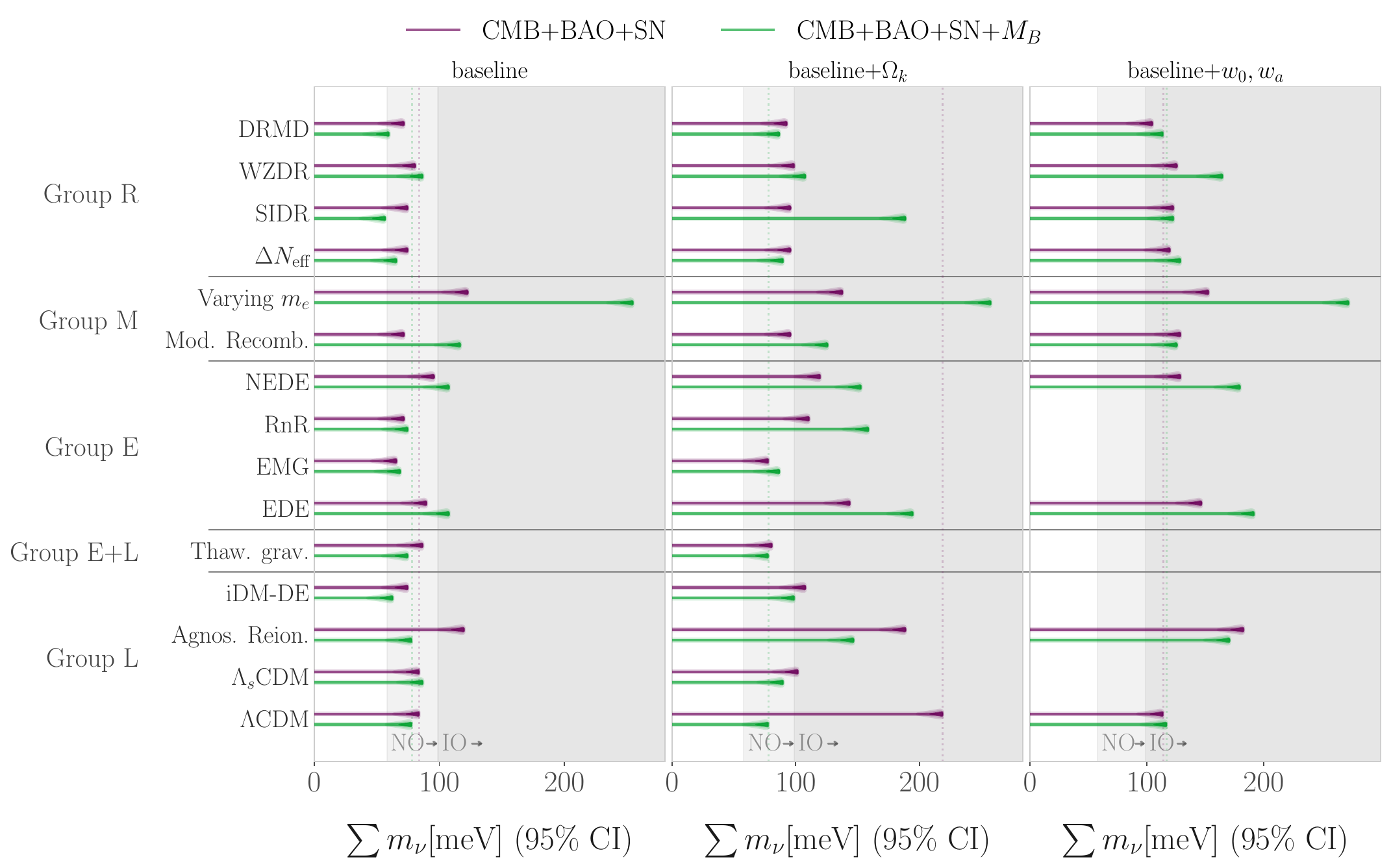}
    \end{minipage}}
    \caption{Comparison of the 95\% CI for the summed neutrino mass. The dotted colored lines correspond to the result in $\Lambda$CDM (+$\nu$), and the neutrino oscillation lower mass sum bounds are shown as vertical bands in dark gray (normal hierarchy, $\sum m_\nu > 59\mathrm{meV}$) and light gray (inverted hierarchy, $\sum m_\nu > 100\mathrm{meV}$). \textit{Left:} The baseline configuration of \cref{ssec:baseline}. \textit{Middle:} the configuration with curvature of \cref{ssec:curved}. \textit{Right:} the configuration with CPL dark energy of \cref{ssec:w0wa}.}
    \label{fig:mnu_comparison}
\end{figure}

\vspace*{0.5\baselineskip}
\textbf{Neutrino mass constraints.} We show the neutrino mass constraints for all models in \cref{fig:mnu_1d,fig:mnu_comparison}. These Bayesian constraints depend on our choice of prior, $\sum m_\nu>0$. We investigate the consequences of priors motivated by neutrino oscillation experiments in \cref{ssec:mnu_norm}. The 95\% credible interval for our data combination in $\Lambda$CDM is $0<\sum m_\nu < 66\mathrm{meV}$. This is mostly compatible with close-to-minimal normal hierarchy scenarios. The upper bound is not significantly different for other models, except for two cases:
\begin{enumerate}
    \item In the case where the \texttt{SROLL2} EE likelihood is dropped in $\Lambda$CDM (see \cref{ssec:nosroll}), the neutrino mass constraints relax significantly to $\sum m_\nu < 120\mathrm{meV}$ (95\% CL). This follows from the well-known degeneracy between the summed neutrino mass and the reionization optical depth $\tau_{\rm reio}$ \cite{Allison:2015qca,Liu:2015txa,Archidiacono:2016lnv,Sailer:2025lxj,Jhaveri:2025neg}. In absence of a low-$\ell$ polarization likelihood, the optical depth is poorly constrained and simultaneous high values of $(\sum m_\nu, \tau_\mathrm{reio})$ are compatible with the data.
    \item In the case where the electron mass varies between the early and late universe, the angular diameter distance to the last-scattering surface is free to change without impacting the CMB anisotropies at first order, as argued in \citealp{Sekiguchi:2020teg,Schoneberg:2024ynd}. This generates a much stronger geometric degeneracy $H_0$-$\sum m_\nu$ that can only be broken by late-time data. The model features a parameter degeneracy in which $H_0$, $\sum m_\nu$ and $m_\mathrm{e}^\mathrm{early}/m_\mathrm{e}^\mathrm{late}$ are all increased without impacting CMB angular power spectra significantly. As a result, the neutrino mass bounds weaken to $\sum m_\nu < 123\mathrm{meV}$ (95\% CL) and is dominated by BAO data -- without BAO data it relaxes to $\sum m_\nu < 291\mathrm{meV}$ (95\% CL), which is the weakest bound of all tested combinations of data and models in this work.
\end{enumerate}
For all other models, the neutrino mass bounds may weaken significantly ($>50\%$) only in combination with free curvature or dynamical dark energy parameters, see \cref{ssec:curved,ssec:w0wa}. The neutrino mass constraints for each model can be found in \cref{app:tables}, \cref{tab:constr_BS,tab:constr_BS_w0wa,tab:constr_BS_Ok}.
\enlargethispage*{2em}

\vspace*{0.5\baselineskip}
\textbf{Other parameter constraints.} All credible intervals are reported in \cref{app:tables}, with \cref{tab:constr_model_BS}
showing the parameters specific to each model and 
\cref{tab:constr_BS,tab:constr_BS_w0wa,tab:constr_BS_Ok} those common to all of them. For the constraints with a prior on $M_B$ added, see \cref{fig:overviews_BSM} in \cref{app:noSBS}.

Like for neutrino masses, most models do not significantly alter the credible intervals for the clustering amplitude $S_8$ and matter fraction $\Omega_{\rm m}$, as illustrated in \cref{fig:overviews}. The stability of $\Omega_{\rm m}$ bounds (c.f. column 2 of \cref{tab:constr_BS}) is driven by BAO data. As a matter of fact, BAO data imposes $\Omega_\mathrm{m} = 0.2995 \pm 0.0035$ (68\%) for $\Lambda$CDM and nearly the same for all other models, while the inclusion of SN data slightly shifts the preferred range to $\Omega_\mathrm{m} = 0.3009 \pm 0.0035$ (as can be checked by comparing \cref{tab:constr_BS} with CMB+BAO+SN data and \cref{tab:constr_B} with CMB+BAO-only data).

One exception is the \idedm{} model, which significantly shifts $S_8$ upward and $\Omega_{\rm m}$ downward, as already found in \cite[Fig.~2]{DiValentino:2019ffd}. Note that in \idedm{} the parameter $\Omega_{\rm m}$ connects less directly to the redshift of the transition from matter to dark energy domination and to observable consequences, due to the energy transfer between dark energy and dark matter in that model. However, using the effective parameter $\Omega_\mathrm{cdm}^\mathrm{eff}$ introduced in \cref{ssec:idmde} in place of $\Omega_\mathrm{cdm}$, we get a bound $\Omega_\mathrm{m}^\mathrm{eff} = 0.2985^{+0.0078}_{-0.0039}$ which essentially agrees with the $\Lambda$CDM one, despite being more asymmetric. Another exception is the \thawing{} model, whose high values of $\Omega_m$ are only compatible with BAO data due to the late modification, resulting in correspondingly high values of $S_8$\,.

\vspace*{0.5\baselineskip}
\textbf{Impact of BAO and SNeIa.} These results can also be compared to the corresponding analyses without supernovae or BAO data (that is, CMB-only or CMB+BAO), shown in \cref{app:noSBS}, \cref{fig:overviews_noBS,fig:overviews_noS}. The main conclusion from these comparisons is that BAO data have the strongest impact on constraints for all model classes, while supernova data often play a subdominant role. Without BAO, models in groups R, M, and E are compatible with broader $\Omega_\mathrm{m}$ ranges, especially towards higher values, and neutrino mass bounds are significantly looser. This opening of the $\Omega_m$ parameter space is strongly limited once BAO data are included.

Note that models relieving the Hubble tension usually feature a negative correlation between $H_0$ and $\Omega_\mathrm{m}$ \cite{Lee:2022gzh,Poulin:2024ken,Pedrotti:2024kpn}. Thus, the fact that DESI BAO data favors smaller values of $\Omega_\mathrm{m}$ tends to help model in reducing the Hubble tension, as already found in \cite{chaussidon:2025,Lynch:2024hzh,Mirpoorian:2025rfp,Poulin:2025nfb,Jhaveri:2026bla}.

\vspace*{0.5\baselineskip}
\textbf{Notable differences with previous literature.} In the 4-parameter recombination model, our results with CMB+BAO+SN data are much more constraining than in previous analyses like \citealp{Mirpoorian:2024fka}. We find a residual tension with SH0ES typically above $5\sigma$. We checked that with the same data combination as in \citealp{Mirpoorian:2024fka}, which does not include ACT and SPT data, we get nearly the same result, $H_0 = (71.23 \pm 0.68) \mathrm{km/s/Mpc}$, as \citealp{Mirpoorian:2024fka}, $H_0 = (71.34 \pm 0.68) \mathrm{km/s/Mpc}$. Thus, the ability to reduce the tension is compromised by the addition of ACT and SPT data. Noticeable is the strong reduction in $\chi^2$ with CMB+BAO+SN data by $\Delta \chi^2 \simeq -11$, suggesting that the model is highly efficient at fitting the data, but penalized by its many parameters as well as its lack of preference for increased $H_0$ values. It is important to stress that a truly flexible recombination model might still show better performance, but checking this is beyond the scope of this work.

The sign-switching $\Lambda_s$CDM model also does not show such a strong performance in reducing the tension as in the previous literature. The reason is both the improved constraining power of the CMB data and the tighter BAO+SNeIa data. We note that our results are nevertheless very similar to \cite[Tab~III]{Ibarra-Uriondo:2026zbp} or \cite[Tab~II]{Akarsu:2021fol}. The performance reported in some papers is enhanced by their use of transversal BAO data from SDSS. However, these data may suffer from radial mode leakage \cite{deCarvalho:2017xye,Pantos:2026rpe} or redshift bin smearing for tophat bins \cite{Ferreira:2025til}. Additionally, they do not capture the full constraining power of DESI BAO. We therefore do not consider transversal BAO data from SDSS here.

The thawing gravity model has been shown in \citealp{Ye:2024zpk} to be able to reduce the Hubble tension in the scaling regime for $|\xi| < 3/16$ for small values of $|\xi \sigma_\mathrm{i}^2/M_\mathrm{pl}^2| \sim 10^{-1.5}$ (c.f. \cite[Fig.~9]{Ye:2024zpk}) and small values of $|\xi| \sim 10^{-2.5}$, requiring $|\sigma_\mathrm{i}/M_\mathrm{pl}| \sim 0.3$. With the CMB+BAO+SN datasets considered here, we find that this parameter space is not preferred, which we attribute to our CMB data, as well as the allowance for larger neutrino mass sums. Instead, the data prefers a parameter space in which the thawing gravity model exhibits phantom crossing at low redshift for the effective dark energy parameter of state \cite{Ye:2024ywg} for all the datasets considered, which is a distinctive signature of the attractor general relativity regime for $|\xi| > 3/16$. Therefore the model has the best $\Delta \chi^2$ for CMB+BAO+SN data among the fourteen models considered ($\Delta \chi^2 \simeq -11.4$), although with a low $H_0$ as happens for $w_0 w_a$CDM when fitting DESI BAO data (which reaches $\Delta \chi^2 \simeq -10$ compared to $\Lambda$CDM). We can recover the scaling regime found in \citealp{Ye:2024zpk} when using only the same Planck CMB data, even when allowing the neutrino mass sum to vary. We conclude that with current Planck+ACT+SPT CMB data the model's capabilities to ease the Hubble tension via early time modifications are not explored, while the late time effects remain relevant.

\begin{figure}[tp]
    \centering
    \includegraphics[width=0.45\linewidth]{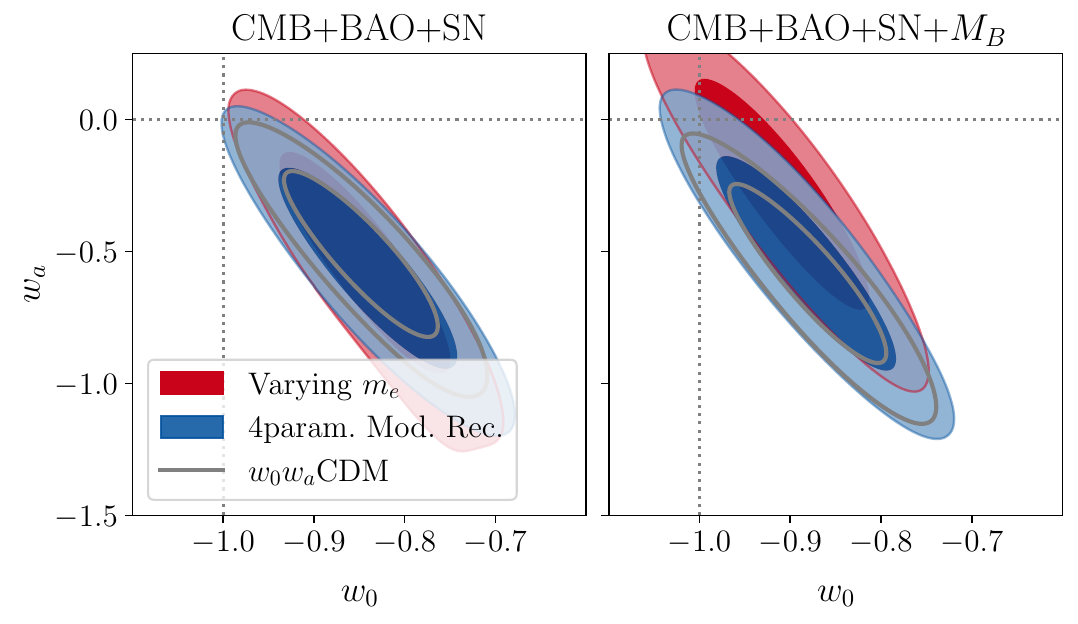}
    \includegraphics[width=0.45\linewidth]{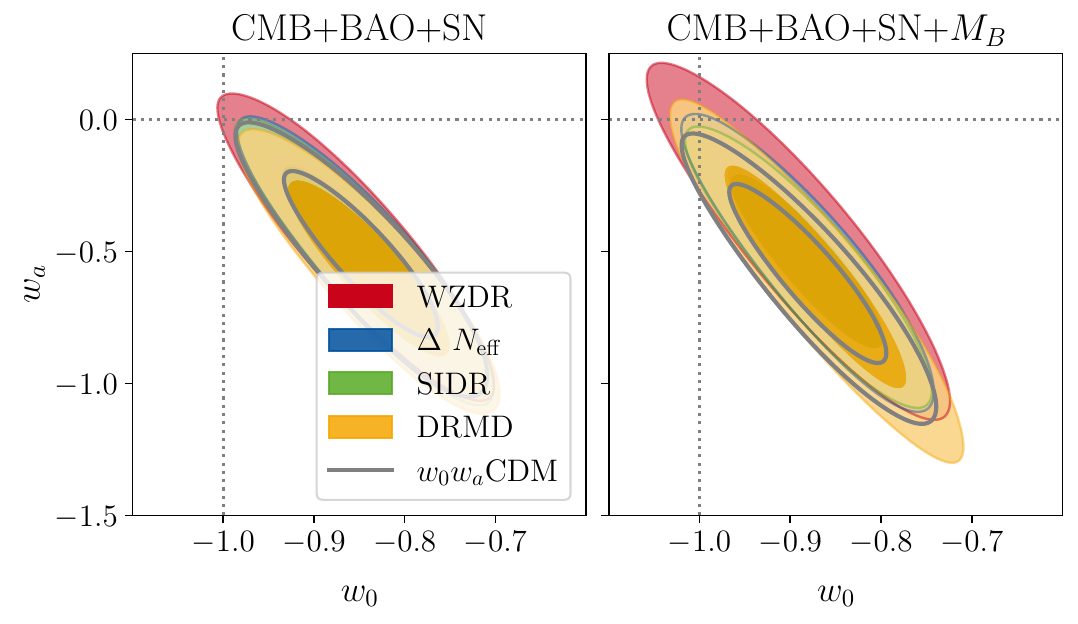}
    \includegraphics[width=0.45\linewidth]{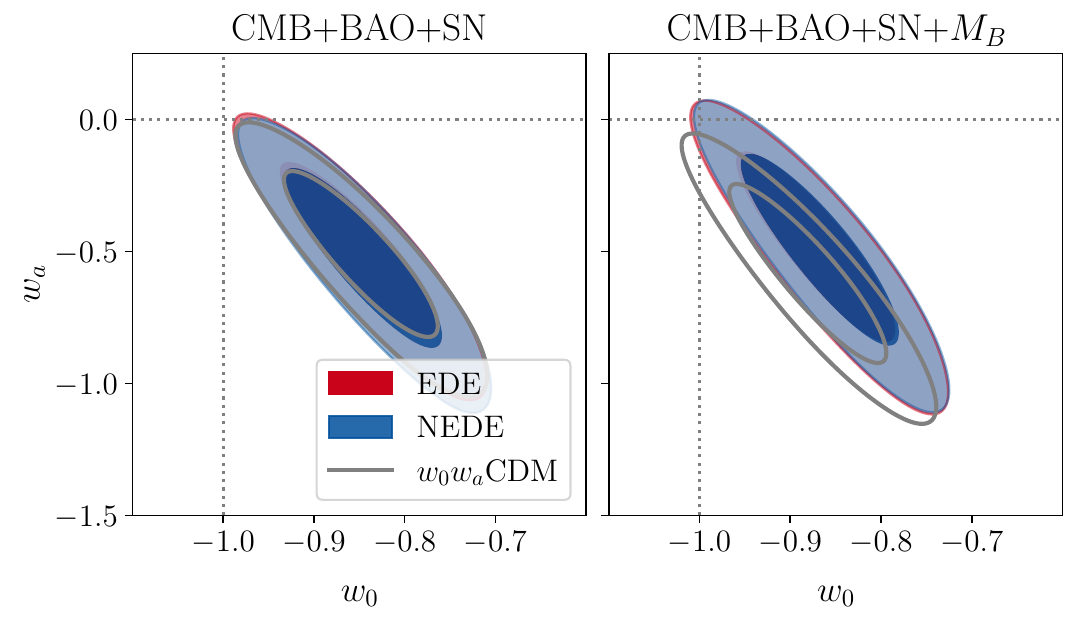}
    \includegraphics[width=0.45\linewidth]{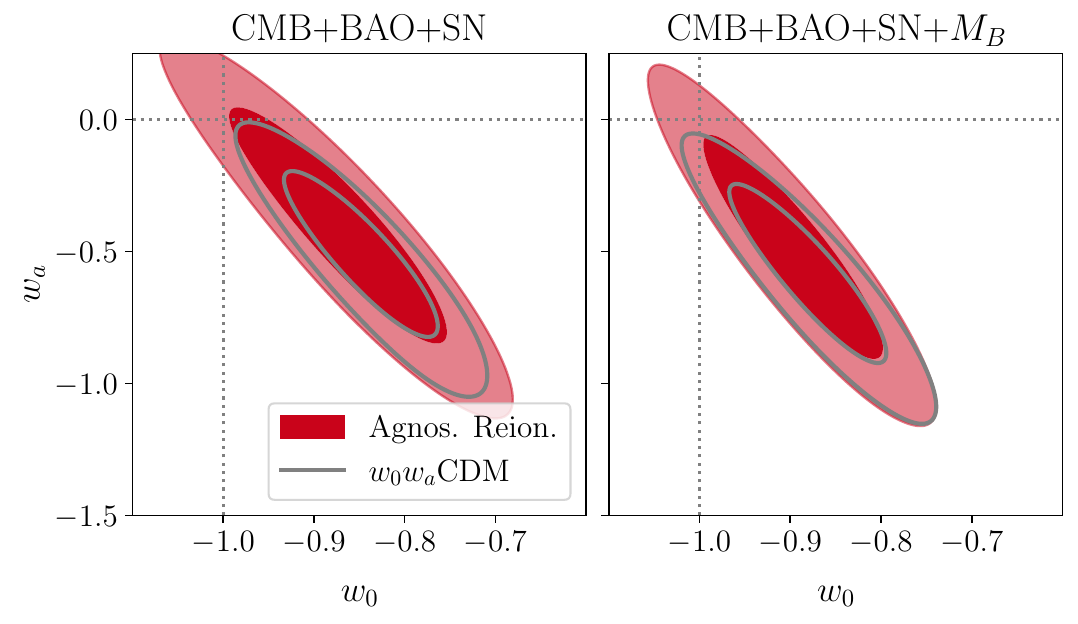}
    \caption{68\% and 95\% credible regions for the models competing in the $H_0$ world cup, when a CPL dark energy is assumed. The constraints are shown for the corresponding parameters $w_0$ and $w_a$\,, where the point $w_0=-1$ and $w_a=0$ (dotted gray lines) corresponds to a cosmological constant. {\it Top left:} Group E. {\it Top right:} Group M. {\it Bottom left:} Group R. {\it Bottom right:} Group L. We include the $w_0 w_a$CDM constraints as gray line contours in all cases for comparison.}
    \label{fig:w0wa_2d}
\end{figure}

\begin{figure}[p]
    \centering
    \includegraphics[width=0.7\linewidth]{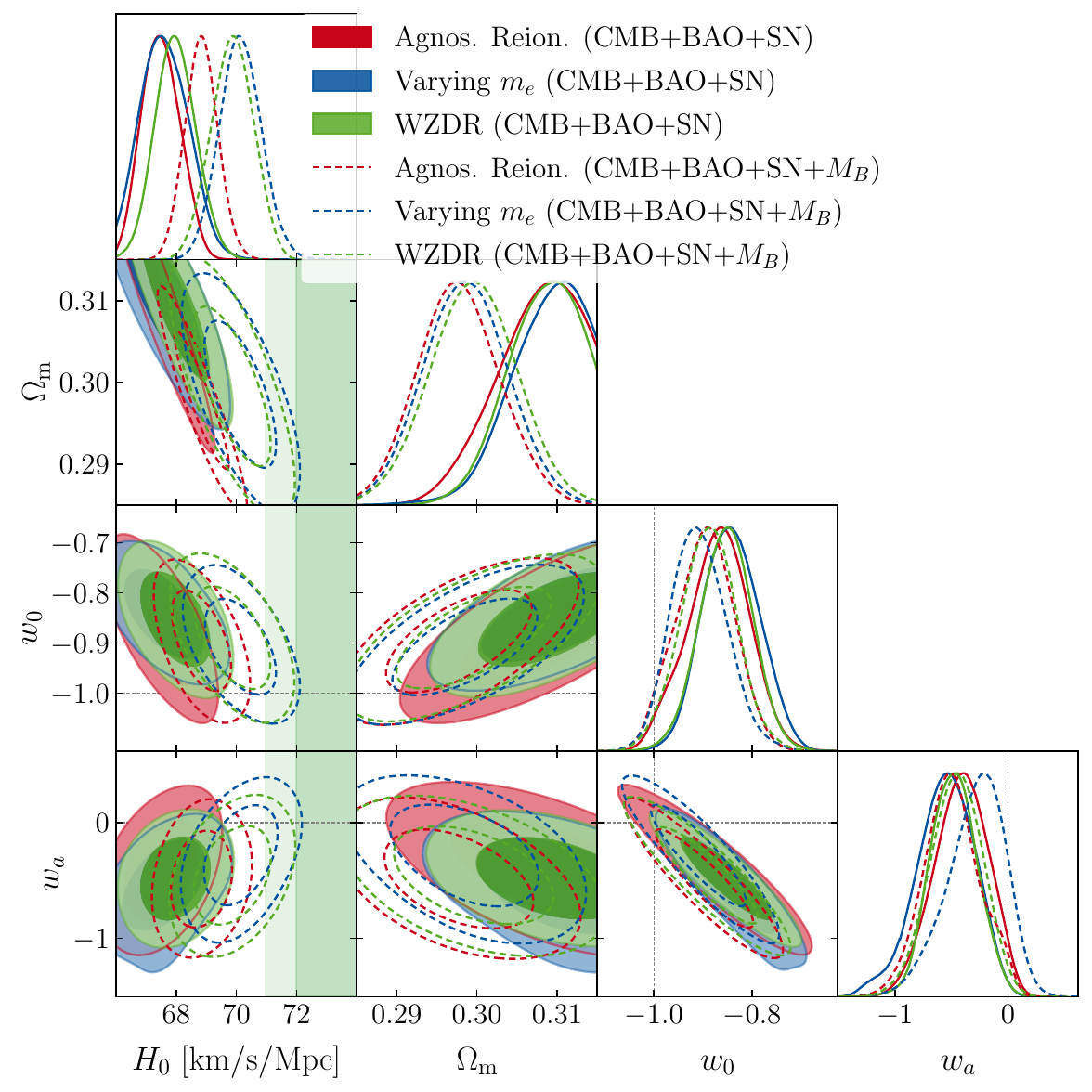}
    \caption{68\% and 95\% credible regions for the \nosroll{}, \me{}, and \wzdr{} models, when a CPL dark energy is assumed. We show the posterior distributions for all relevant cosmological parameters, highlighting the strong shift in the $w_0$,$w_a$ constraints when a prior on $M_B$ is imposed, which forces the corresponding mechanism to ease the Hubble tension of the respective model to be employed, resulting in further shifts in the other parameters as well.}
    \label{fig:special_w0wa_reduction}
\end{figure}
\begin{table}[tp]
    \centering
    \begin{tabular}{c|c c}
        \multirow{2}{*}{Base model name} & Tension in $w_0,w_a$ & Tension in $w_0,w_a$ \\ 
        & (CMB+BAO+SN) &(CMB+BAO+SN+$M_B$ prior)\\\midrule  \arrayrulecolor[HTML]{CCCCCC}
        $w_0w_a$CDM & $2.28\sigma$ & $2.66\sigma$\\
        \nosroll & $1.34\sigma$ & $1.18\sigma$\\ \midrule
        \me & $2.27\sigma$ & $1.23\sigma$\\
        \modrec & $2.46\sigma$ & $1.93\sigma$\\ \midrule
        \ede & $2.25\sigma$ & $1.88\sigma$\\
        \coldnede & $2.43\sigma$ & $1.95\sigma$\\ \midrule
        \wzdr & $2.18\sigma$ & $1.51\sigma$\\
        \neff  & $2.02\sigma$ & $2.02\sigma$ \\
        \sidr & $2.31\sigma$ & $2.31\sigma$\\
        \hotnede & $2.62\sigma$ & $2.01\sigma$ \\ \arrayrulecolor{black}\bottomrule
    \end{tabular}
    \caption{
    Tension in the $w_0$-$w_a$ parameters of CPL dark energy compared to a cosmological constant ($w_0=-1, w_a=0$). The tension is computed according to \cref{eq:parameter_shift_prob} using the corresponding posteriors in $w_0, w_a$ and defining $\Delta \theta=(w_0+1, w_a)$ such that $\Delta \theta=0$ corresponds to the cosmological constant case.}
    \label{tab:w0wa_tensions}
\end{table}

\subsection{Baseline in a universe with evolving dark energy}\label{ssec:w0wa}
The cosmological constant provides the simplest description of late-time acceleration, but current low-redshift data also motivate tests of more general dark-energy dynamics. In particular, recent DESI BAO \cite{DESI:2024mwx,DESI:2025zgx,DESI:2024aqx,DESI:2025fii} and full-shape \cite{DESI:2024hhd,Forero-Sanchez:2026zfb,Lu:2025gki} data combined with CMB and SN measurements favor departures from $\Lambda$CDM in the CPL parametrization, with a preference at the $3$--$4\sigma$ level for an evolving equation of state that crosses into the phantom regime \cite{Shlivko:2026jxa}. We find that for the considered CMB dataset, this preference is typically reduced to slightly above $2\sigma$ for the vanilla model -- i.e. the model that reduces to $\Lambda$CDM in the limit $(w_0,w_a)\rightarrow(-1,0)$, see \cref{tab:w0wa_tensions}. The constraints in the $w_0, w_a$ parameter space for all models are shown in \cref{fig:w0wa_2d}, while the full contours can be found in \cref{app:permodel}. The only model for which this preference is significantly reduced is the the \nosroll{} case of \cref{ssec:nosroll}, in which only a $1.2\sigma$ residual preference remains. Such a reduction is expected from the results of \citealp{Sailer:2025lxj}, which we confirm with up-to-date data.

\enlargethispage*{1em}
However, it is interesting to point out that by imposing the $M_B$ prior on top of the CMB+BAO+SN data combination (such that the degeneracy between $H_0$ and model parameters is further explored), the preference is generally reduced, as seen in \cref{tab:w0wa_tensions}. For the \me{} and \wzdr{} models this shift is particularly strong, reducing the preference for dynamical dark energy below $1.5\sigma$ when the $M_B$ prior is enforced. For the \me{} model the preference is reduced from $2.3\sigma$ to $1.2\sigma$ ($-1.1\sigma$), whereas for the \wzdr{} model it is reduced from $2.2\sigma$ to $1.5\sigma$ ($-0.7\sigma$), see \cref{tab:w0wa_tensions}. The reason that forcing a reduction of the Hubble tension coincides with a reduction for preference of dynamical dark energy in these models is that they more strongly reduce $\Omega_\mathrm{m}$ when increasing $H_0$ compared to other solutions -- noting that the discrepancy of the $\Omega_\mathrm{m}$ value between BAO and CMB data appears to be partially driving the preference for dynamical dark energy. The constraints for these two models (together with the \nosroll{} model) are highlighted in \cref{fig:special_w0wa_reduction}.

Finally, none of the models benefit from the inclusion of an evolving dark energy equation of state in their ability to ease the Hubble tension. For all models the tension metrics increase by up to $0.5\sigma$ -- the only exclusion from this rule is the DRMD model for which tension metrics remain about the same (see \cref{tab:baseline_tension,tab:baseline_QDMAP}). For all models, the comparison metrics ($\Delta$AIC and Bayes Factor) worsen when dynamical dark energy is considered since the marginal improvement in overall fit is outweighed by the cost of the two additional parameters, see \cref{tab:baseline_chi2,tab:baseline_evidence}. 

In summary, while the reduction of $\Omega_\mathrm{m}$ seen in most models that ease the Hubble tension is lowering the preference for dynamical dark energy observed in the DESI BAO data (especially when combined with SNeIa data), the allowance for dynamical dark energy does not help in mitigating the Hubble tension since the latter tends to lead to a decrease in the late time expansion rate.
\begin{figure}[t]
    \centering\makebox[\textwidth][c]{%
    \begin{minipage}{1.1\textwidth}
    \centering
    \includegraphics[scale=0.38]{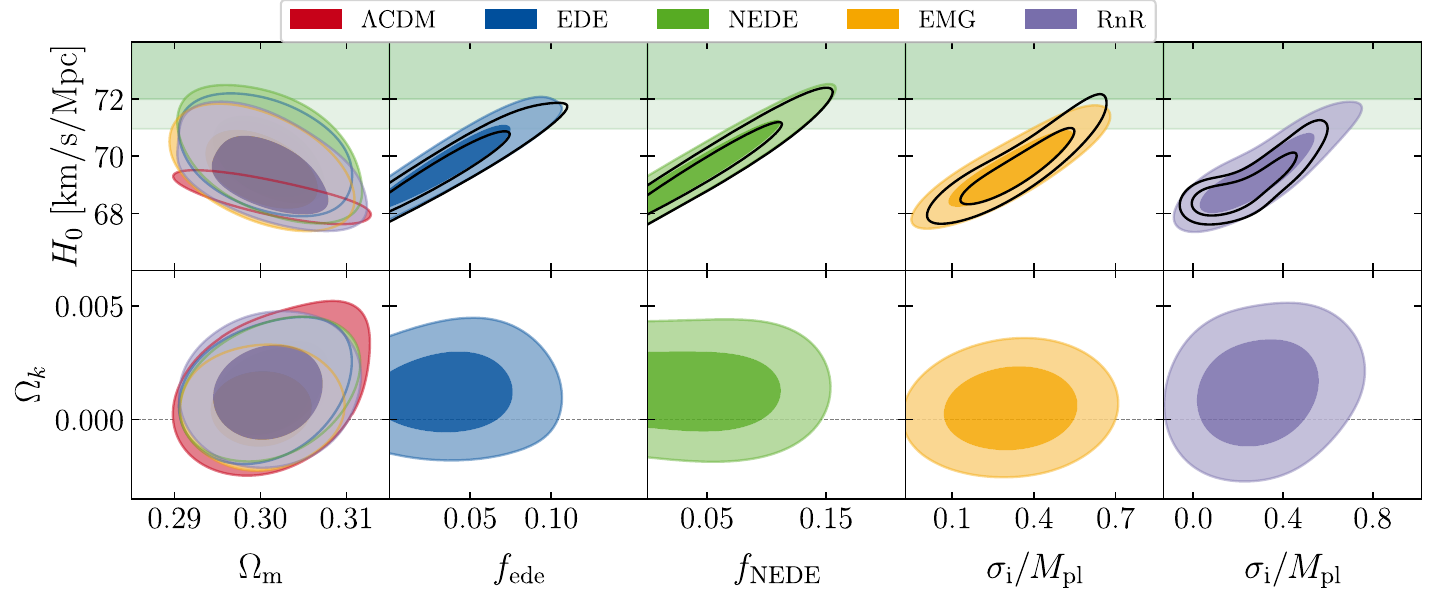}
    \includegraphics[scale=0.38]{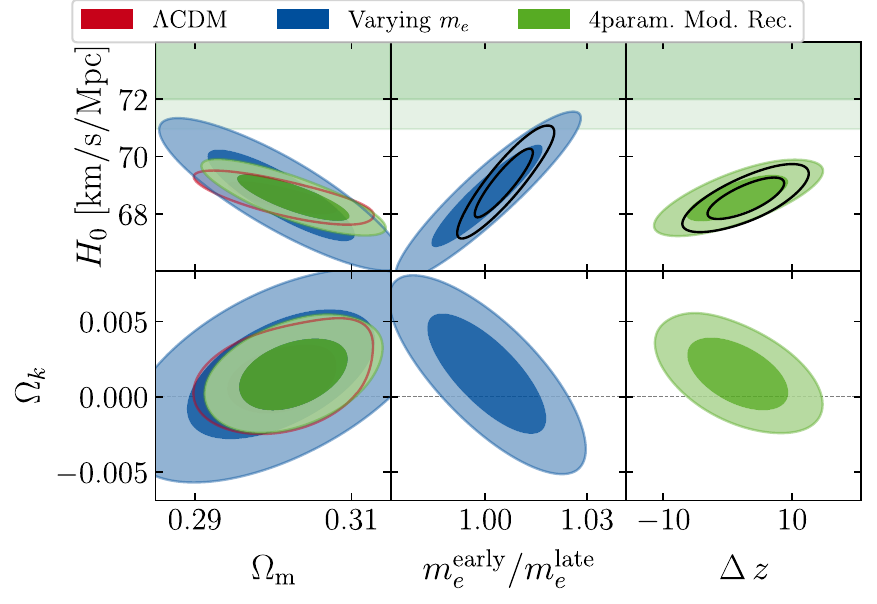} \\
    \includegraphics[scale=0.38]{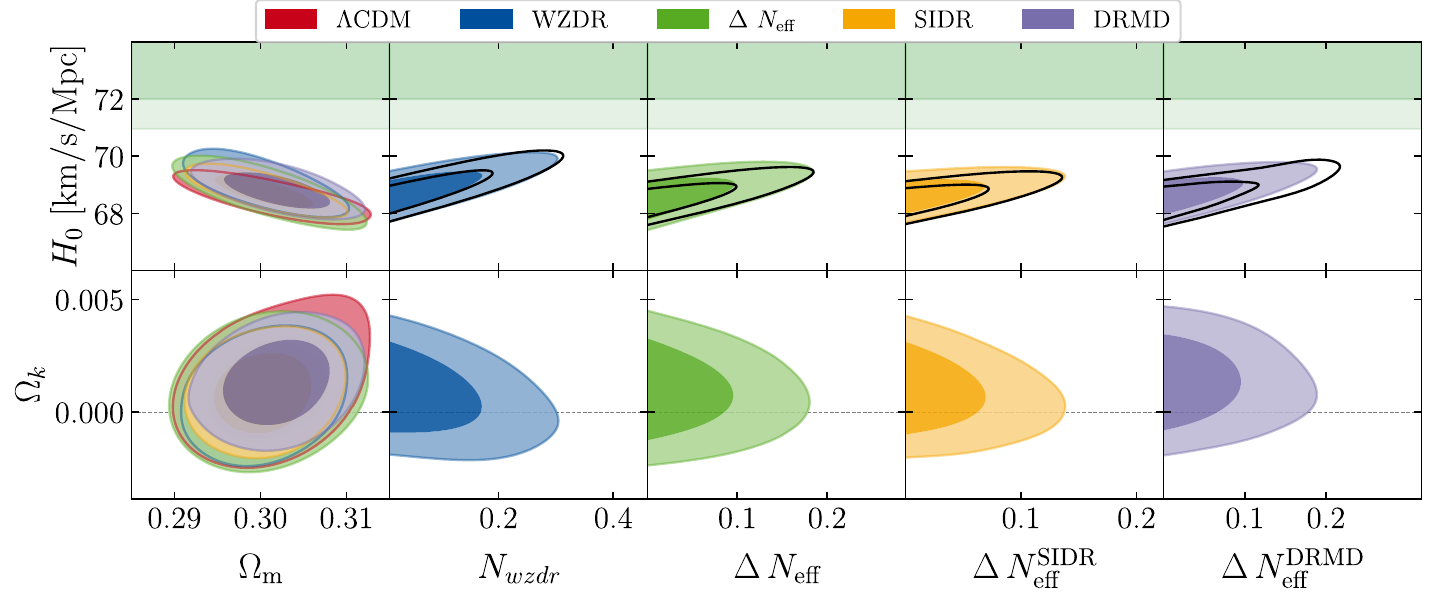}
    \includegraphics[scale=0.38]{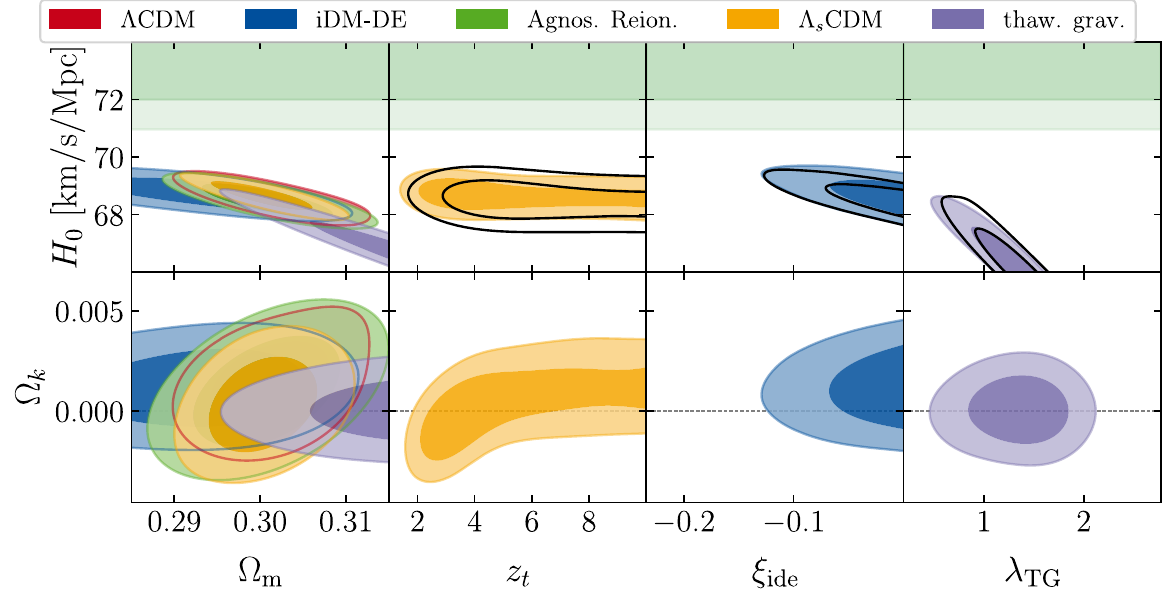}
    \end{minipage}
    }
    \caption{68\% and 95\% credible regions for the models competing in the $H_0$ world cup, with (solid) and without (black lines) assuming spatial curvature parametrized by $\Omega_k$\,. The flat case corresponds to $\Omega_k=0$ (dotted gray lines). {\it Top left:} Group E. {\it Top right:} Group M. {\it Bottom left:} Group R. {\it Bottom right:} Groups L and E+L. We include the $\Lambda$CDM+$\Omega_k$ model in all cases for comparison. The $x$-axis of each panel includes $\Omega_\mathrm{m}$ and the parameters most relevant for the Hubble tension in each model. The black lines show the model performance without curvature for comparison.}
    \label{fig:onepar_curv}
\end{figure}

\subsection{Baseline in a curved universe}\label{ssec:curved}

Another well motivated late-time extension of $\Lambda$CDM is spatial curvature, parametrized by $\Omega_k$\,. Similarly to dynamical dark energy, it was recently found to provide a better fit to CMB+DESI data \cite{Chen:2025mlf}. It is also known that models affecting recombination such as the varying electron-mass model further reduce the Hubble tension if curvature is nonzero \cite{Sekiguchi:2020teg,Schoneberg:2021qvd,Poulin:2024ken,Schoneberg:2024ynd}. Indeed, the best-performing model in \citealp{Schoneberg:2021qvd} was the combination of varying $m_\mathrm{e}$ and nonzero curvature, which although phenomenologically well-understood \cite{Sekiguchi:2020teg,Schoneberg:2021qvd, Poulin:2024ken,Schoneberg:2024ynd}, lacks a deeper theoretical motivation as to why electron mass variations should be connected to curvature. 
Here, we follow a more systematic approach and
check for all models whether allowing $\Omega_k$ to be nonzero changes their ability to ease the tension or improve the fit to the data. This comparison is shown in \cref{fig:onepar_curv}, where we display the constraints on the parameters most directly relevant to the Hubble tension, with and without curvature. Overall, we find that the impact of curvature is minor; see also \cref{tab:baseline_chi2,tab:baseline_QDMAP,tab:baseline_tension,tab:baseline_tension_2,tab:baseline_evidence}.

As found in \citealp{Schoneberg:2021qvd}, the largest effect is still obtained for the varying electron mass model. Allowing curvature increases the $-\Delta$AIC from $12.6$ to $14.5$, lowers $\Delta_\mathrm{DMAP}$ from $4.3\sigma$ to $3.5\sigma$, and increases the log-Bayes factor from 3.5 (very strong evidence) to 4.9 (decisive evidence). The other Bayesian tension metrics are essentially unchanged.
In the present analysis, the combined CMB dataset constrains curvature much more tightly, $\Omega_k = (1.5 \pm 1.6)\times10^{-3}$, compared to $(-3.2 \pm 5.0)\times10^{-3}$ in \citealp{Schoneberg:2021qvd}. As a result, it provides less freedom to improve the varying-electron-mass fit than it did in the previous analysis. 

The other model that benefits substantially from nonzero curvature is the RnR model, for which the inclusion of curvature reduces the tension by about $-0.4\sigma$ to $-0.7\sigma$ depending on the metric, while increasing the log-Bayes factor from 7.7 to 10.8. This appears to be due to larger values of $\sigma_\mathrm{ini}$ (which is most strongly correlated with $H_0$ in this model) being more compatible with the CMB data when slightly larger $\Omega_\mathrm{m}$ values are considered -- in this case the effect on late time observables can be compensated by a larger neutrino mass sum $\sum m_\nu$ together with positive curvature $\Omega_k > 0$. This can be checked in \cref{app:fig:RnR_Ok}.

\begin{table}[tp]
    \centering
    \begin{subtable}{0.56\textwidth}
        \centering
        \resizebox{\textwidth}{!}{
        \begin{tabular}{ccc}
        \toprule
        Model & $\ln Z$ & $\ln$ BF \\
        \midrule \arrayrulecolor[HTML]{CCCCCC}
\lcdm & $-2935.94\pm0.02$ & $0.00\pm0.02$ \\
\switch & $-2934.21\pm0.11$ & $1.73\pm0.11$ \\
\idedm & $-2939.09\pm0.03$ & $-3.15\pm0.03$ \\ \midrule
\thawing & $-2929.66\pm0.35$ & {\cellcolor[HTML]{FFB3B3}} $6.28\pm0.35$ \\ \midrule
\ede & $-2925.73\pm0.01$ & {\cellcolor[HTML]{FFB3B3}} $10.20\pm0.02$ \\
\emg & $-2929.06\pm0.07$ & {\cellcolor[HTML]{FFB3B3}} $6.88\pm0.07$ \\
\rnr & $-2927.92\pm0.02$ & {\cellcolor[HTML]{FFB3B3}} $8.01\pm0.03$ \\
\coldnede & $-2929.99\pm0.03$ & {\cellcolor[HTML]{FFB3B3}} $5.94\pm0.04$ \\ \midrule
\modrec & $-2936.68\pm0.07$ & $-0.74\pm0.07$ \\
\me & $-2927.92\pm0.06$ & {\cellcolor[HTML]{FFB3B3}} $8.02\pm0.06$ \\ \midrule
\neff & $-2926.61\pm0.03$ & {\cellcolor[HTML]{FFB3B3}} $9.33\pm0.03$ \\
\sidr & $-2926.40\pm0.02$ & {\cellcolor[HTML]{FFB3B3}} $9.54\pm0.02$ \\
\wzdr & $-2926.03\pm0.03$ & {\cellcolor[HTML]{FFB3B3}} $9.91\pm0.03$ \\
\hotnede & $-2931.06\pm0.09$ & {\cellcolor[HTML]{FFB3B3}} $4.88\pm0.09$ \\
        \arrayrulecolor{black}\bottomrule
        \end{tabular}
        }
        \caption{\label{tab:baseline_noACT_evidence} Log-evidence $\ln Z$ and log-Bayes-Factor $\ln \mathrm{BF}$ as defined in \cref{ssec:BF} for the contending models, when no ACT data is included. The uncertainties are the ones reported by MCEvidence, though through tests of small analysis variations of the chains (e.g. thinning), we found that the uncertainty is likely at least $\pm 0.5$.}
    \end{subtable}
    \hfill 
    \begin{subtable}{0.4\textwidth}
        \centering
        \resizebox{\textwidth}{!}{
        \begin{tabular}{ccc}
        \toprule
        Model & $\Delta_\mathrm{shift}$ & $\Delta_\mathrm{Gauss}$ \\
        \midrule \arrayrulecolor[HTML]{CCCCCC}
        \lcdm & $5.8 \sigma$ & $5.8 \sigma$ \\
        \switch & $5.5 \sigma$ & $5.5 \sigma$ \\
        \nosroll & $5.7 \sigma$ & $5.7 \sigma$ \\
        \idedm & $5.5 \sigma$ & $5.5 \sigma$ \\ \midrule
        \thawing & $5.3 \sigma$ & $5.4 \sigma$ \\ \midrule
        \ede & {\cellcolor[HTML]{FFB3B3}} $3.3 \sigma$ & $3.5 \sigma$ \\
        \emg & $3.5 \sigma$ & $3.9 \sigma$ \\
        \rnr & $3.7 \sigma$ & $4.0 \sigma$ \\
        \coldnede & {\cellcolor[HTML]{FFB3B3}} $2.3 \sigma$ & {\cellcolor[HTML]{FFB3B3}} $2.3 \sigma$ \\ \midrule
        \modrec & $5.4 \sigma$ & $5.5 \sigma$ \\
        \me & $3.6 \sigma$ & $3.7 \sigma$ \\ \midrule
        \neff & $3.5 \sigma$ & $3.7 \sigma$ \\
        \sidr & $3.5 \sigma$ & $3.7 \sigma$ \\
        \wzdr & {\cellcolor[HTML]{FFB3B3}} $3.3 \sigma$ & {\cellcolor[HTML]{FFB3B3}} $3.3 \sigma$ \\
        \hotnede & {\cellcolor[HTML]{FFB3B3}} $2.8 \sigma$ & {\cellcolor[HTML]{FFB3B3}} $2.9 \sigma$ \\
        \arrayrulecolor{black}\bottomrule
        \end{tabular}
        }
        \caption{\label{tab:baseline_noACT_tension} Same as \cref{tab:baseline_tension_2}, but without ACT data (and therefore also without the curvature and CPL dark energy extensions).}
    \end{subtable}
    \caption{Tables summarizing the Bayesian performance without ACT data.}
\end{table}

\begin{figure}[tp]
    \centering\makebox[\textwidth][c]{%
    \begin{minipage}{1.2\textwidth}
    \centering
    \includegraphics[scale=0.5]{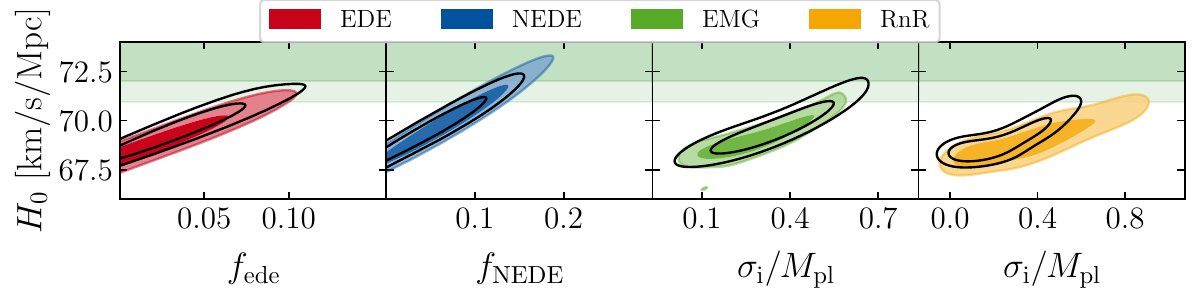}
    \includegraphics[scale=0.5]{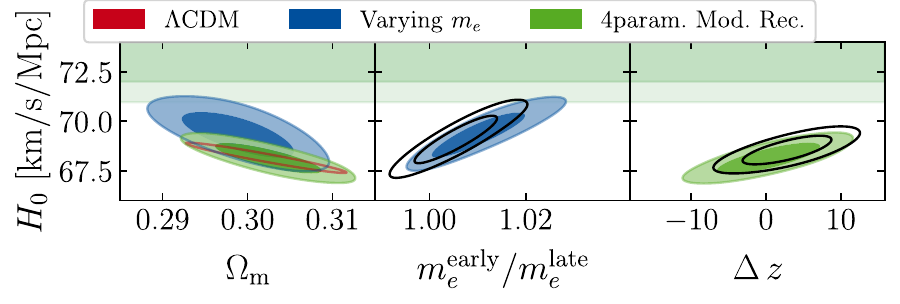} \\
    \includegraphics[scale=0.5]{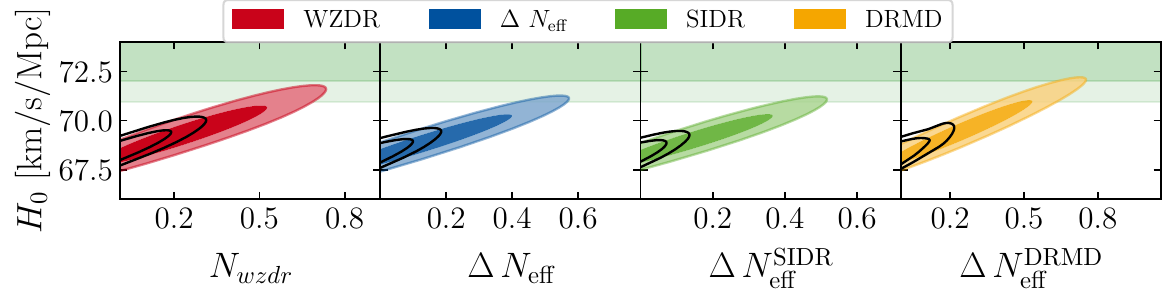}
    \includegraphics[scale=0.5]{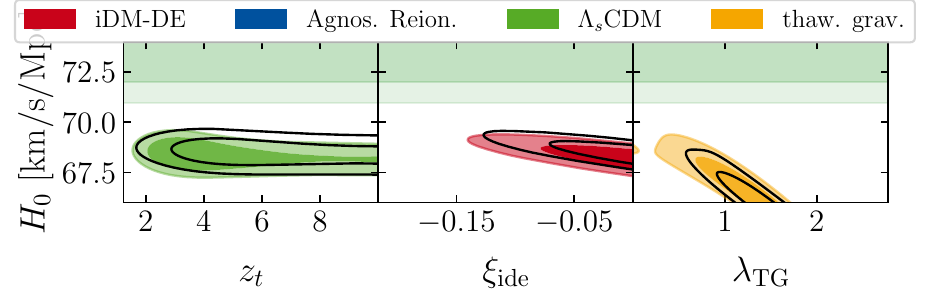}
    \end{minipage}
    }
    \caption{68\% and 95\% credible regions for the models competing in the $H_0$ world cup, without (solid) and with (black lines) ACT data. 
    {\it Top:} Group E. {\it Upper:} Group M. {\it Lower:} Group R. {\it Bottom:} Groups L and E+L. The $x$-axis of each panel includes the parameters most relevant for the Hubble tension in each model.}
    \label{fig:overview_noACT}
\end{figure}

\subsection{No ACT runs} \label{ssec:noACT}

\begin{figure}[tp]
    \centering
    \includegraphics[width=0.95\linewidth]{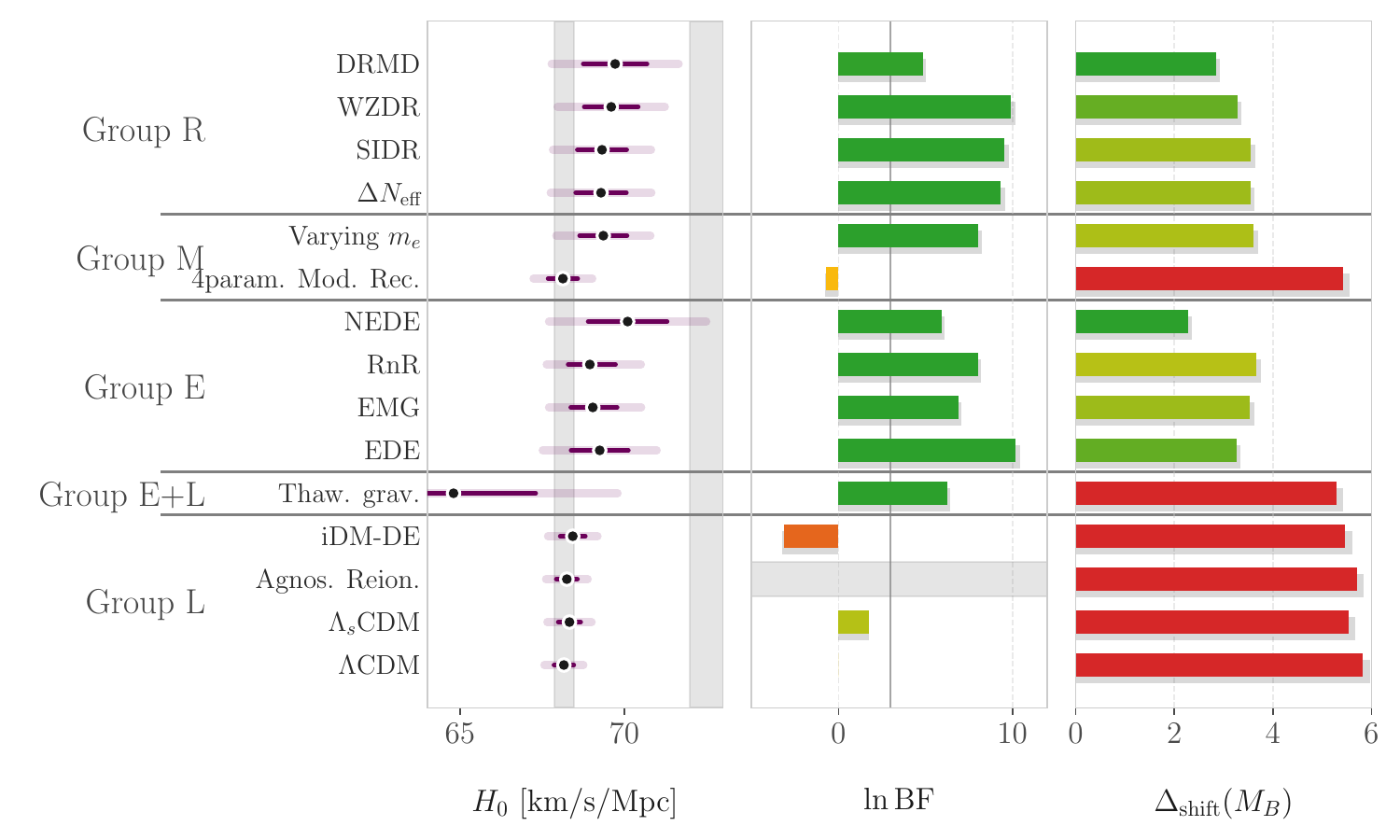}
    \caption{Same as \cref{fig:summary_bayesian}, without ACT data.}
    \label{fig:summary_noACT_bayesian}
\end{figure}

In this competition, many models are subject to significantly tighter constraints than what one would expect by plain extrapolation of increased uncertainty based on previous analyses like \citealp{Schoneberg:2021qvd} that relied solely on Planck CMB data without taking into account the different multipole scales accessed.\footnote{A notable exception is early dark energy models, for which the tension has decreased compared to pre-ACT and pre-DESI data, when it was at the $3.5\sigma$ level \cite{Efstathiou:2023fbn,Chatrchyan:2024xjj}.} In most cases, the reason for this discrepancy is the interplay between ACT and Planck data leading to extremely strong constraints on deviations from $\Lambda$CDM, especially towards DR. The level of consistency of the ACT and Planck+SPT data has been a matter of debate, given the $2-3\sigma$ tension between the value of $N_\mathrm{eff}$ inferred from these CMB datasets \cite{SPT-3G:2025bzu,Tristram:2025you}. It is therefore instructive to assess the impact of ACT on the constraints and test their robustness when ACT data are excluded. We run analyses for the CMB+BAO+SN combination without any ACT data (lensing and primary CMB), while keeping the truncated Planck $\ell$-range. In this case we find much more permissive constraints, as shown in \cref{fig:overview_noACT,fig:summary_noACT_bayesian} as well as in \cref{tab:baseline_noACT_evidence,tab:baseline_noACT_tension}. 

We observe that group R is benefiting the most from the removal of ACT data, followed by group M, with the other groups showing relatively stable results. The metrics for group R decrease from around $5\sigma$ with ACT data to $3-3.5\sigma$ without it -- the biggest improvement being for the DRMD model. In group M, the tension also improves from roughly $4\sigma$ to $3.6\sigma$ for the varying electron mass model. This confirms that ACT data are preventing several models from efficiently reducing the tension, especially when they involve DR.

The tension metrics $\Delta_{\rm shift}$ and $\Delta_{\rm Gauss}$ remain roughly at the 5.5$\sigma$ level for $\Lambda$CDM and models of groups L and E+L, whereas the Bayes factor improves quite a bit for the thawing gravity model. While the significantly lower mean value of $H_0$ for thawing gravity prevents it from strongly reducing the tension in this case, it does generate a good fit over a wide range of $H_0$ values, reflecting in the large uncertainty and higher Bayes factor.

A dedicated analysis is needed to pin down the reason of the large impact of ACT data on models of groups R and M. We note that ACT has a slight preference for higher power at high $\ell$, c.f. \citealp[Figs.~41,42]{ACT:2025blo} and \citealp[Fig.~3]{Poulin:2025nfb}. We therefore surmise that its impact can be understood from the impact of each model on the high-$\ell$ tail of the CMB power spectra (see in particular ~\citealp{Smith:2025zsg}). It is well known that models with enhanced radiation lead to a decrease of power on small scales \cite{Hou:2011ec,Lesgourgues:2018ncw,Smith:2025zsg}, while EDE instead shows a strong upward curvature towards large~$\ell$ \cite{Poulin:2023lkg,Smith:2023oop,Smith:2025arq}. The varying electron mass model \enquote{sits in between}, leaving the high-$\ell$ tail similar to $\Lambda$CDM \cite{Sekiguchi:2020teg,Schoneberg:2024ynd,Smith:2025arq} 
The larger high-$\ell$ tail in ACT data is therefore more beneficial for group E models than for group R and group M.

As shown in \citealp{Garny:2026gcs} a simple way to alter the high-$\ell$ spectra and affect ACT constraints is to change the shape of the primordial power spectrum. We further discuss the impact of running and running of running on the other models in \cref{ssec:running} below.
\footnote{Another possibility suggested in the literature to weaken constraints on model with extra radiation is to exploit the \enquote{mirror world} symmetry of CMB observables \cite{Cyr-Racine:2021oal,Ge:2022qws,Greene:2023cro,Greene:2024qis}. We leave investigating this possibility for future work.}

\subsection{Running of the primordial power spectrum} \label{ssec:running}

\begin{figure}[tp]
    \centering
    \includegraphics[width=0.48\linewidth]{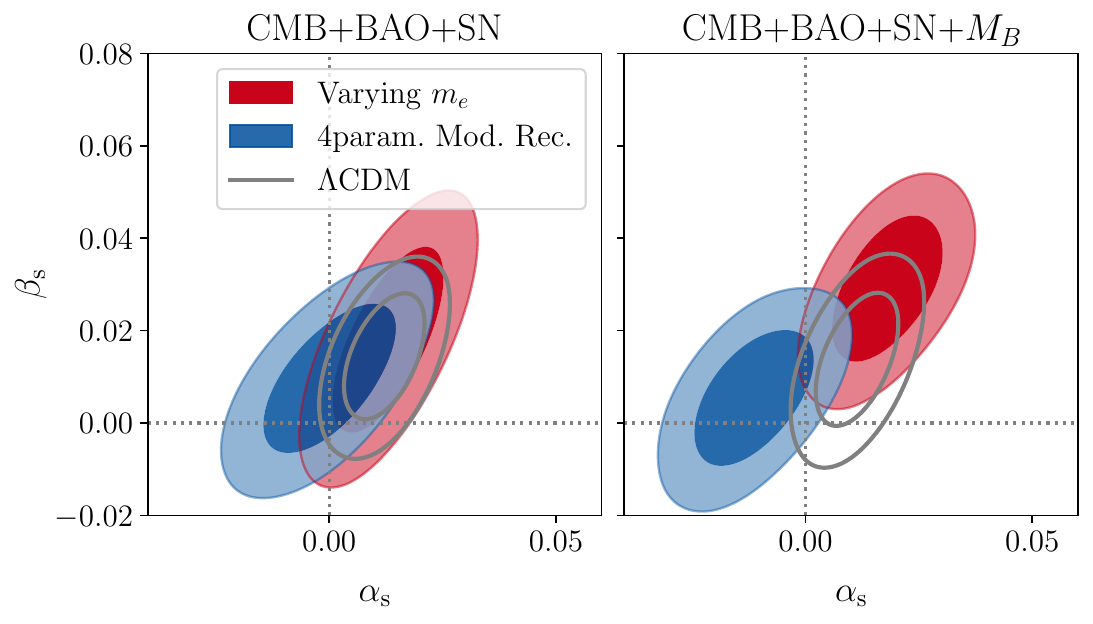}
    \includegraphics[width=0.48\linewidth]{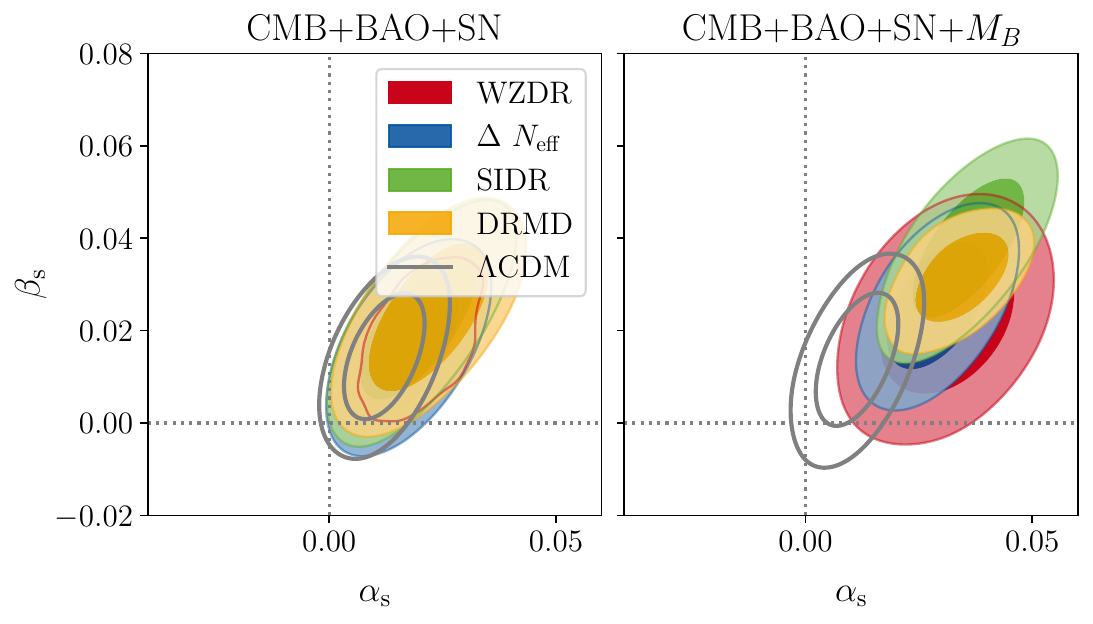}
    \includegraphics[width=0.48\linewidth]{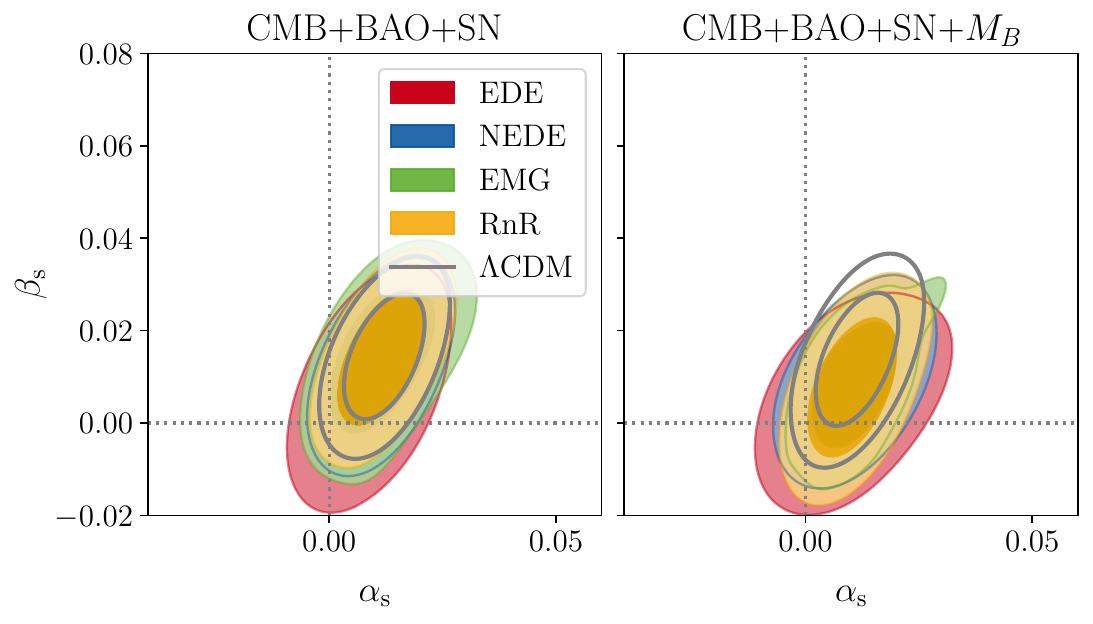}
    \includegraphics[width=0.48\linewidth]{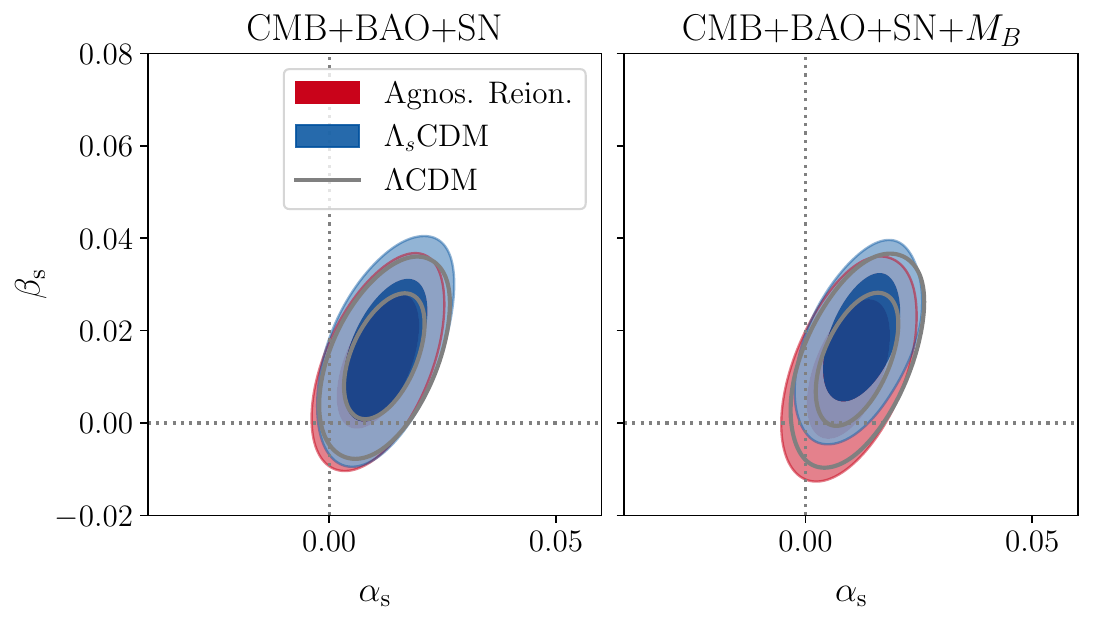}
    \caption{68\% and 95\% credible regions for the models competing in the $H_0$ world cup, when running and running of the running of the primordial power spectrum is assumed. The constraints are shown for the corresponding parameters $\alpha_s$ and $\beta_s$\,, where the point $\alpha_s=0$ and $\beta_s=0$ (dotted gray lines) corresponds to the baseline case. {\it Top left:} Group E. {\it Top right:} Group M. {\it Bottom left:} Group R. {\it Bottom right:} Group L. We include the $\Lambda$CDM+$\alpha_s,\beta_s$ constraints as gray line contours in all cases for comparison.}
    \label{fig:ab_2d}
\end{figure}

\begin{figure}[tp]
    \centering\makebox[\textwidth][c]{%
    \begin{minipage}{1.2\textwidth}
    \centering
    \includegraphics[scale=0.42]{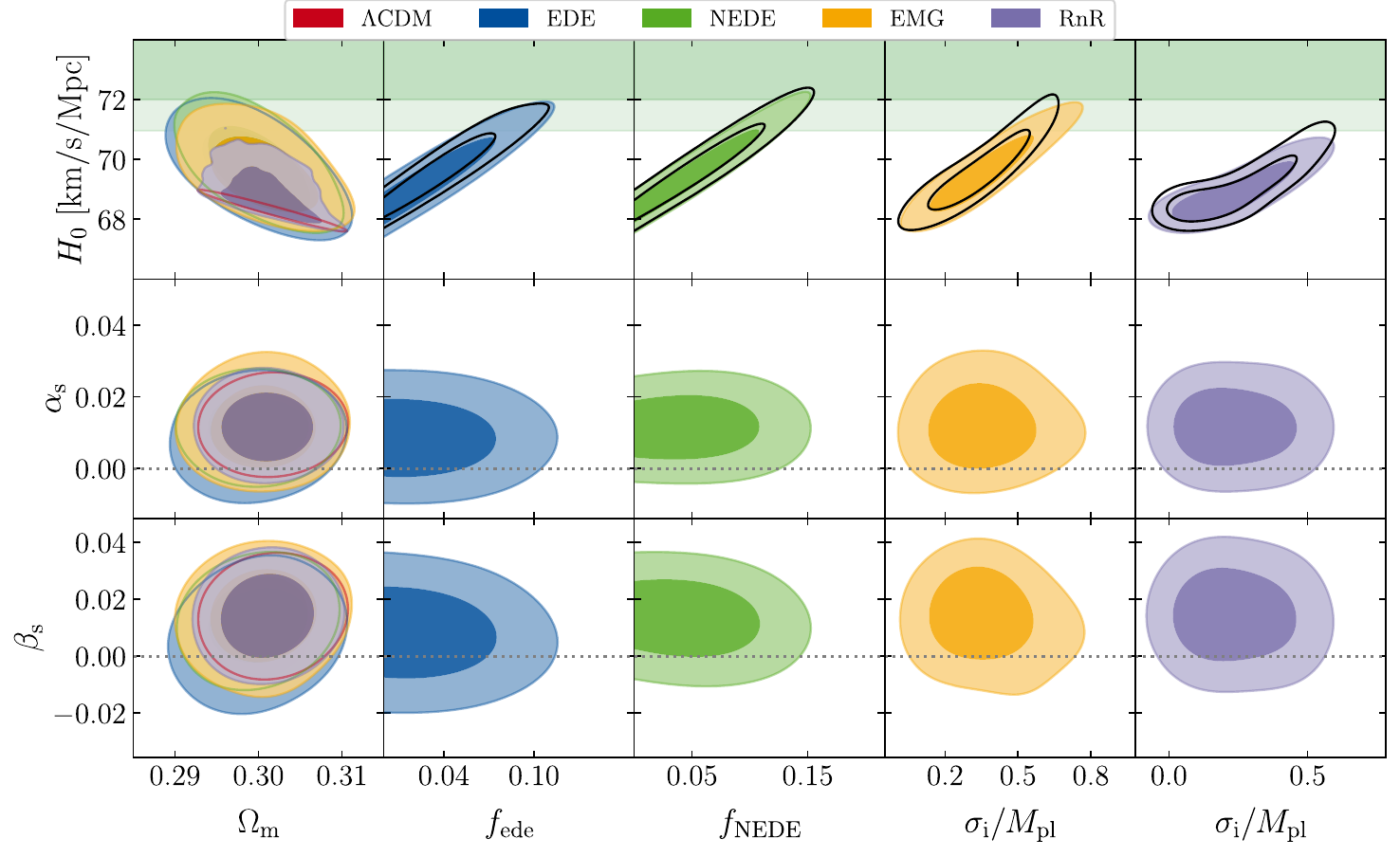}
    \includegraphics[scale=0.42]{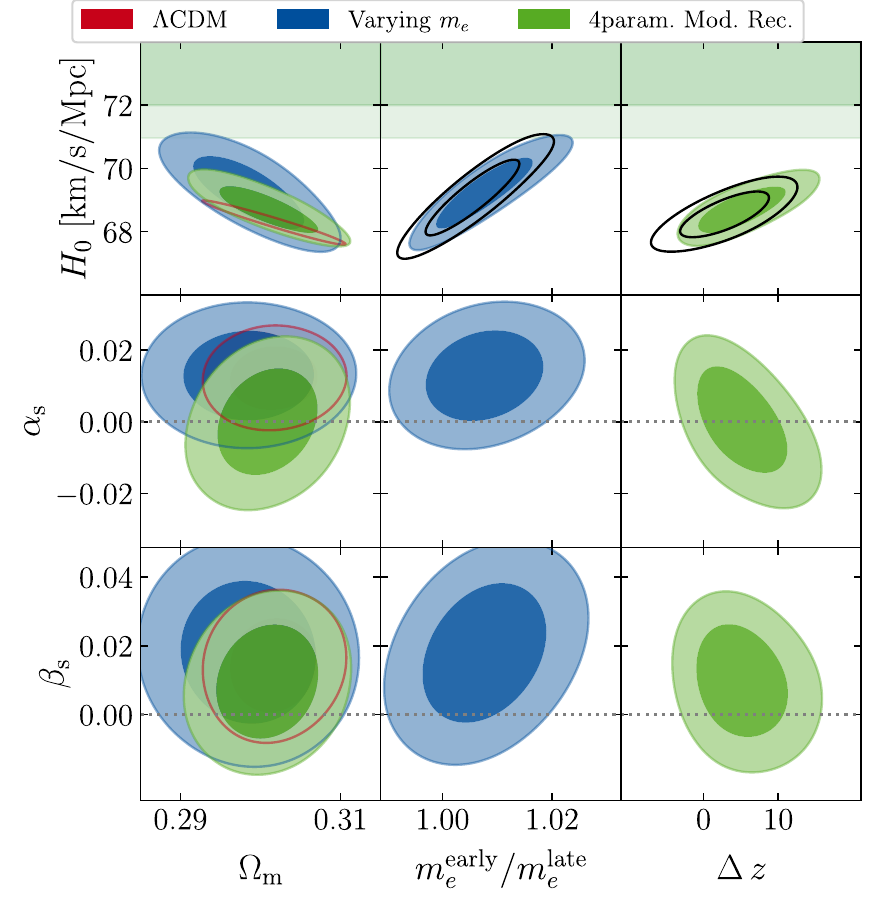} \\
    \includegraphics[scale=0.42]{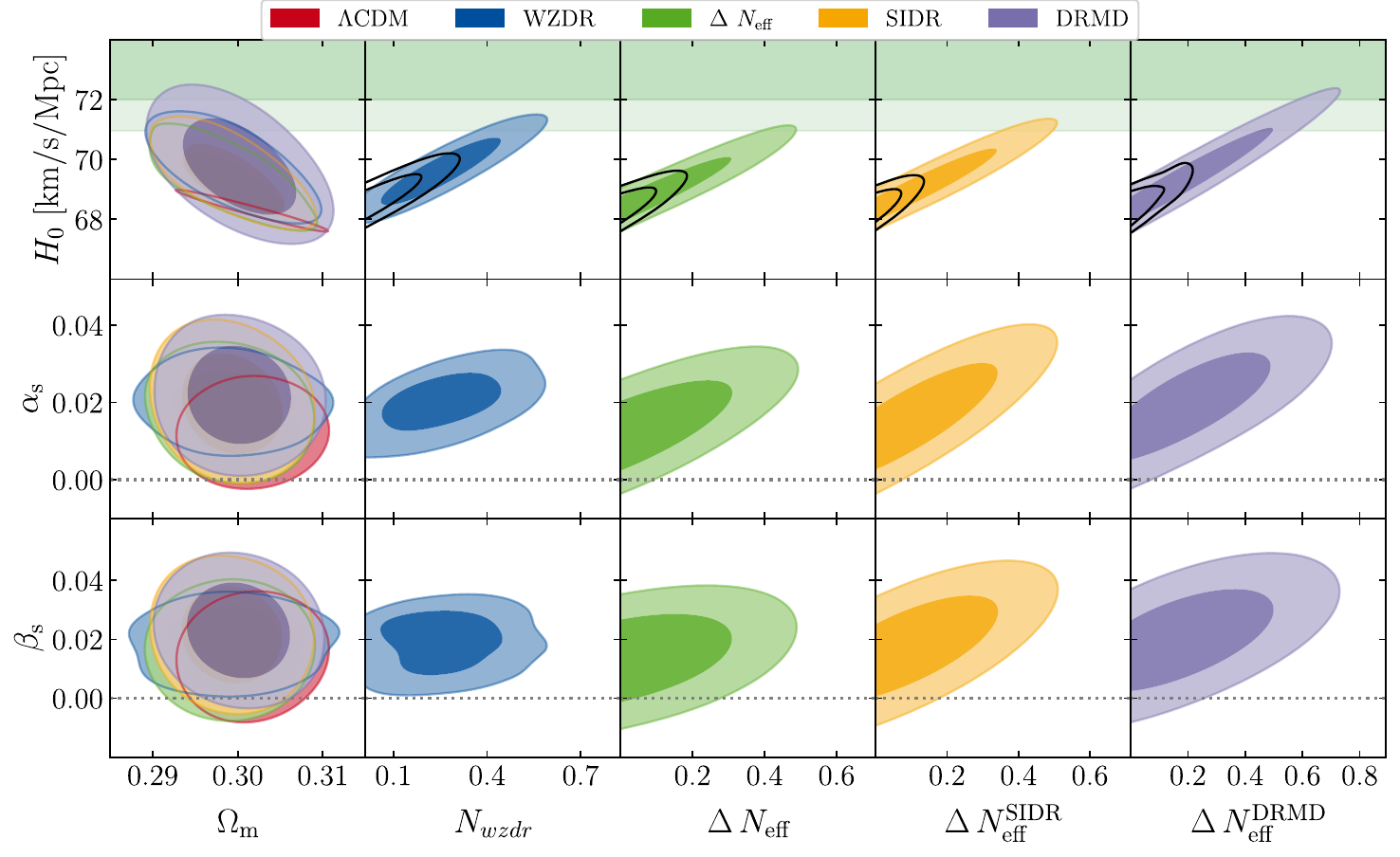}
    \includegraphics[scale=0.42]{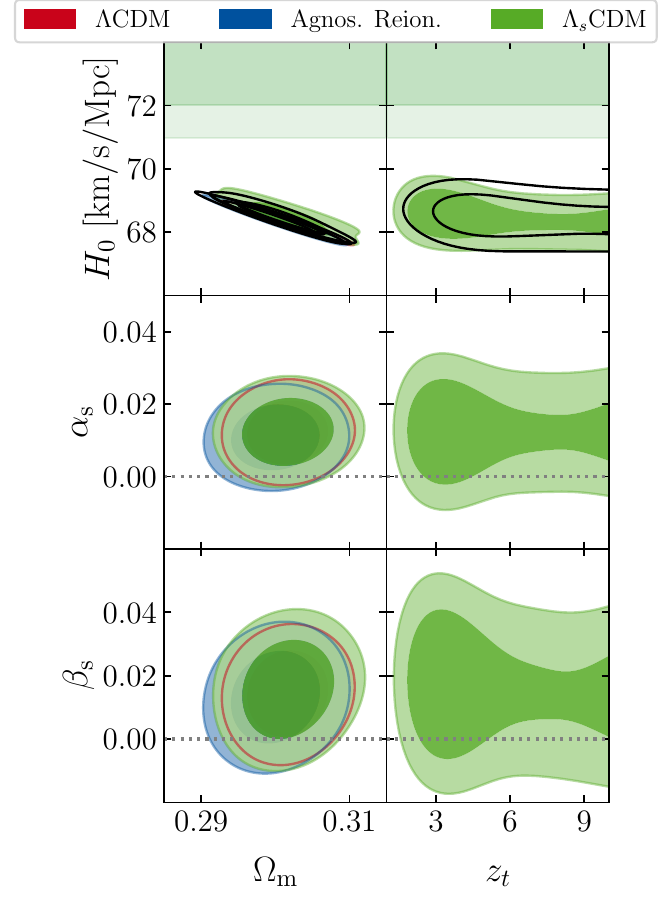}
    \end{minipage}
    }
    \caption{68\% and 95\% credible regions for the models competing in the $H_0$ world cup, with (solid) and without (black lines) assuming a running and running of running in the primordial power spectrum for the CMB+BAO+SN dataset. The baseline case corresponds to $\alpha_s=\beta_s=0$ (dotted gray lines). {\it Top left:} Group E. {\it Top right:} Group M. {\it Bottom left:} Group R. {\it Bottom right:} Group L. We include the $\Lambda$CDM+$\alpha_s,\beta_s$ model in all cases for comparison. The $x$-axis of each panel includes $\Omega_\mathrm{m}$ and the parameters most relevant for the Hubble tension in each model. The black lines show the baseline model performance for comparison.}
    \label{fig:onepar_ab}
\end{figure}

Most inflationary models predict a deviation from a primordial power spectrum $P_\mathcal{R}(k)$ described by a simple power-law, which is the result of the simple single-field paradigm to lowest order in a slow-roll expansion. Higher orders in the slow-roll expansion naturally lead to a running of the spectral index ($\alpha_s$), or a running of the running of this index ($\beta_s$)~\cite{Kosowsky:1995aa}. 

\noindent In general, one can write down a primordial power spectrum expansion of the form
\begin{equation}
    \ln P_\mathcal{R}(k)=\ln A_s+(n_s-1) \ln\left(\frac{k}{k_*}\right)+\frac{1}{2!} \alpha_s \ln\left(\frac{k}{k_*}\right)^2 + \frac{1}{3!} \beta_s \ln\left(\frac{k}{k_*}\right)^3 + \ldots~,
\end{equation}
with $k_* = 0.05/\mathrm{Mpc}$ the conventional reference scale. In typical slow-roll inflation, higher-order terms are expected to be suppressed by additional powers of the small slow-roll parameters, leading to a natural hierarchy $\beta_s \ll \alpha_s \ll n_s-1$~\cite{Easther:2006tv}, such that these additional corrections would be negligible. However, if slow roll inflation is driven by multiple fields (see e.g. ~\cite{Bjorkmo:2017nzd}) or if there are deviations from the slow-roll conditions (see e.g. \cite{Motohashi:2017kbs,Kobayashi:2010pz}), the higher order terms can become relevant. We truncate the series at third order and check if the data prefers significant deviations from the traditional slow roll limit, which we approximate as the point at which $\alpha_s=\beta_s=0$.

\enlargethispage*{1em}
More general power spectrum modifications have been shown to lead to an increase in the inferred value of $H_0$ from CMB data \cite{Hazra:2022rdl,Antony:2022ert,Lee:2025yah,Calderon:2025xod}, often involving oscillatory modifications of the primordial power spectrum. We leave analyses with such oscillatory features in the primordial power spectrum to future work.

\begin{table}[tp]
    \centering
    \begin{subtable}{0.60\textwidth}
        \centering
        \resizebox{\textwidth}{!}{
\begin{tabular}{ccc}
\toprule
Model & $\ln Z$ & $\ln$ BF \\
\midrule \arrayrulecolor[HTML]{CCCCCC}
\lcdm & $-3023.49\pm0.07$ & $0.00\pm0.10$ \\
\switch & $-3022.11\pm0.25$ & $1.38\pm0.26$ \\
\idedm & --- & --- \\ \midrule
\thawing & $-3034.39\pm0.87$ & $-10.90\pm0.87$ \\ \midrule
\ede & $-3014.06\pm0.29$ & {\cellcolor[HTML]{FFB3B3}} $9.43\pm0.30$ \\
\emg & $-3016.89\pm0.12$ & {\cellcolor[HTML]{FFB3B3}} $6.60\pm0.14$ \\
\rnr & $-3017.09\pm0.16$ & {\cellcolor[HTML]{FFB3B3}} $6.40\pm0.18$ \\ \midrule
\coldnede & $-3021.48\pm0.06$ & $2.00\pm0.09$ \\
\modrec & $-3026.13\pm0.20$ & $-2.64\pm0.21$ \\
\me & $-3016.63\pm0.26$ & {\cellcolor[HTML]{FFB3B3}} $6.86\pm0.27$ \\ \midrule
\neff & $-3015.59\pm0.05$ & {\cellcolor[HTML]{FFB3B3}} $7.89\pm0.09$ \\
\sidr & $-3014.84\pm0.04$ & {\cellcolor[HTML]{FFB3B3}} $8.65\pm0.08$ \\
\wzdr & $-3014.52\pm0.10$ & {\cellcolor[HTML]{FFB3B3}} $8.97\pm0.12$ \\
\hotnede & $-3018.27\pm0.27$ & {\cellcolor[HTML]{FFB3B3}} $5.22\pm0.28$  \\
\arrayrulecolor{black}\bottomrule
\end{tabular}
        }
        \caption{\label{tab:baseline_ab_evidence} Same as \cref{tab:baseline_noACT_evidence} but for a model with $\alpha_s$ and $\beta_s$ parameters (and with the usual ACT data).}
    \end{subtable}
    \hfill 
    \begin{subtable}{0.36\textwidth}
        \centering
        \resizebox{\textwidth}{!}{
\begin{tabular}{ccc}
\toprule
Model & $\Delta_\mathrm{shift}$ & $\Delta_\mathrm{Gauss}$ \\
\midrule \arrayrulecolor[HTML]{CCCCCC}
\lcdm & $5.7 \sigma$ & $5.7 \sigma$ \\
\switch & $5.3 \sigma$ & $5.4 \sigma$ \\
\nosroll & $5.6 \sigma$ & $5.6 \sigma$ \\
\idedm & --- & --- \\ \midrule
\thawing & $5.1 \sigma$ & $5.4 \sigma$ \\ \midrule
\ede & {\cellcolor[HTML]{FFB3B3}} $3.1 \sigma$ & {\cellcolor[HTML]{FFB3B3}} $3.2 \sigma$ \\
\emg & {\cellcolor[HTML]{FFB3B3}} $3.1 \sigma$ & {\cellcolor[HTML]{FFB3B3}} $3.3 \sigma$ \\
\rnr & $4.0 \sigma$ & $4.3 \sigma$ \\
\coldnede & {\cellcolor[HTML]{FFB3B3}} $2.9 \sigma$ & {\cellcolor[HTML]{FFB3B3}} $3.0 \sigma$ \\ \midrule
\modrec & $5.0 \sigma$ & $4.9 \sigma$ \\
\me & $3.9 \sigma$ & $3.9 \sigma$ \\ \midrule
\neff & $3.6 \sigma$ & $3.9 \sigma$ \\
\sidr & {\cellcolor[HTML]{FFB3B3}} $3.5 \sigma$ & $3.6 \sigma$ \\
\wzdr & {\cellcolor[HTML]{FFB3B3}} $3.4 \sigma$ & $3.5 \sigma$ \\
\hotnede & {\cellcolor[HTML]{FFB3B3}} $2.9 \sigma$ & {\cellcolor[HTML]{FFB3B3}} $3.1 \sigma$ \\
\arrayrulecolor{black}\bottomrule
\end{tabular}
        }
        \caption{\label{tab:baseline_ab_tension} Same as \cref{tab:baseline_noACT_tension}, but with an extension of running with the $\alpha_s$ and $\beta_s$ parameters.}
    \end{subtable}
    \caption{Tables summarizing the Bayesian performance when running parameterized by $\alpha_s$ and $\beta_s$ is allowed.}
\end{table} 
This beyond slow-roll analysis is especially interesting in relation to the aforementioned preference for more power on small scales in the ACT data. For example, in the $\Lambda$CDM+$\alpha_s$+$\beta_s$ model the CMB+BAO+SN data combination yields $\alpha_s = 0.0122 \pm 0.0057$ and $\beta_s = 0.0144 \pm 0.088$, clearly indicating a preference for an upturn at large $k$, corresponding to large $\ell$. In fact, it has been found that in the SIDR and DRMD models the combination of Planck+ACT data favor non-zero $\alpha_s$ and $\beta_s$, while allowing for significantly larger $H_0$ \cite{Garny:2026gcs}. Here, we obtain similar results, $\alpha_s = 0.0198 \pm 0.0081$, $\beta_s = 0.022 \pm 0.010$, and $H_0 = (69.40 \pm 0.74) \mathrm{km/s/Mpc}$ where \citealp{Garny:2026gcs} with only Planck+ACT data finds $\alpha_s = 0.024 \pm 0.0082$, $\beta_s = 0.0264^{+0.012}_{-0.061}$ and $H_0 = 69.94^{+0.84}_{-1.1}$, suggesting that SPT data does not significantly alter these conclusions apart from slightly improving the constraint on the deviation from $\Lambda$CDM and the corresponding ability to ease the Hubble tension.

A preference for deviation from the slow-roll expectation ($\alpha_s\approx 0$, $\beta_s\approx 0$) is also seen for other models, as shown in \cref{fig:ab_2d,fig:onepar_ab}, including for $\Lambda$CDM+$\alpha_s+\beta_s$\,. As visible in \cref{fig:ab_2d}, models of group R favor even larger values of $\alpha_s$ and $\beta_s$ than in the $\Lambda$CDM+$\alpha_s$+$\beta_s$ case. The varying $m_e$ also shows a similar (albeit smaller) preference. In contrast, models in groups E and L yield results that are more compatible with a simple power-law than the $\Lambda$CDM+$\alpha_s$+$\beta_s$ case when a prior on $M_B$ is imposed. 
The \modrec{} case has entirely different contours in this case, both with and without a prior on $M_B$, remaining roughly compatible with $\alpha_s=0$ and preferring only slightly positive values of $\beta_s$. 

Correspondingly, as visible in \cref{fig:onepar_ab}, models of group R especially benefit from the expansion through $\alpha_s$ and $\beta_s$, in good agreement with \citealp{Garny:2026gcs}. In particular, \cref{fig:onepar_ab} highlights that the strong reduction in tension (up to the $3.5\sigma$ level) for group R reported in \cref{tab:baseline_ab_tension} is only compatible with simultaneously strong deviations of $\alpha_s$ and $\beta_s$ from zero. \Cref{fig:onepar_ab} also highlights that the other models do not benefit from these additional degrees of freedom in terms of their ability to reduce the tension.

Finally, it is worth pointing out that the $\alpha_s$/$\beta_s$ solutions preferred here do not significantly alter the constraints on $S_8$ -- without $\alpha_s$/$\beta_s$\,, the average $S_8$ is 0.829, while with $\alpha_s$/$\beta_s$ the average is $S_8$ is 0.826. The shift is small compared to the average uncertainty of $0.007$. However, the small-scale sensitivity of galaxy clustering \cite{Ivanov:2019pdj,DAmico:2019fhj,DESI:2024hhd} and the Lyman-$\alpha$ forest \cite{Palanque-Delabrouille:2019iyz,Goldstein:2023gnw,Rogers:2023upm,Fernandez:2023grg,Walther:2024tcj,Chaves-Montero:2026hqd} on models with an enhancement or suppression of the power spectrum could restrict these models further.  We leave a dedicated analysis for future work.

\subsection{Selection of finalists}\label{sec:results}

Let us recall that we select the finalist models according to thresholds defined in \cref{ssec:thresholds}, which require a significant improvement in the combined fit (with the $M_B$ prior). A model passes to the knockout round whenever $-\Delta \mathrm{AIC} > 10$ or $\ln {\rm BF} > 3$.

It is clear from \cref{sec:competition} and especially \cref{ssec:baseline} that group E models outperform other groups according to the model preference metrics. The varying electron mass model from group M is the only other model which barely passes the thresholds. Finally, no models in groups R, L, or E+L pass the threshold. Nevertheless, we have seen in \cref{ssec:noACT} and \cref{fig:summary_noACT_bayesian} that group R models would pass the thresholds in absence of ACT data or when supplemented with a running of the power spectrum. To illustrate how they perform in the additional tests of the knockout round, we select as a `wild card' and for reference, the best-performing model from this group, which is the WZDR model.

An important question that arises is whether the tests of the knockout round should be carried out on the finalist models considered only in their simplest form (with a cosmological constant and assuming flatness), or also in their extended versions (with CPL dark energy, spatial curvature, or a deviation from a power law primordial power spectrum). We decide to focus on models in their simplest form to reduce computational complexity. However, we make an exception for the \enquote{gold medal winner} of \citealp{Schoneberg:2021qvd}, the varying $m_e$ model with curvature, which is shown for reference. Extended models that we mention as potential finalists but which are not included are the curved EMG and RnR models, as well as all group R models with $\alpha_s/\beta_s$. Note that for all models, the ability to mitigate the Hubble tension remains unchanged when considering an evolving DE with a CPL parametrization.

\section{The knockout round: additional tests on the finalist models}\label{sec:golden}

\begin{figure}[tp]
    \centering\makebox[\textwidth][c]{%
    \includegraphics[width=1.1\linewidth]{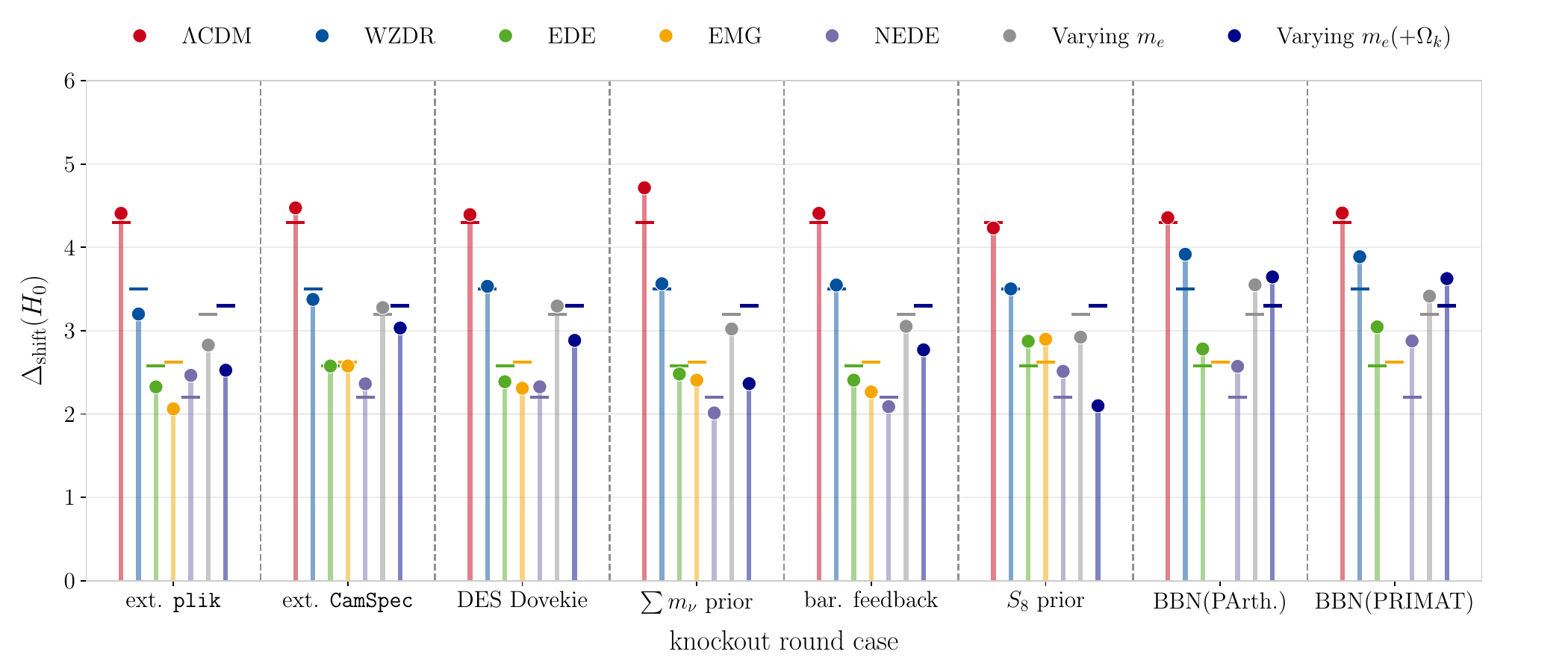}
    }
    \caption{Summary of the results of \cref{sec:golden} in terms of the shift metric $\Delta_\mathrm{shift}$ for the parameter $H_0$\,. The small horizontal lines correspond to the results 
    in the baseline analysis of \cref{ssec:baseline} and the dots to the additional tests of the knockout round.}
    \label{fig:golden_summary}
\end{figure}

We perform additional tests and validations of the finalist models, using new data combinations listed in \cref{ssec:data_golden} below. Our goal is to assess the extent to which the conclusions drawn in \cref{ssec:baseline} hold up when subject to well-founded variations, such as adding or slightly changing the used datasets, changing the neutrino mass priors, or marginalizing over the impact of baryonic feedback.  For simplicity and brevity, we gauge the impact of these different choices solely on the Bayesian tension metric $\Delta_{\rm shift}$, reported in \cref{fig:golden_summary,tab:golden_summary}.

\subsection{Data for the knockout round}\label{ssec:data_golden}

\begin{table}[!htp]
    \centering
\resizebox{\textwidth}{!}{
\begin{tabular}{cccccccc}
\toprule
DataSet & $\Lambda$CDM & WZDR & EDE & EMG & NEDE & Varying $m_e$ & Varying $m_e$(+$\Omega_k$) \\
\midrule
baseline & $5.6\sigma (4.3\sigma)$ & $4.5\sigma (3.5\sigma)$ & $3.0\sigma (2.6\sigma)$ & $2.9\sigma (2.6\sigma)$ & $2.7\sigma (2.2\sigma)$ & $4.0\sigma (3.2\sigma)$ & $4.1\sigma (3.3\sigma)$ \\
ext. \texttt{plik} & $5.6\sigma (4.4\sigma)$ & $4.0\sigma (3.2\sigma)$ & $2.8\sigma (2.3\sigma)$ & $2.5\sigma (2.1\sigma)$ & $3.0\sigma (2.5\sigma)$ & $3.6\sigma (2.8\sigma)$ & $3.0\sigma (2.5\sigma)$ \\
ext. \texttt{CamSpec} & $5.7\sigma (4.5\sigma)$ & $4.1\sigma (3.4\sigma)$ & $3.2\sigma (2.6\sigma)$ & $3.2\sigma (2.6\sigma)$ & $2.9\sigma (2.4\sigma)$ & $4.2\sigma (3.3\sigma)$ & $3.6\sigma (3.0\sigma)$ \\
DES Dovekie & --- $ (4.4\sigma)$ & --- $ (3.5\sigma)$ & --- $ (2.4\sigma)$ & --- $ (2.3\sigma)$ & --- $ (2.3\sigma)$ & --- $ (3.3\sigma)$ & --- $ (2.9\sigma)$ \\
$\sum m_\nu$ prior & $5.9\sigma (4.7\sigma)$ & $4.3\sigma (3.6\sigma)$ & $3.1\sigma (2.5\sigma)$ & $2.9\sigma (2.4\sigma)$ & $2.5\sigma (2.0\sigma)$ & $3.8\sigma (3.0\sigma)$ & $2.9\sigma (2.4\sigma)$ \\
bar. feedback & $5.6\sigma (4.4\sigma)$ & $4.5\sigma (3.5\sigma)$ & $3.0\sigma (2.4\sigma)$ & $2.8\sigma (2.3\sigma)$ & $2.6\sigma (2.1\sigma)$ & $3.8\sigma (3.1\sigma)$ & $3.5\sigma (2.8\sigma)$ \\
$S_8$ prior & $5.5\sigma (4.2\sigma)$ & $4.5\sigma (3.5\sigma)$ & $3.5\sigma (2.9\sigma)$ & $3.6\sigma (2.9\sigma)$ & $3.1\sigma (2.5\sigma)$ & $3.7\sigma (2.9\sigma)$ & $2.7\sigma (2.1\sigma)$ \\
BBN(PArth.) & $5.6\sigma (4.4\sigma)$ & $5.1\sigma (3.9\sigma)$ & $3.5\sigma (2.8\sigma)$ & --- $ ($---$)$ & $3.2\sigma (2.6\sigma)$ & $4.5\sigma (3.6\sigma)$ & $4.5\sigma (3.6\sigma)$ \\
BBN(PRIMAT) & $5.7\sigma (4.4\sigma)$ & $4.9\sigma (3.9\sigma)$ & $3.7\sigma (3.0\sigma)$ & --- $ ($---$)$ & $3.5\sigma (2.9\sigma)$ & $4.3\sigma (3.4\sigma)$ & $4.5\sigma (3.6\sigma)$ \\
\bottomrule
\end{tabular}
}
    \caption{Bayesian shift metric on $M_B$ (on $H_0$) for the models in the knockout round, evaluated for the different cases investigated in \cref{sec:golden}. The DES Y5 Dovekie case has a different SNeIa calibration, and hence no $M_B$ tension can be computed here. Similarly, the EMG case is excluded from the BBN tests, see \cref{ssec:bbn}.}
    \label{tab:golden_summary}
\end{table}

\begin{itemize}
    \item \textbf{Extended Planck variations:}
    \begin{itemize}
        \item In the `extended \texttt{CamSpec}' variation, we use the very same likelihoods, but we include Planck data until $\ell=1800$ in TT, $\ell=1050$ in TE, and $\ell=600$ in EE, cutting the ACT and SPT data correspondingly. These transition values are chosen such that Planck and ACT have roughly similar constraining power at the corresponding multipoles. Thus, this choice maximizes the overall constraining power from the CMB.
        \item In the `extended \texttt{plik}' variation, we use the same transition multipoles as in the `extended \texttt{CamSpec}' variation, but we replace the \texttt{CamSpec} likelihood by the \texttt{plik} one, marginalizing over all nuisance parameters with the priors described in \citealp{Planck:2019nip}.
    \end{itemize}
    \item \textbf{DES Y5 Dovekie}: For uncalibrated SN data, we substitute Pantheon+ with the Dovekie-recalibrated DES Y5 dataset~\cite{DES:2025sig}. 
    \item $\boldsymbol{S_8}$ \textbf{prior}: We add a Gaussian prior on $S_8= 0.789 \pm 0.012$ corresponding to the result of the DES Y6 3x2pt photometric analysis \cite{DES:2026fyc}. 
    \item \textbf{BBN}: We use the \texttt{PRyMordial} code to compute the light element abundances in a grid of ($\Omega_\mathrm{b} h^2$, $\Delta N_\mathrm{eff}$ or $m_\mathrm{e}^\mathrm{early}/m^\mathrm{late}_\mathrm{e}$) using either the \texttt{PArthENoPE} or \texttt{PRIMAT} reaction rates. These abundances are then compared to \citealp{Cooke:2017cwo} which measures the deuterium fraction $D/H = (2.527\pm 0.030) \cdot 10^{-5}$ and \citealp{Aver:2026dxv} which measures the Helium abundance $Y_p = 0.2458 \pm 0.0013$. 
    Similarly to \citealp{Garramone:2026evc}, we set \texttt{tau\_n\_flag=False} and let the \texttt{PRyMordial} code self-consistently determine the neutron lifetime.
\end{itemize}

\subsection{Extended Planck multipoles}\label{ssec:full_planck}

The cuts in CMB multipoles in our baseline analysis of \cref{ssec:baseline} allow for easier comparison to previous literature such as \citealp{SPT-3G:2025bzu,ACT:2025blo}, but are not tuned to optimize the overall CMB constraining power. Indeed, they were chosen to include all ACT and/or SPT data and avoid covariance with Planck, which is hard to estimate. Here, instead, we attempt to build a combined CMB dataset that (approximately) maximizes constraining power. We do so by comparing the binned uncertainties from each experiment and choosing a multipole cut that approximately coincides with the transition from Planck-dominated to ground-based-experiment-dominated error bars, similarly to the approach of \citealp{Antony:2024vrx}. Consequently, these constraints incorporate more of the Planck CMB data compared to the baseline cuts. 

\begin{figure}[tp]
    \centering
    \includegraphics[width=0.99\linewidth]{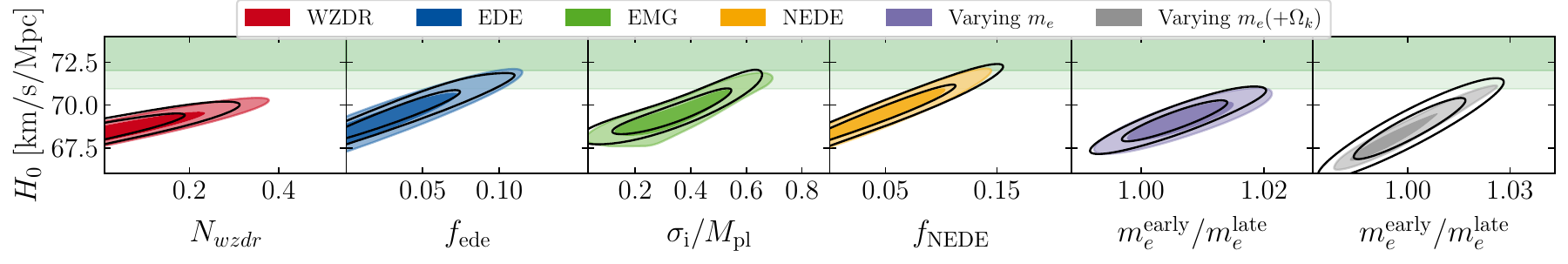}
    \caption{68\% and 95\% parameter constraints for the knockout round models when the extended \texttt{CamSpec} configuration is used. The black lines correspond to the CMB+BAO+SN constraints from the baseline analysis of \cref{ssec:baseline}. The green bands show the 68\% and 95\% constraints on $H_0$ from SH0ES.}
    \label{fig:full_npipe}
\end{figure}

\begin{figure}[tp]
    \centering
    \includegraphics[width=0.99\linewidth]{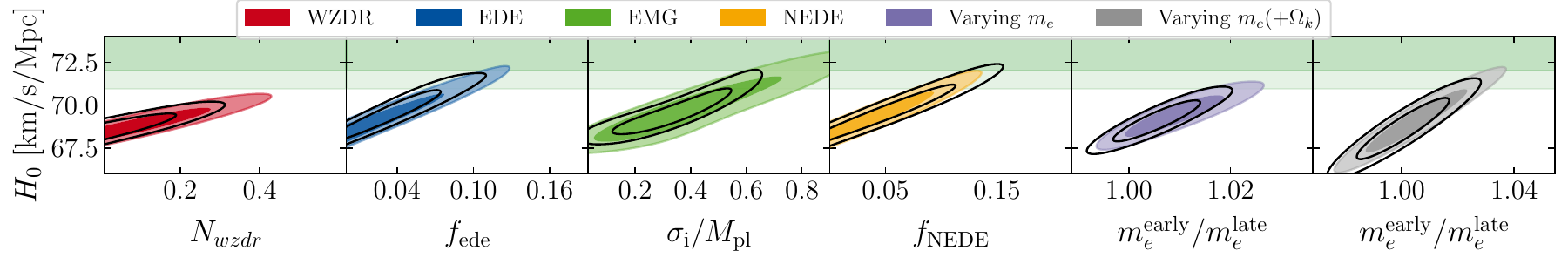}
    \caption{Same as \cref{fig:full_npipe}, but for the extended \texttt{plik} configuration.}
    \label{fig:full_plik}
\end{figure}

In general, the constraints do not shift very significantly, as shown in \cref{fig:full_npipe,fig:full_plik}.
Between the baseline and `extended \texttt{CamSpec}' analysis, the tension level fluctuates by at most $\pm 0.4\sigma$. All models except for the WZDR and \me{}($+\Omega_k$) models see an increase in the tension by about $0.2\sigma$, consistent with an increase in the constraining power of the `extended \texttt{CamSpec}' compared to the `baseline' configuration. Instead, the WZDR model shows a decrease in the tension from $4.5\sigma$ to $4.1\sigma$ in $M_B$, consistent with the preference in ACT data against a large $N_\mathrm{wzdr}$, while the \me{}($+\Omega_k$) model also shows a reduction from $4.1\sigma$ to $3.6\sigma$. 

The `extended \texttt{plik}' configuration, on the other hand, decreases the tension almost universally (by roughly $-0.5\sigma$), with the only exception being the NEDE model, for which the tension increases slightly. As before, the WZDR and \me($+\Omega_k$) models show the largest change ($-0.6\sigma$ and $-1.1\sigma$, respectively), though the EMG model and the varying $m_e$ model also see a significant decrease in the tension (both $-0.4\sigma$).

\subsection{DES-Dovekie supernovae} \label{ssec:dovekie}

In our baseline analysis we employ the Pantheon+ SNeIa compilation. While the impact of the SN compilation on the final constraints is less than that of the CMB or BAO data, the use of different SN compilations or analysis pipelines can in principle alter our conclusions.

This is because the SNeIa datasets have an impact mainly as an additional probe of the fractional matter density $\Omega_\mathrm{m}$. In particular, they prefer slightly larger values than CMB and DESI BAO data. Given that the $\Omega_m$ parameter is impacted differently by the competing models, SNeIa data can be quite important for model selection. For example, the initial analysis of the DES Y5 dataset in \citealp{DES:2024jxu} found comparatively higher values of $\Omega_\mathrm{m} = 0.352 \pm 0.017$ (68\%) in $\Lambda$CDM than the Pantheon+ value $\Omega_\mathrm{m} = 0.334\pm 0.018$. Instead,  the recent DES Dovekie reanalysis of \citealp{DES:2025sig} found that an improved photometric cross-calibration and a retraining of the light-curve model yield smaller values, $\Omega_\mathrm{m} = 0.330 \pm 0.015$, in better agreement with Pantheon+. The DES Dovekie error bar is however smaller, restricting both low and high values of $\Omega_\mathrm{m}$.

Due to parameter degeneracies (e.g. \citealp{Sekiguchi:2020teg,Schoneberg:2024ynd}), there is a systematic trend of models from group M to reach high values of $H_0$ only at the expense of significantly lowering $\Omega_\mathrm{m}$\,. The trend is less pronounced for models from groups E and R. We thus expect the switch from Pantheon+ to DES Dovekie data to cut into the region of parameter space of group M models with the lowest $\Omega_\mathrm{m}$\,, and to limit their ability to reduce the Hubble tension.

\begin{figure}[tp]
    \centering
    \includegraphics[width=0.99\linewidth]{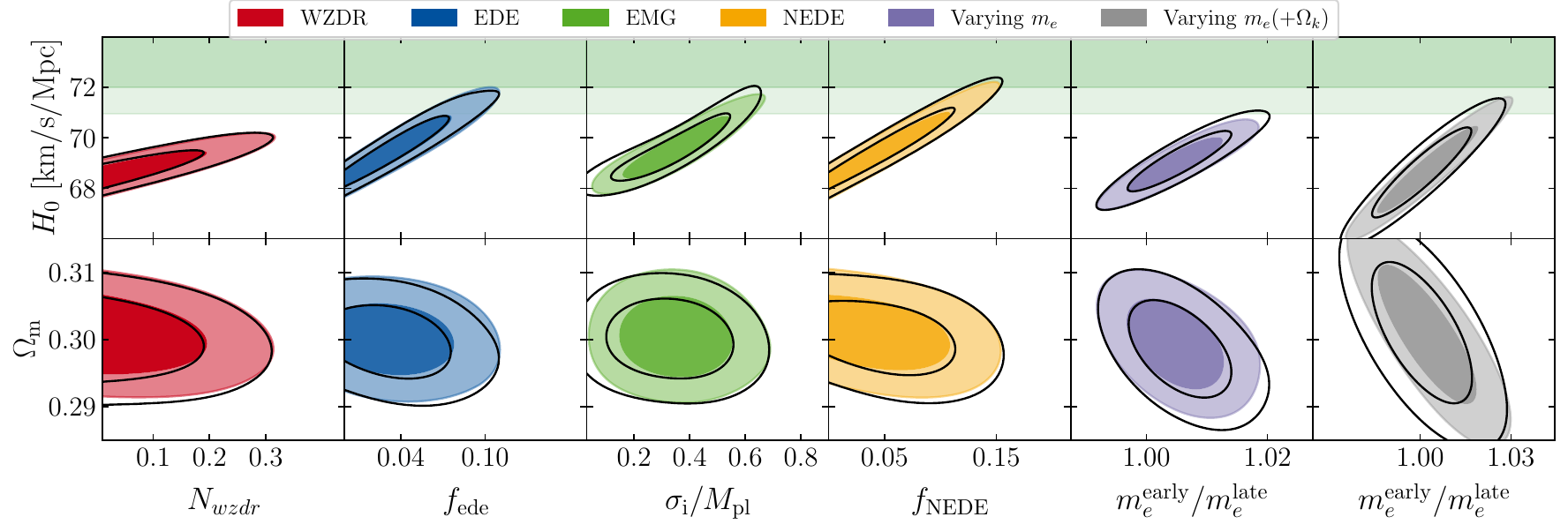}
    \caption{Same as \cref{fig:full_npipe}, but with the DES Dovekie SN data instead of Pantheon+.}
    \label{fig:desdov}
\end{figure}

Our results are shown in \cref{fig:desdov}. As expected, the varying-electron-mass model is noticeably affected by the DES Dovekie SN data, since it produces the largest reduction in $\Omega_{\mathrm m}$.  Nevertheless, the effect remains small, shifting the constraints from $H_0 = ({69.10}^{+0.71}_{-0.68})\mathrm{km/s/Mpc}$ in the baseline analysis to $H_0 = (68.97 \pm 0.65)\mathrm{km/s/Mpc}$, corresponding to a mere $0.14\sigma$ shift in the mean. Correspondingly, the tension in $H_0$ only worsens by $0.2\sigma$. In contrast, the \me{}(+$\Omega_k$) model shows barely any change due to the large decrease in the SNeIa constraining power for the $\Omega_m$ parameter once curvature is introduced.

\subsection{Neutrino mass prior}\label{ssec:mnu_norm}

In the baseline analysis presented in \cref{sec:results}, we have used a flat prior on the neutrino mass sum that requires it to be larger than 0.
However, values of $\sum m_\nu$ below $58\,$meV are already in $>2\sigma$ tension with neutrino oscillation experiments even assuming normal mass ordering \cite{Esteban:2024eli}. Since in many cases the sum of neutrino masses correlates with other parameters, in particular $H_0$ and extended model parameters, it is crucial to investigate whether an oscillation-motivated prior changes the ability of these models to reduce the Hubble tension. 

\begin{figure}[tp]
    \centering
    \includegraphics[width=0.99\linewidth]{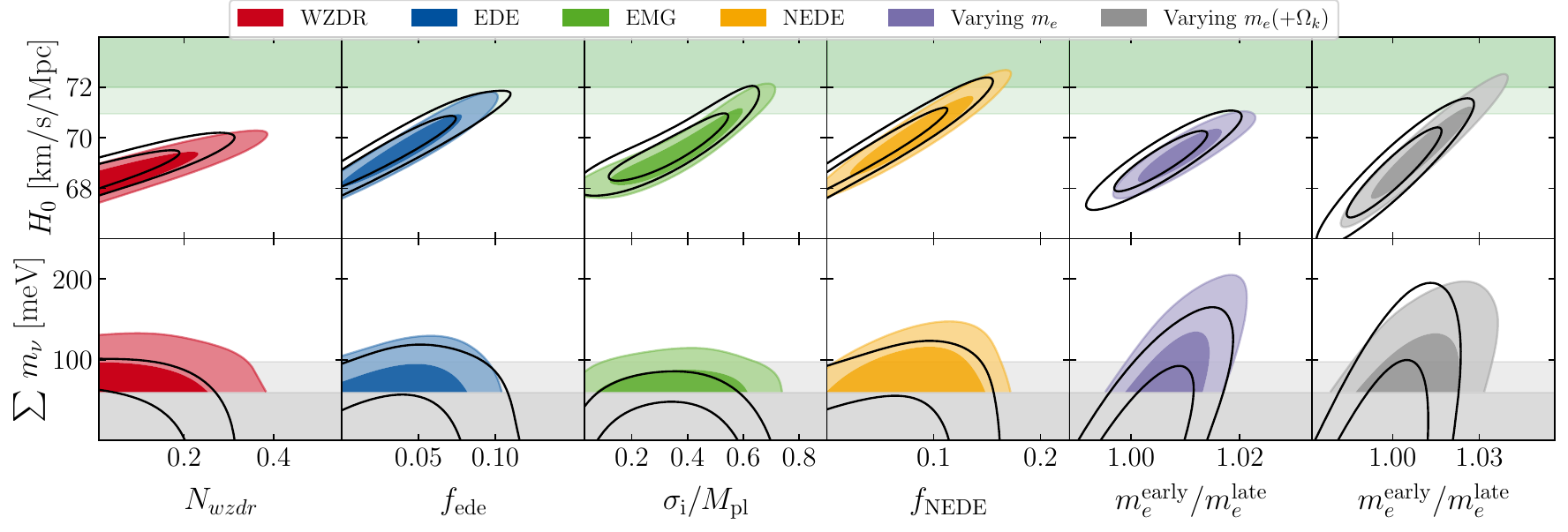}
    \caption{Same as \cref{fig:full_npipe}, but with a prior on $\sum m_\nu > 60\mathrm{eV}$ instead. The prior excludes the region shaded in dark gray, while the light gray region shows the inverted hierarchy bounds for reference.
    }
    \label{fig:mnu_norm}
\end{figure}

We thus repeat our CMB+BAO+SN analysis with a new prior $\sum m_\nu \geq 58\,$meV. The results are summarized in \cref{fig:mnu_norm}. Also in this case, the new results do not depart significantly from those of the baseline analysis. The tension level in $M_B$ remains very stable for most models, and decreases at most by $-0.2\sigma$ for the WZDR, NEDE, and varying $m_e$ models. Exceptionally, the \me{}($+\Omega_k$) model shows a reduction of the tension by $-1.4\sigma$ in this case, down to $2.9\sigma$, making it competitive with group E models in this case.
\pagebreak[5]
\subsection{Baryonic feedback}

\begin{figure}[tp]
    \centering
    \includegraphics[width=0.99\linewidth]{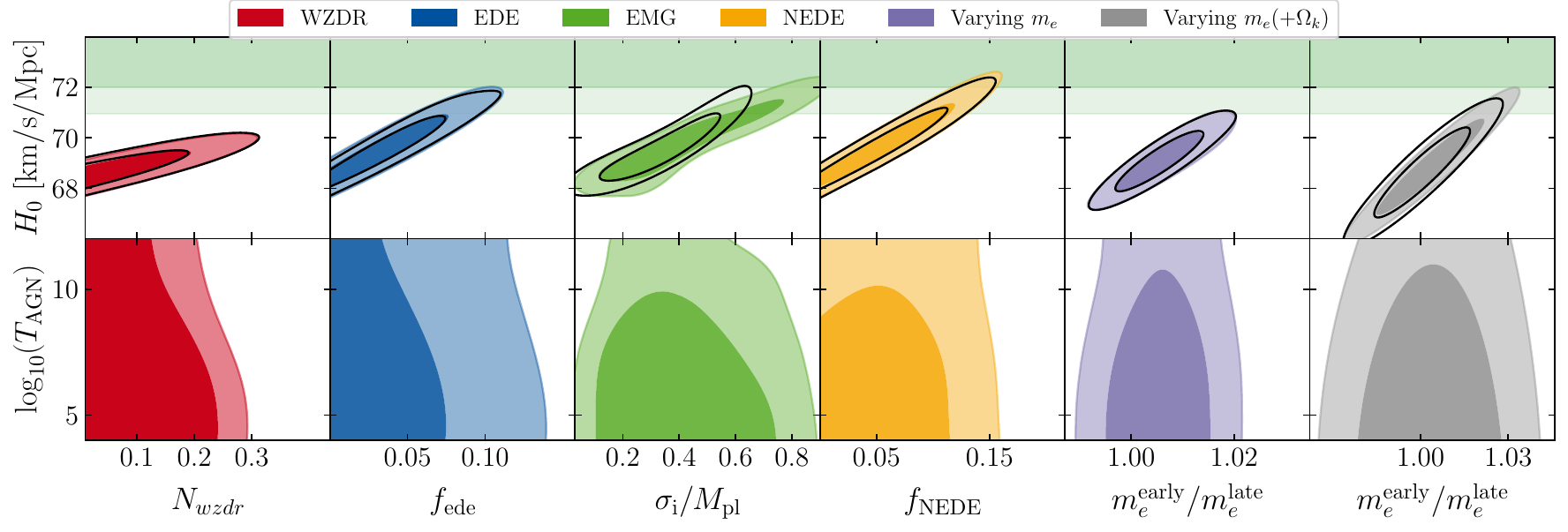}
    \caption{Same as \cref{fig:full_npipe}, but with a marginalization over the baryonic feedback parameter $T_\mathrm{AGN}$ of \texttt{HMcode 2020}.}
    \label{fig:tagn}
\end{figure}

Non-linear structure formation affects the high-$\ell$ CMB lensing power spectrum and therefore the lensed TT,TE,EE power spectra. It is thus very relevant for analyses performed in this work, given the large multipole range considered here.
One important question is whether the impact of high-$\ell$ damping-tail measurements on the Hubble tension, already discussed in \cref{ssec:noACT,ssec:running}, might be affected by the details of non-linear modeling. For this purpose, we investigate the consequence of marginalizing over the parameter $T_\mathrm{AGN}$ of \texttt{HMcode 2020} \cite{Mead:2020vgs}, which provides a proxy for the amplitude of baryonic feedback effects on the non-linear matter spectrum (baryonic feedback was neglected in the baseline analysis). We show the impact of this marginalization on the main model parameters in \cref{fig:tagn}. No significant deviation is found with respect to the baseline analysis, and according to \cref{fig:golden_summary} the tension level does not fluctuate by more than $\pm 0.2\sigma$. This is consistent with the expectation that the uncertainty of current CMB experiments is sufficient to necessitate a non-linear treatment of the CMB lensing power spectrum, but not to distinguish between different baryonic feedback models, as illustrated in \citealp{Trendafilova:2025dce,Smith:2025arq}. Notably, while much larger ranges of $\sigma_\mathrm{i}/M_\mathrm{pl}$ are allowed for the \emg{} model, this does not result in much wider contours for $H_0$\,, reducing the tension in $H_0$ only by $\sim -0.3\sigma$.

\subsection{Clustering amplitude \texorpdfstring{$S_8$}{S8}}\label{ssec:S8}

\begin{figure}[tp]
    \centering
    \includegraphics[width=0.99\linewidth]{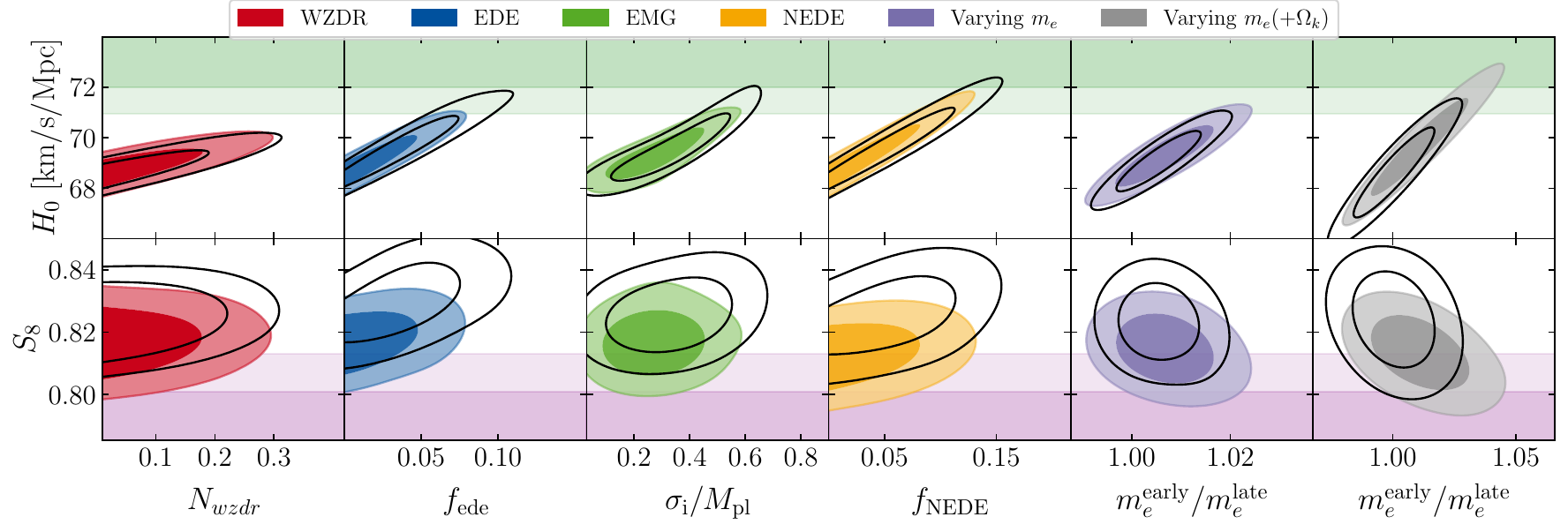}
    \caption{Same as \cref{fig:full_npipe}, but with an additional prior on $S_8=0.789 \pm 0.012$ from the DES Y6 3x2pt large scale structure analysis.}
    \label{fig:S8}
\end{figure}

As is well known (e.g. \cite{Poulin:2024ken,Pedrotti:2024kpn}), models resolving the Hubble tension can lead to a larger $S_8$, in particular those in group E \cite{Hill:2020osr,DAmico:2020ods,Ivanov:2020ril,Smith:2020rxx,Murgia:2020ryi, Herold:2021ksg}, as was discussed in \cref{sec:competition}. In this section we investigate the impact of data constraining $S_8$ on the ability of finalist models to reduce the Hubble tension. For this purpose, we use the combined $S_8$ measurement from the DES-Y6 3x2pt analysis\footnote{We expect that using KiDS legacy results with $S_8 = 0.815^{+0.016}_{-0.021}$ \cite{Wright:2025xka} instead would lead to constraints closer to the baseline analyses presented previously.} \cite{DES:2026fyc}, $S_8=0.789\pm0.012$, which is the tightest currently available measurement of the clustering of matter. For simplicity, we treat the $S_8$ parameter derived under $\Lambda$CDM from the full analysis as a Gaussian likelihood. The resulting constraints are shown in \cref{fig:S8}. As expected, most of group E models are more constrained once $S_8$ data are included and, according to \cref{tab:golden_summary}, the tension level in $M_B$ raises by up to $0.7\sigma$ for these models (for $H_0$ only $+0.3\sigma$). Instead, the WZDR model from group R is unaffected, and the varying electron mass model from group M actually performs better ($-0.3\sigma$), in particular once curvature is included ($-1.4\sigma$, down to $2.7\sigma$ tension). We already mentioned that models from group M ease the tension at the expense of smaller values of $\Omega_\mathrm{m}$, which naturally lead to a smaller $S_8=\sigma_8\sqrt{\Omega_\mathrm{m}/0.3}$ at constant small-scale clustering amplitude $\sigma_8$\,.

\subsection{Big Bang Nucleosynthesis constraints}\label{ssec:bbn}

Finally, we investigate whether the addition of BBN constraints coming from the measurement of light element abundances significantly impact our results. It  has been pointed out already in \citealp{Giovanetti:2026aku} and \citealp{Garramone:2026evc} that the $\Omega_\mathrm{b} h^2$ constraints provided by deuterium-to-hydrogen ratio measurements can significantly affect models that ease the Hubble tension. The degeneracies exploited by group E models typically favor higher values of $\Omega_\mathrm{b} h^2$, potentially bringing them into tension with BBN constraints.

Since the EMG model involves non-trivial modifications of gravitational couplings even at early times relevant for nucleosynthesis, specific modifications of BBN theory codes are required to accurately predict primordial element abundances in this case. This goes beyond the scope of our work. Thus, we exclude the EMG and Thawing Gravity models from this analysis, but note that tight bounds such as those derived in \citealp{Alvey:2019ctk} are likely to apply also here.

\enlargethispage*{1em}
There are currently two main approaches to model the BBN nuclear reaction rates, exemplified in the \texttt{PArthENoPE} \cite{Pisanti:2020efz,Gariazzo:2021iiu} and \texttt{PRIMAT} \cite{Pitrou:2018cgg,Pitrou:2020etb} codes. As these are known to yield significantly different values for the deuterium fraction, we perform two sets of analyses, using either models. 

\begin{figure}[tp]
    \centering
    \includegraphics[width=0.99\linewidth]{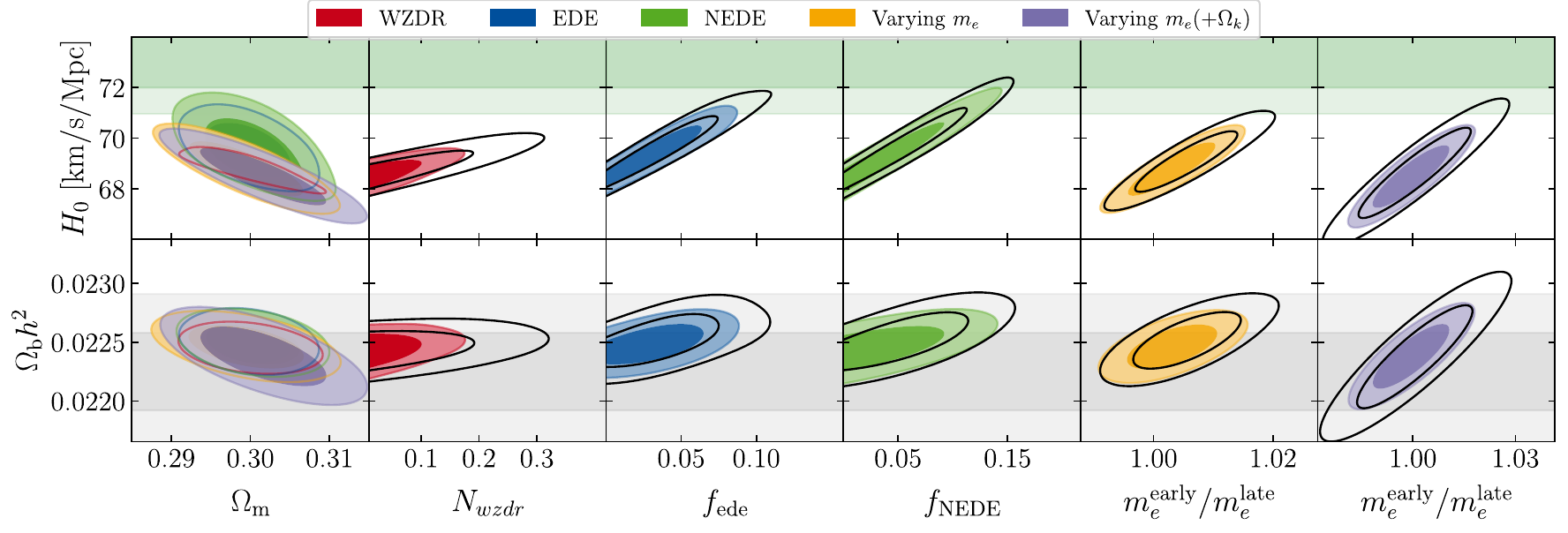}
    \caption{Same as \cref{fig:full_npipe}, but using a BBN likelihood based on the \texttt{PArthENoPE} nuclear reaction rates. The gray bands show the 68\% and 95\% constraints on $\Omega_\mathrm{b} h^2$ derived from primordial element abundances in the flat $\Lambda$CDM case.}
    \label{fig:bbn_parth}
\end{figure}

\begin{figure}[tp]
    \centering
    \includegraphics[width=0.99\linewidth]{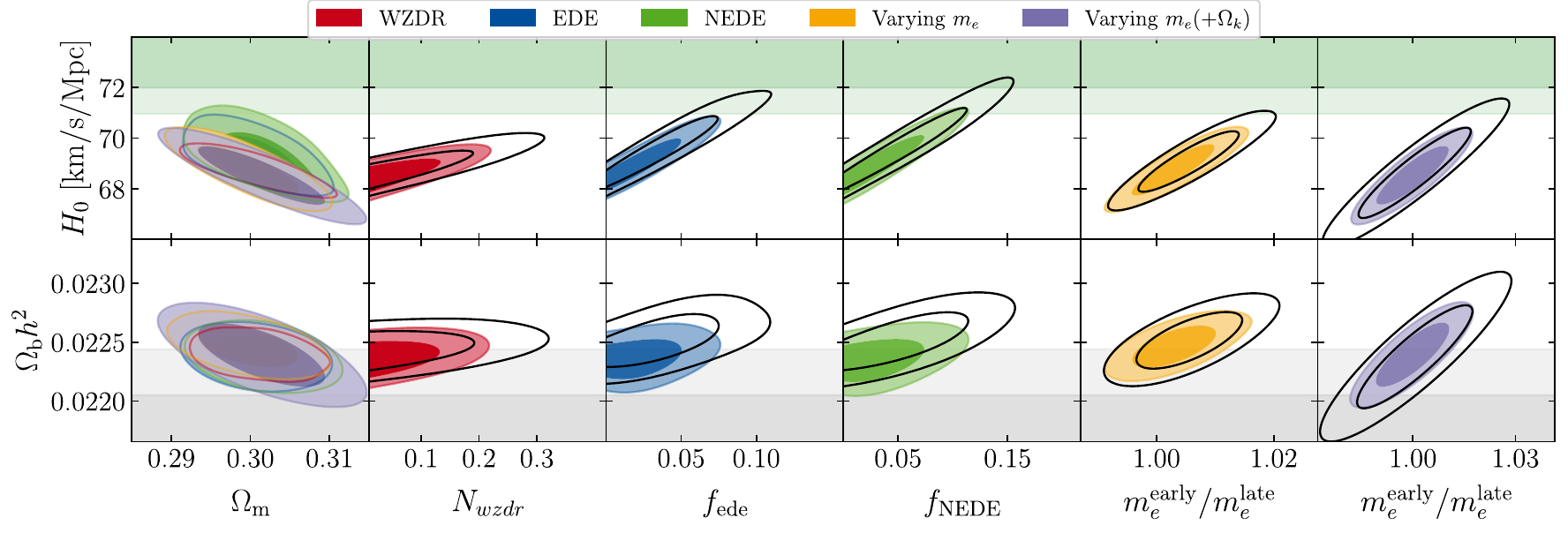}
    \caption{Same as \cref{fig:bbn_parth}, but with the \texttt{PRIMAT} nuclear reaction rates.\vspace*{-1em}}
    \label{fig:bbn_primat}
\end{figure}

We find that regardless of the code used, the parameter constraints are significantly stronger than in the baseline analysis for all the finalist models. With the \texttt{PArthENoPE} rates (see \cref{fig:bbn_parth}), the level of tension increases by $+0.5\sigma$ in $M_B$ uniformly. With the \texttt{PRIMAT} rates (see \cref{fig:bbn_primat}), $\Omega_\mathrm{b} h^2$ is constrained to lower values by primordial deuterium measurements (resulting in a $\sim 2\sigma$ tension with the baryon density inferred from Planck in $\Lambda$CDM). This further strengthens the parameter constraints and increases the tension level for all group E models -- for instance, by $+0.7\sigma$ for NEDE, while causing a $\sim +0.4\sigma$ overall tension level increase for the other models.

\enlargethispage*{3em}
\subsection{Summary of the knockout round}
The knockout round allowed us to check the robustness of our main results against several relevant variations in data and assumptions, summarized in \cref{tab:golden_summary}. The overall conclusion is that the baseline results of the group stage analysis are extremely stable, with the most significant change coming from the inclusion of BBN constraints or the inclusion of an $S_8$ prior from DES Y6 3x2pt data. BBN constraints universally worsen the outcomes (increasing the tension in $M_B$ by about $\sim 0.4-0.7\sigma$), while the $S_8$ constraints raise the tension strongly for group E models (by around $+0.5\sigma$) but reduce the tension for the varying $m_e$ model ($-0.3\sigma$ and $-1.4\sigma$ with curvature), making the curved varying $m_e$ model competitive with group E in this case. The curved varying $m_e$ model also has a strong reduction in tension when a physical prior on the neutrino mass is imposed, by $-1.2\sigma$. In most other cases studied here, the tension level typically varies by at most $\pm 0.3\sigma$.

\section{Conclusions}\label{sec:conclusions}

Using current CMB, BAO, and Type Ia supernova data, we have compared fourteen proposed solutions to the Hubble tension with a common analysis framework, which we dub the \enquote{$H_0$ World Cup}. The contenders are representative of five broad categories: models with late-time modifications of the background evolution (group L), modifications of recombination (group M), additional radiation before recombination (group R), and early dark energy injection or early modified gravity (group E). We have assessed the performance of each model using complementary Frequentist and Bayesian measures of both the residual tension and the improvement in the fit relative to $\Lambda$CDM. The two statistical approaches yield a broadly consistent ranking, and models based on the same physical mechanism generally exhibit similar performance.

This work builds upon the original \enquote{$H_0$ Olympics} which established a common framework for benchmarking proposed solutions using the cosmological data and model implementations then available \cite{Schoneberg:2021qvd}. This approach was already updated in \citealp{Khalife:2023qbu} with newer CMB and BAO data (ACT-DR4, SPT-3G 2018, eBOSS). The current work uses even more recent and constraining data in its baseline analysis (ACT-DR6, SPT-3G D1, DESI DR2, Pantheon+) and in the extended analyses of \cref{sec:golden} (DES Y5 Dovekie, DES Y6 3x2pt, recent primordial abundances). It incorporates additional models and mechanisms proposed recently. It complements the Frequentist criteria of \citealp{Schoneberg:2021qvd} with Bayesian parameter-shift and model-evidence metrics. It systematically tests the impact of curvature, CPL dark energy, and a free neutrino mass, before subjecting the finalists to dedicated robustness tests involving alternative CMB likelihoods and multipole cuts, supernova samples, large-scale-structure information, and BBN. 

As in the original analysis, the models considered here should be understood primarily as representative of phenomenological mechanisms rather than as complete fundamental theories, since several of them lack a compelling particle-physics embedding or a known ultraviolet completion. Our updated ranking therefore identifies promising physical directions and clarifies which conclusions depend on particular datasets or analysis choices; it should not be interpreted as evidence that any specific implementation is the correct theory.

\vspace*{0.5\baselineskip}
\noindent \textbf{Baseline competition.} For our baseline CMB+BAO+SN dataset, $\Lambda$CDM exhibits a discrepancy above $5.5\sigma$ in the supernova calibration parameter $M_B$. We use $M_B$, rather than $H_0$ alone, because it consistently incorporates the calibration information while accounting for the small $\Omega_\mathrm{m}$ tension emerging between CMB, DESI BAO, and uncalibrated supernova data. The main baseline results, summarized in \cref{tab:baseline_chi2,tab:baseline_evidence,tab:baseline_QDMAP,tab:baseline_tension,tab:baseline_tension_2}, are:
\begin{itemize}
    \item \textbf{Group E gives the best overall performance.} These models reduce the residual tension to approximately $3\sigma$ and are strongly favored over $\Lambda$CDM by the model-comparison metrics.
    \item \textbf{Group M provides an intermediate improvement.} Its best-performing model reduces the residual tension to approximately $4\sigma$ and receives moderate support relative to $\Lambda$CDM.
    \item \textbf{Group R performs only marginally better than $\Lambda$CDM.} Its best model leaves a residual tension of approximately $4.5\sigma$, with little or no model-comparison preference over $\Lambda$CDM.
    \item \textbf{Groups L and E+L perform worst overall.} The late-time and mixed late+early-time models considered here do not significantly reduce the tension once confronted with the full baseline dataset.
\end{itemize}
The DESI DR2 BAO data adds significant constraining power on $\Omega_\mathrm{m}$ for all models and tend to shift the contours in a direction beneficial to the Hubble tension, except for the \idedm{} and \me{} models.
Apart from $H_0$ and $\Omega_\mathrm{m}$, the standard cosmological parameters have stable constraints under the inclusion of new physics, with the exception of the summed neutrino mass which may reach higher values in the \me{} model or in the \nosroll{} case.

\vspace*{0.5\baselineskip}
\noindent \textbf{Comparison with $H_0$ Olympics results.} The updated ranking differs most clearly for groups R and M. Radiation-based solutions, which remained viable in the previous competition, are no longer competitive in the baseline analysis and do not meet the thresholds to be selected as finalists. The WZDR model is retained in the knockout round only as a reference. The \me{} model also performs substantially worse than in the original Olympics and barely passes the threshold to be among the finalists.
These changes are driven primarily by the inclusion of ACT DR6 data, which tightly constrain the CMB signatures produced by these mechanisms.

As a matter of fact, we have checked that removing ACT substantially changes the relative ranking and improves the Bayes factor almost universally. In the no-ACT analysis, most models in groups R, M, and E reduce the residual tension to approximately $3$--$3.5\sigma$ and achieve comparable Bayesian model preferences; cold \coldnede{} reaches approximately $2.3\sigma$. ACT data are therefore central to the stronger baseline constraints on groups R and M and to the clearer preference for group E.

\vspace*{0.5\baselineskip}
\noindent \textbf{Extension of each model.} Next, we have tested whether adding late-time freedom to each contender in the form of spatial curvature or CPL dark energy changes the ranking:
\begin{itemize}
    \item \textbf{Spatial curvature} has little impact on most models. The clearest improvement occurs for group M models and the Rock-n-Roll model, which benefit from the additional freedom in the angular-diameter distance and from the weaker late-time determination of $\Omega_{\mathrm m}$ when curvature is allowed.
    \item \textbf{CPL dark energy} does not improve the ability of the models to reach the local calibration from SH0ES. Variations in $(w_0,w_a)$ generally counteract the pre-recombination modifications and shift the inferred $H_0$ downward. However, introducing models that ease the Hubble tension tends to reduce the preference for a departure from a cosmological constant. This reduction is strongest for \me{} and \wzdr{}, while the \nosroll{} case yields the broadest constraints.
    \item \textbf{A deviation from a power-law primordial power spectrum}, parametrized by $\alpha_s$ and $\beta_s$\,, strongly improves the performance of all group R models as well the varying electron mass model. This is because of the role that ACT plays in severely constraining those models.  
    We find a preference for $\alpha_s\neq0$ and $\beta_s\neq 0$, in agreement with \citealp{Garny:2026gcs}, and a reduction of the Hubble tension to the $3-3.5\sigma$ level.
\end{itemize}

\vspace*{0.5\baselineskip}
\noindent\textbf{Robustness of the finalists.} After selecting the best-performing baseline models, we subject them to the additional tests described in \cref{sec:golden}. The main conclusions are:
\begin{itemize}
    \item \textbf{CMB likelihoods:} the overall conclusions are stable over the studied variations, although the \texttt{plik} likelihood is generally more permissive than \texttt{CamSpec}. WZDR and varying $m_e$($+\Omega_k$) benefit the most from shifting the transition from Planck to ACT data to higher multipoles, consistent with the strong ACT constraints on dark-radiation models. Allowing for baryonic feedback effects in the nonlinear modeling of CMB lensing does not significantly alter the constraints.
    \item \textbf{Supernova samples:} the tighter $\Omega_{\mathrm m}$ constraint from the Dovekie-recalibrated DES Y5 sample weakens the constraints on the \me{} model by a very small amount, but has a negligible impact on the other finalists.
    \item \textbf{Neutrino-mass prior:} imposing the lower bound on $\Sigma m_\nu$ motivated by the normal hierarchy shifts most models toward slightly higher $H_0$, with the exceptions of \ede{} and \lcdm{}. The strongest shift is seen for the varying $m_e$($+\Omega_k$) model, making it competitive with group E models in this case.
    \item \textbf{Clustering amplitude:} adding the DES Y6 $S_8$ constraint weakens group E solutions because their increase in $H_0$ is generally accompanied by a larger $S_8$\,. WZDR is nearly unaffected, whereas \me{} benefits mildly and \me{}($+\Omega_k$) strongly from the additional constraint because its lower preferred $\Omega_{\mathrm m}$ also lowers $S_8$\,. A complete assessment including full-shape large-scale-structure information is left for future work.
    \item\textbf{BBN constraints:} light-element abundances can restrict models in groups R, M, and E, but the strength of this conclusion depends somewhat on whether the \texttt{PArthENoPE} or \texttt{PRIMAT} nuclear reaction rates are adopted. Nuclear-rate uncertainties therefore remain an important caveat when assessing these models.
\end{itemize}
\enlargethispage*{2em}
Overall, no model fully resolves the Hubble tension in the full baseline analysis, but the comparison reveals a clear hierarchy among physical mechanisms: early dark energy injection (minimally and non-minimally coupled to gravity) currently performs the best, while recombination and dark-radiation solutions remain more sensitive to the choice of CMB and low-redshift datasets. The ranking is robust to several analysis choices, but $S_8$ and BBN information expose important complementary tests and theoretical challenges. Deviations from a pure power-law primordial power spectrum substantially improve the ability of models with extra radiation (group R) or a varying electron mass to address the tension, bringing them close to the performance of group E models. A varying electron mass model also benefits from curvature, although not as substantially as previously found. Progress will require continued scrutiny of observational systematics, to strengthen the underlying theoretical basis of many proposed mechanisms, and to look for new mechanisms that can ease multiple tensions simultaneously. The time is ripe for revolutionary models in cosmology and the search for a solution beyond the $\Lambda$CDM cosmological standard model is now more important than ever.

\section*{Acknowlegements}

We thank Lennart Balkenhol, Gia Dvali, Ali Rida Khalife, Florian Niederman, Martin Sloth, and Gen Ye for very useful discussions. We thank Tristan L. Smith for initial contributions as well as useful discussions throughout. The authors acknowledge the use of Anthropic's Claude AI, Google's Gemini 3.5, and OpenAI Codex as supplementary research-assistance tools for plotting, data analysis, and text editing. The scientific ideas, analysis choices, interpretation of the results, as well as the manuscript writing are the work of the authors. A repository containing the analysis outputs and reproducibility notebooks is available at XXX. We acknowledge the use of the Python packages \texttt{NumPy}, \texttt{SciPy}, \texttt{Matplotlib}, \texttt{classy}, \texttt{MontePython}, \texttt{OLE}, \texttt{GetDist}, \texttt{Cobaya}, \texttt{tensiometer}, \texttt{liquidcosmo}, \texttt{MCevidence}, \texttt{prospect}, and \texttt{procoli}.
NS acknowledges support from the Excellence Cluster ORIGINS which is funded by the Deutsche Forschungsgemeinschaft (DFG, German Research Foundation) under Germany’s Excellence Strategy - EXC-2094/2 - 390783311, as well as the funding through a Fraunhofer-Schwarzschild
Fellowship at the LMU. VP acknowledges the European Union's Horizon Europe research and innovation programme under the Marie Sk\l odowska-Curie Staff Exchange grant agreement no.\ 101086085 -- ASYMMETRY. This work received funding support from the European Research Council (ERC) under the European Union's HORIZON-ERC-2022 (grant agreement no.\ 101076865). JL and MM acknowledge support from the DFG grant LE 3742/8-1. 
AGF, FF, LM acknowledge the Open Physics Hub project (hosted by the University of Bologna) and INFN for granting access to their computational resources. FF and LM acknowledge partial financial support from the Progetti di Astrofisica Fondamentale INAF 2023, from INFN InDark initiative, from the contract ASI/INAF for the Euclid mission n. 2018-23-HH.0 and the ASI grant 2020-9-HH.0 (participation in LiteBIRD phase A). RKS
thanks the Alexander von Humboldt Foundation for their
support.
This publication is based upon work from the COST Action CA21136 ``Addressing observational tensions in cosmology with systematics and fundamental physics'' (CosmoVerse), supported by COST (European Cooperation in Science and Technology). The results obtained in this paper were computed through resources from the Universe and Particles Laboratory of Montpellier (LUPM). We thank LUPM for providing the technical support, computing and storage facilities. The authors gratefully acknowledge the computing time provided to them at the NHR Center NHR4CES at RWTH Aachen University (project number p0021792).

\appendix

\section{Results without BAO or SNeIa}\label{app:noSBS}

\begin{figure}[tp]
    \centering
    \includegraphics[width=0.48\linewidth]{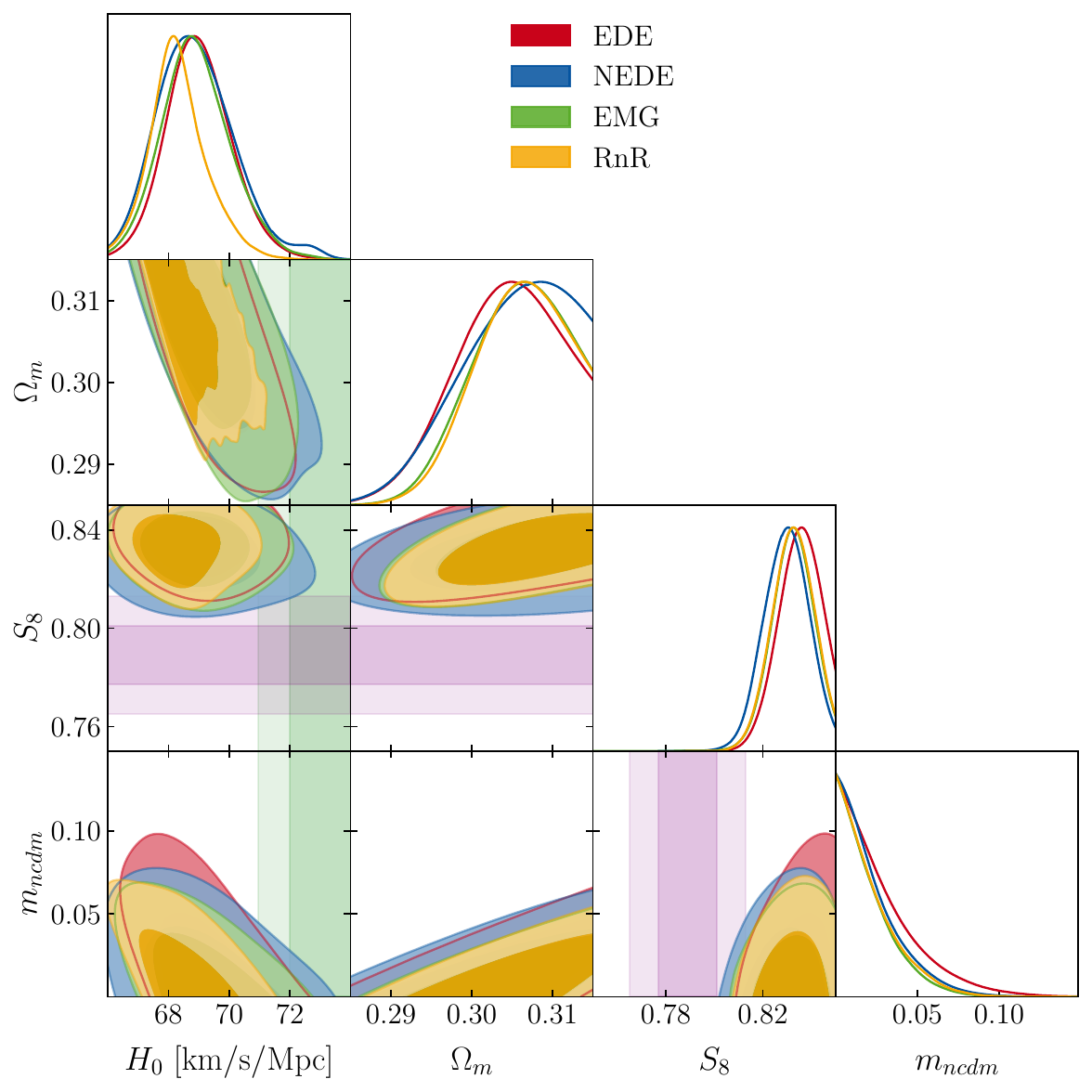}
    \includegraphics[width=0.48\linewidth]{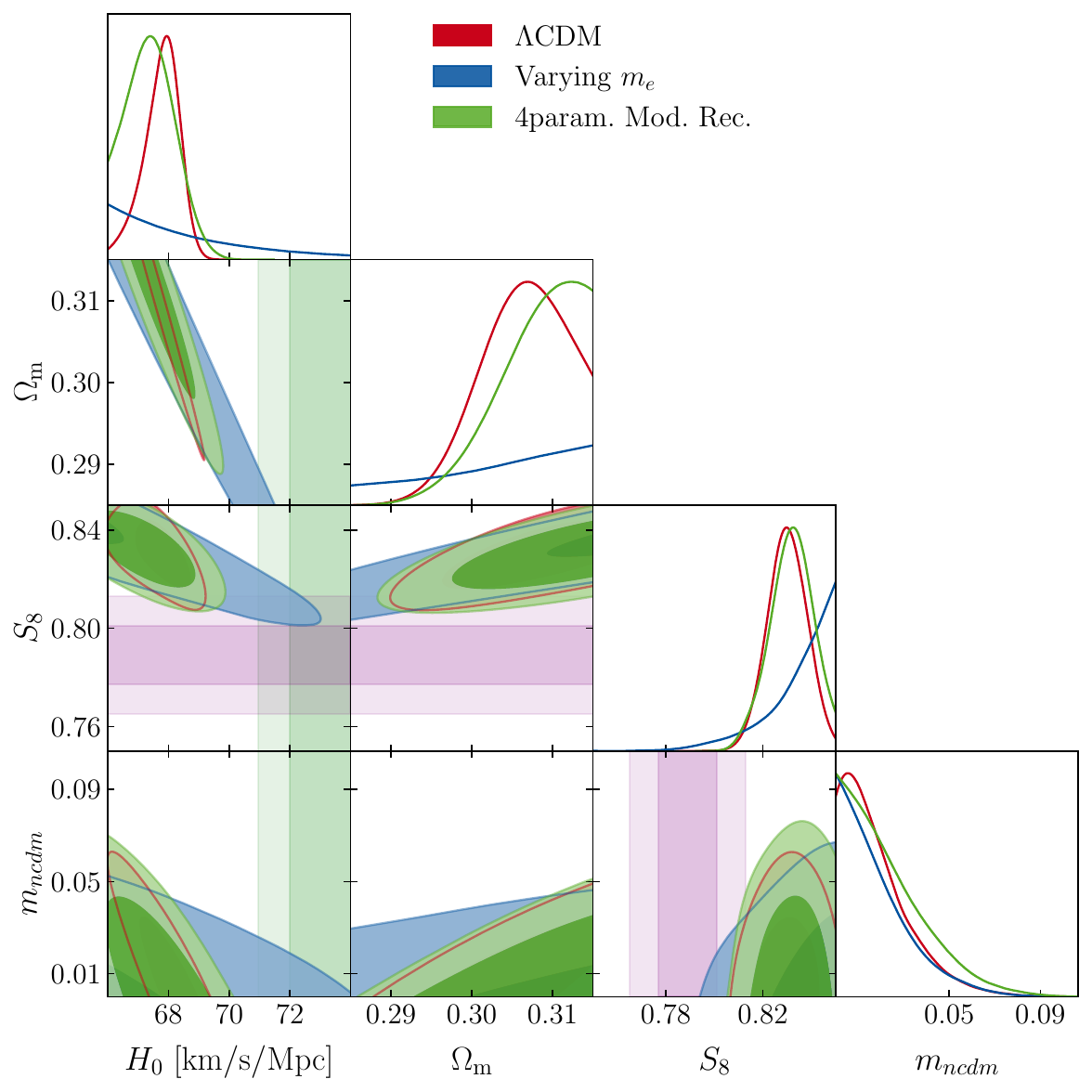}
    \includegraphics[width=0.48\linewidth]{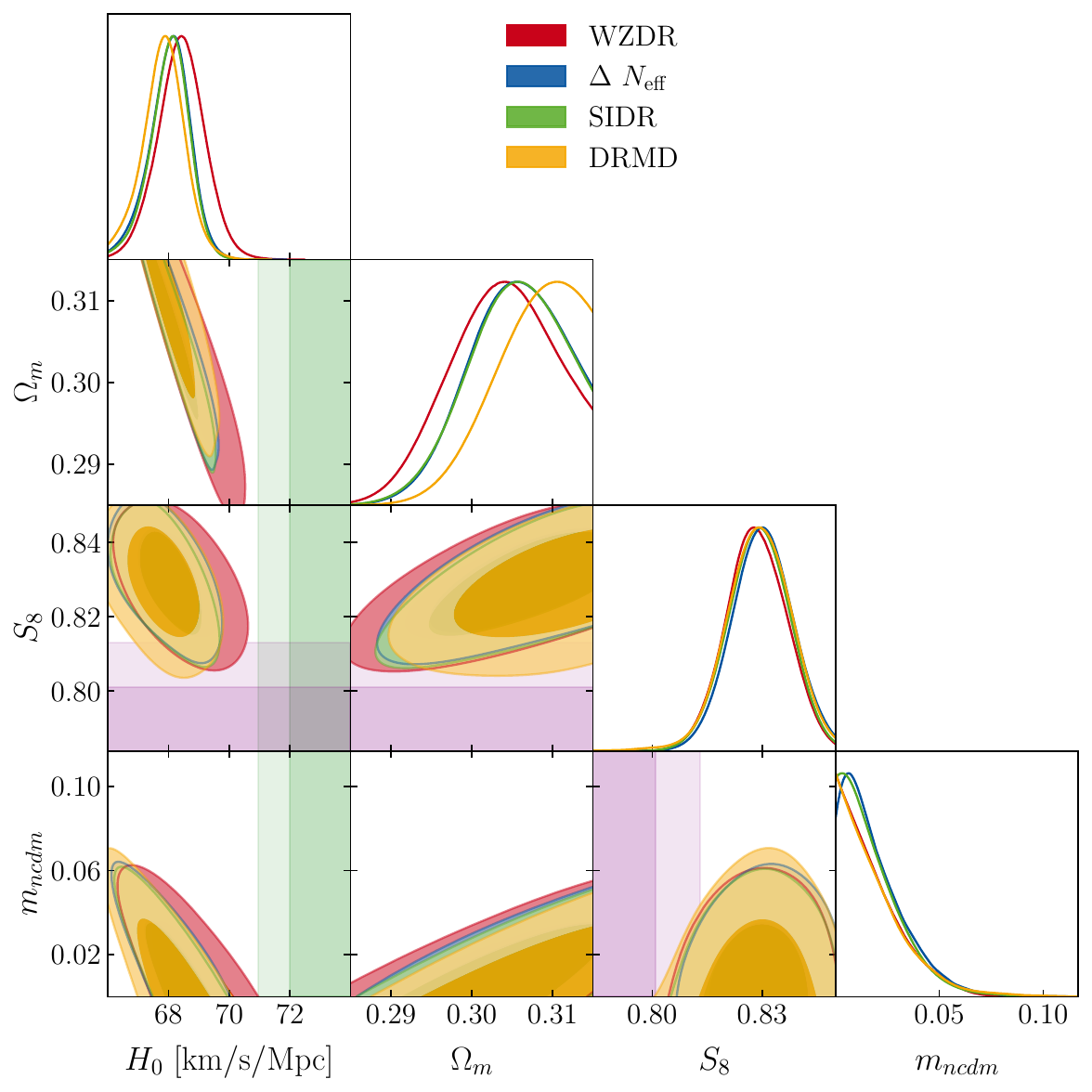}
    \includegraphics[width=0.48\linewidth]{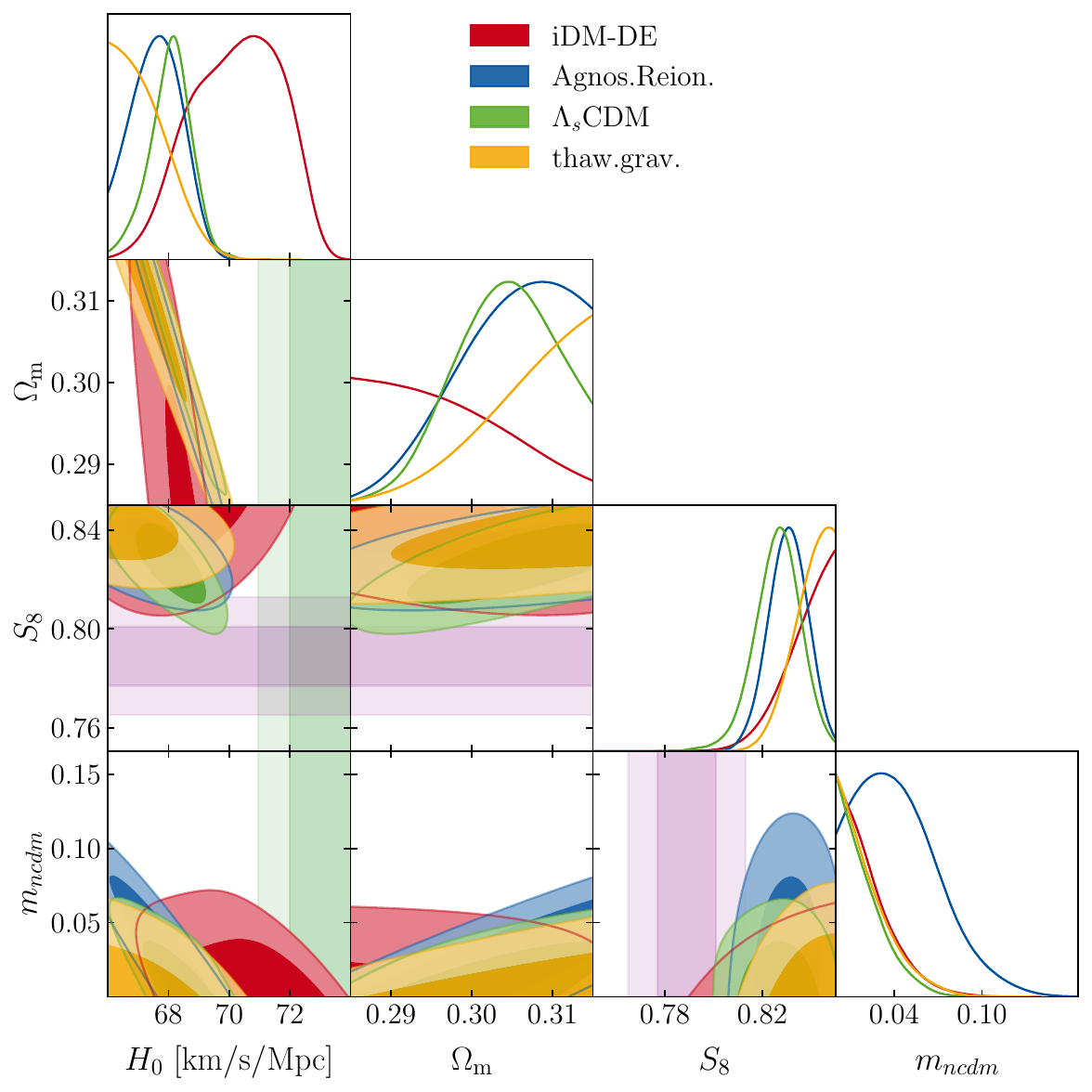}
    \caption{Same as \cref{fig:overviews}, but for CMB data without BAO or SNeIa. For comparisons between different datasets for a given individual model, see \cref{app:permodel}.}
    \label{fig:overviews_noBS}
\end{figure}

\begin{figure}[tp]
    \centering
    \includegraphics[width=0.48\linewidth]{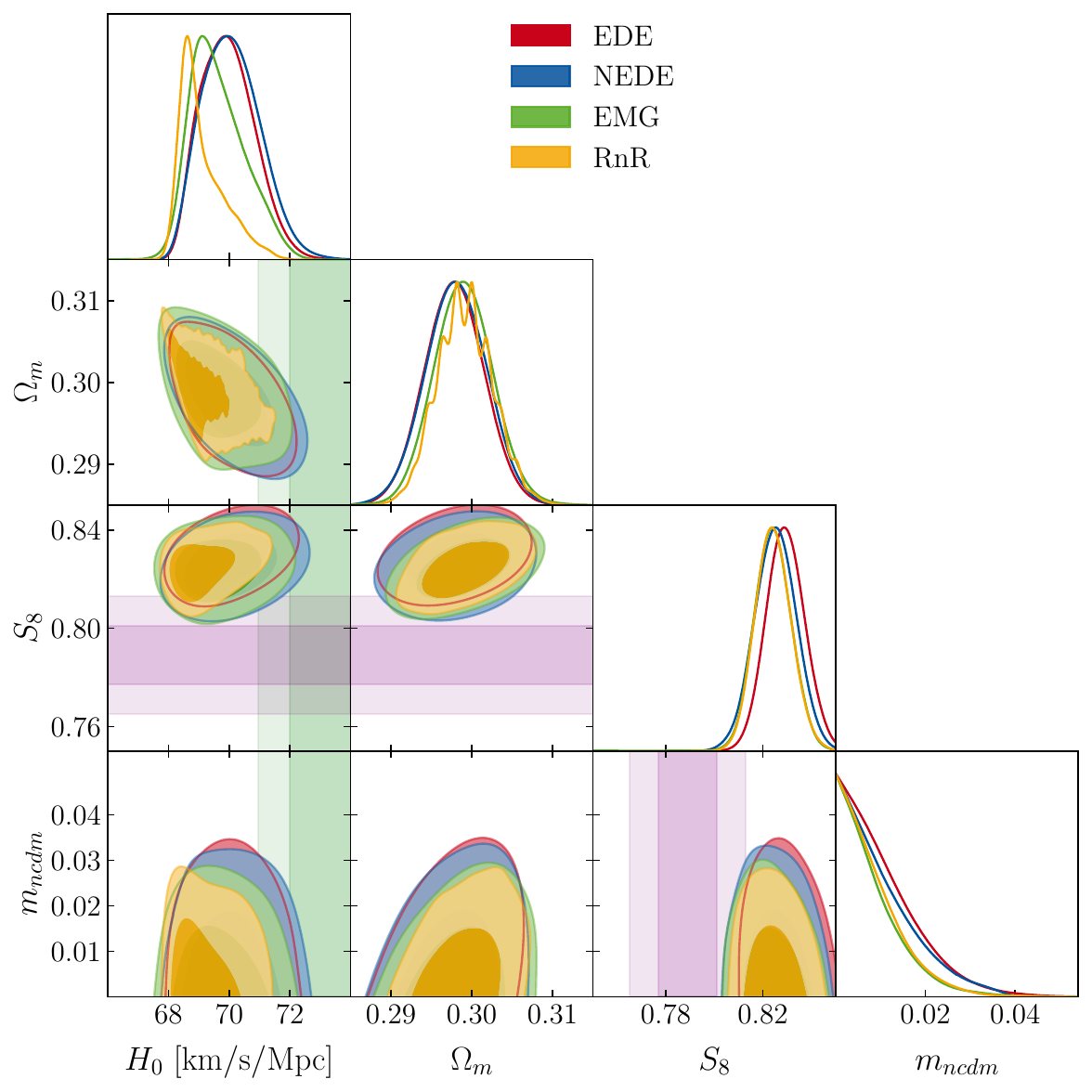}
    \includegraphics[width=0.48\linewidth]{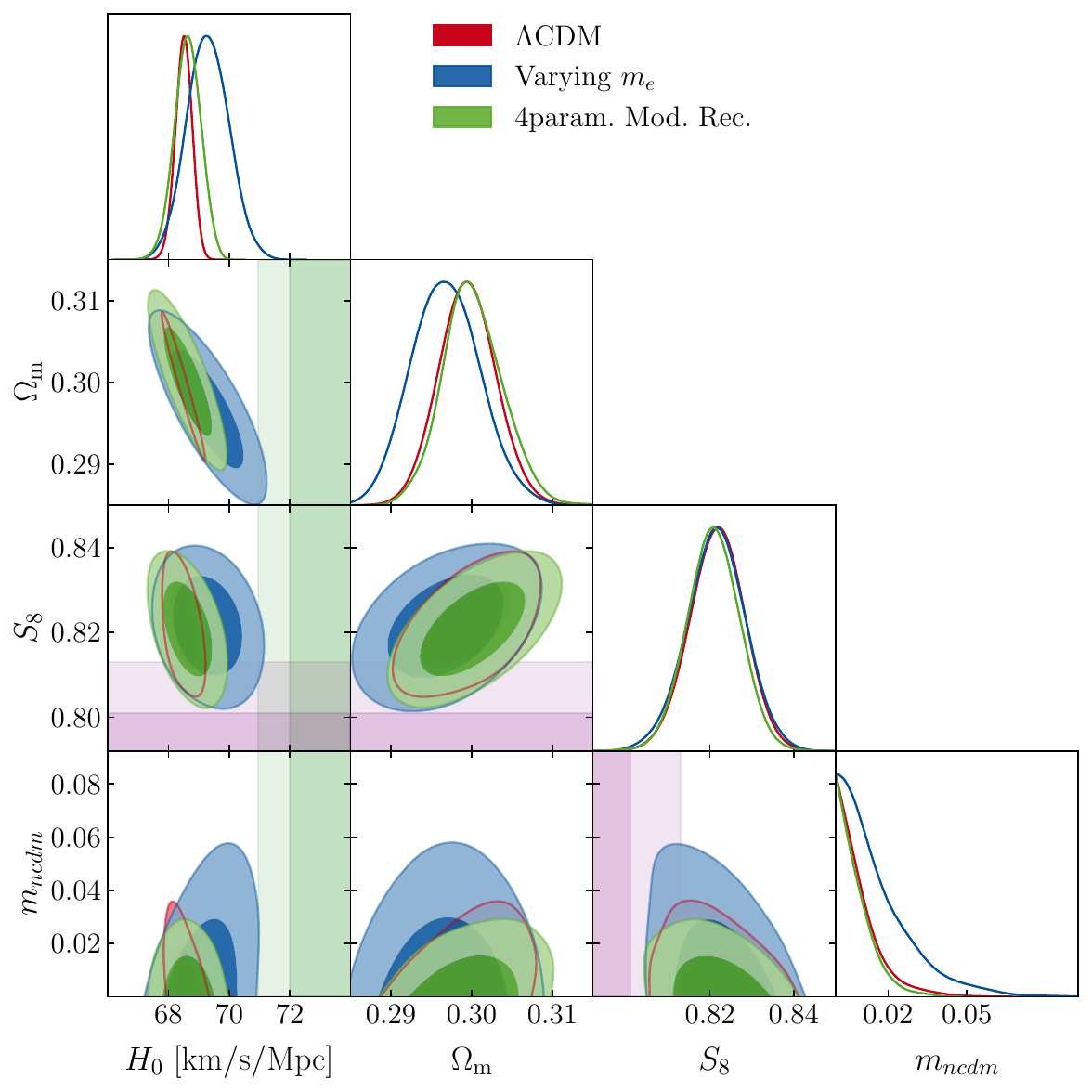}
    \includegraphics[width=0.48\linewidth]{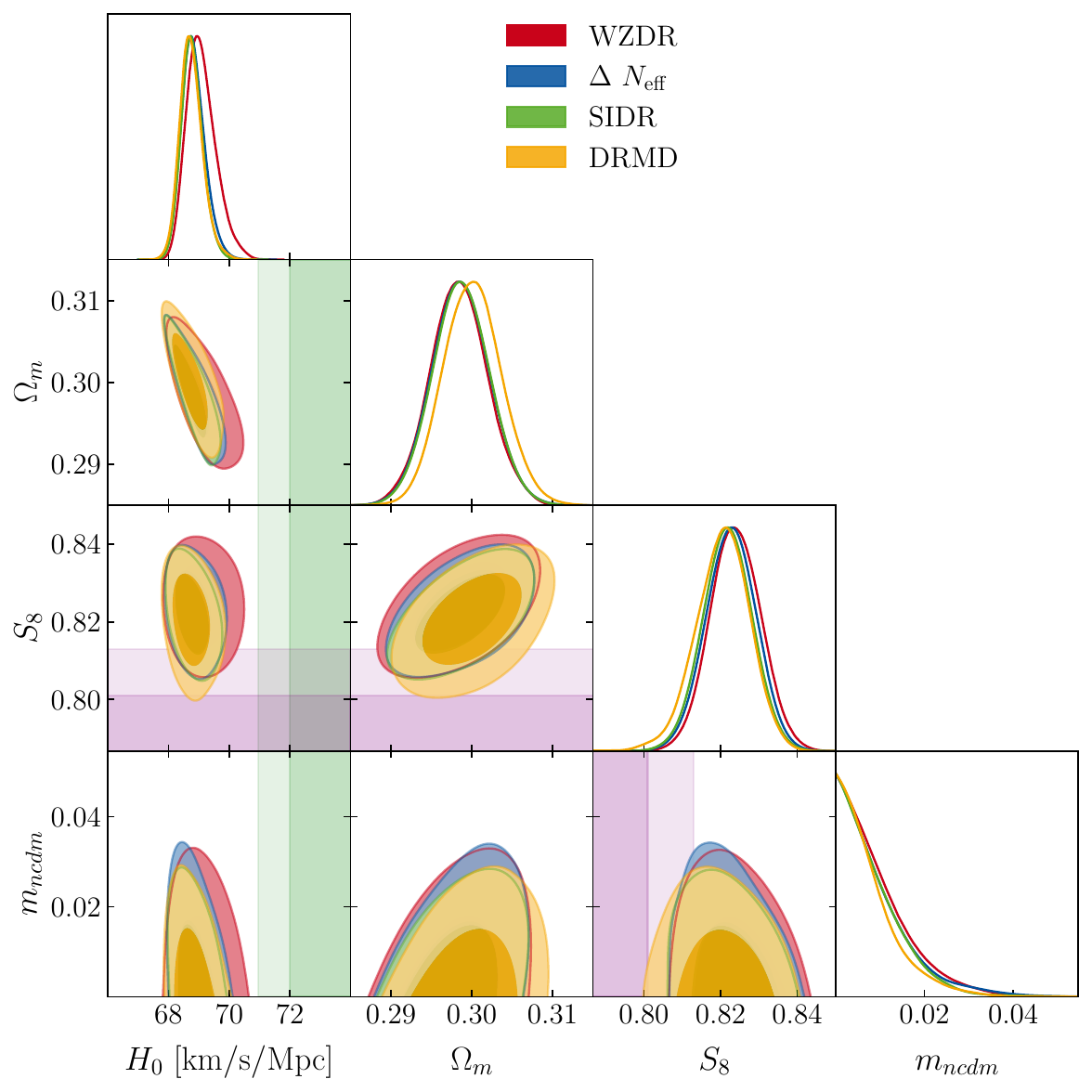}
    \includegraphics[width=0.48\linewidth]{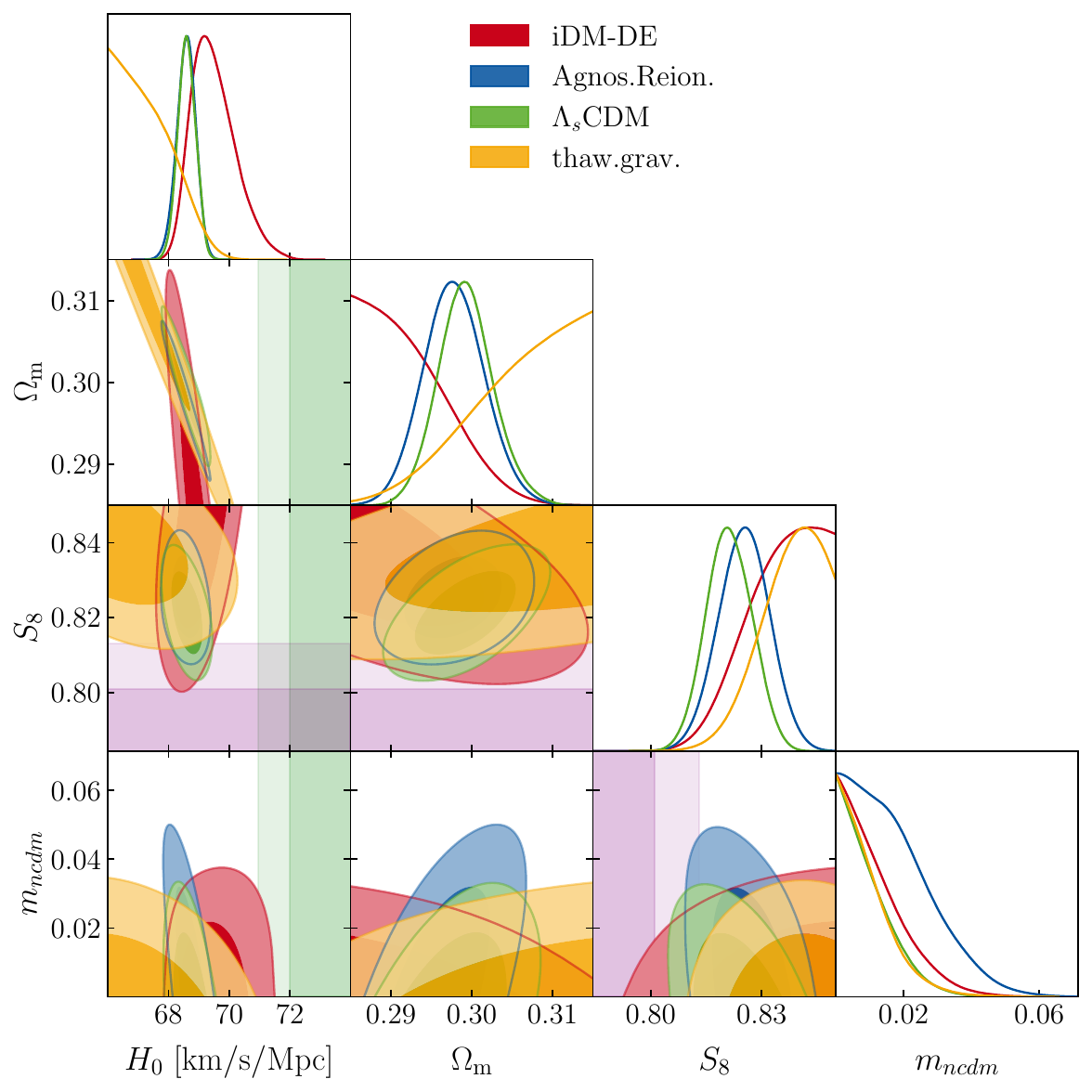}
    \caption{Same as \cref{fig:overviews}, but for CMB+BAO data without SNeIa. For comparisons between different datasets for a given individual model, see \cref{app:permodel}.}
    \label{fig:overviews_noS}
\end{figure}

In \cref{fig:overviews_noBS,fig:overviews_noS} we show the constraints for the CMB dataset alone and for the CMB+BAO datasets. Comparing especially the results between \cref{fig:overviews_noS} and \cref{fig:overviews}, we can see that the BAO data have a very significant impact on the preferred parameter regions, while the addition of the SNeIa data causes just slight shifts in parameter space for most models.


\begin{figure}[tp]
    \centering
    \includegraphics[width=0.48\linewidth]{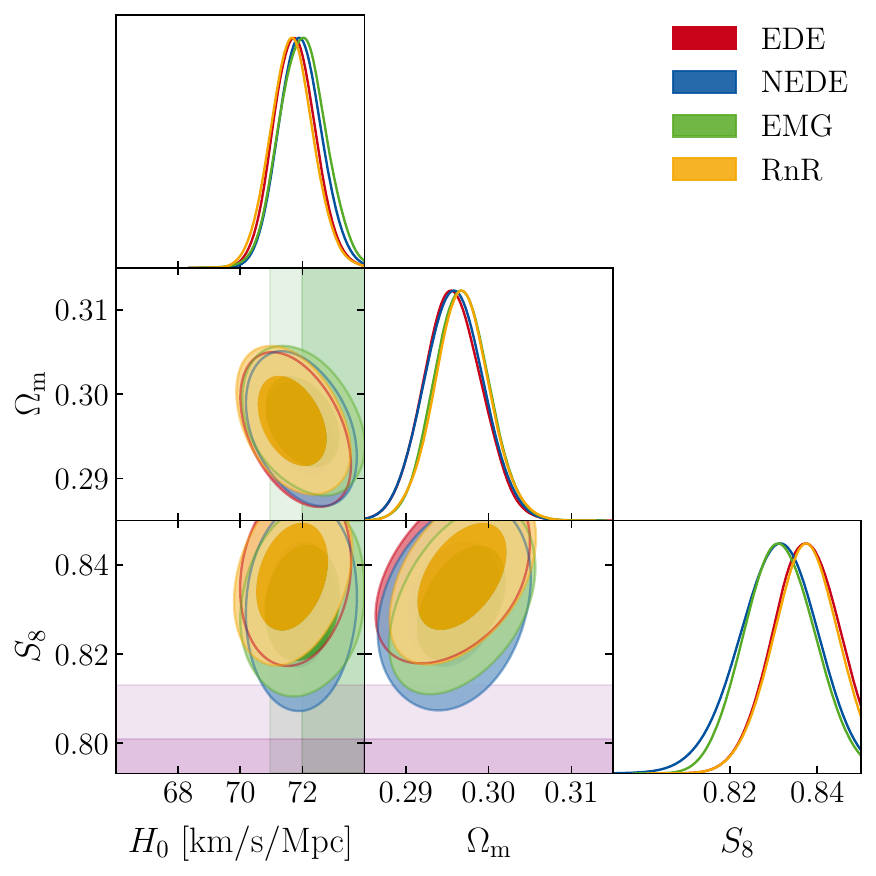}
    \includegraphics[width=0.48\linewidth]{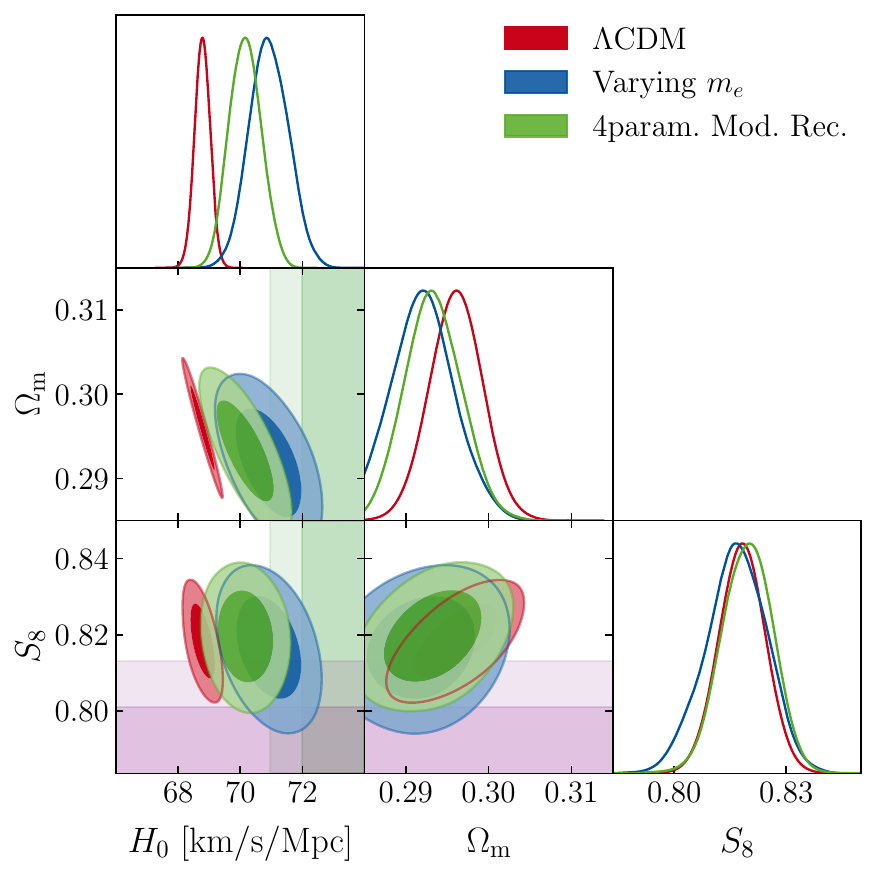}
    \includegraphics[width=0.48\linewidth]{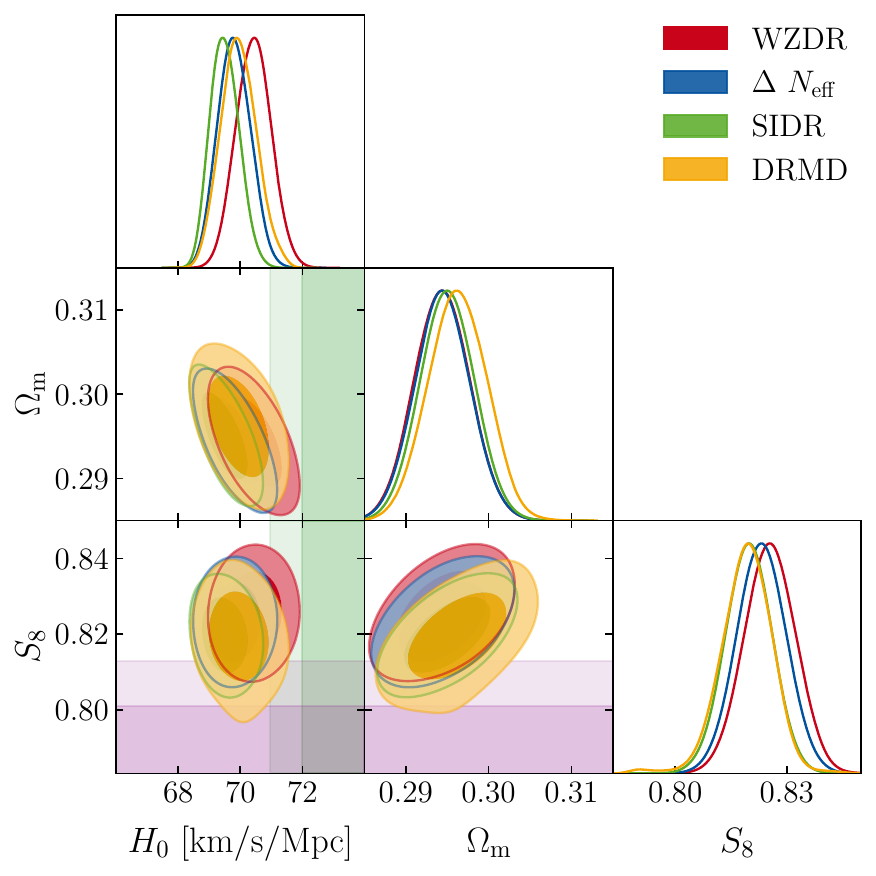}
    \includegraphics[width=0.48\linewidth]{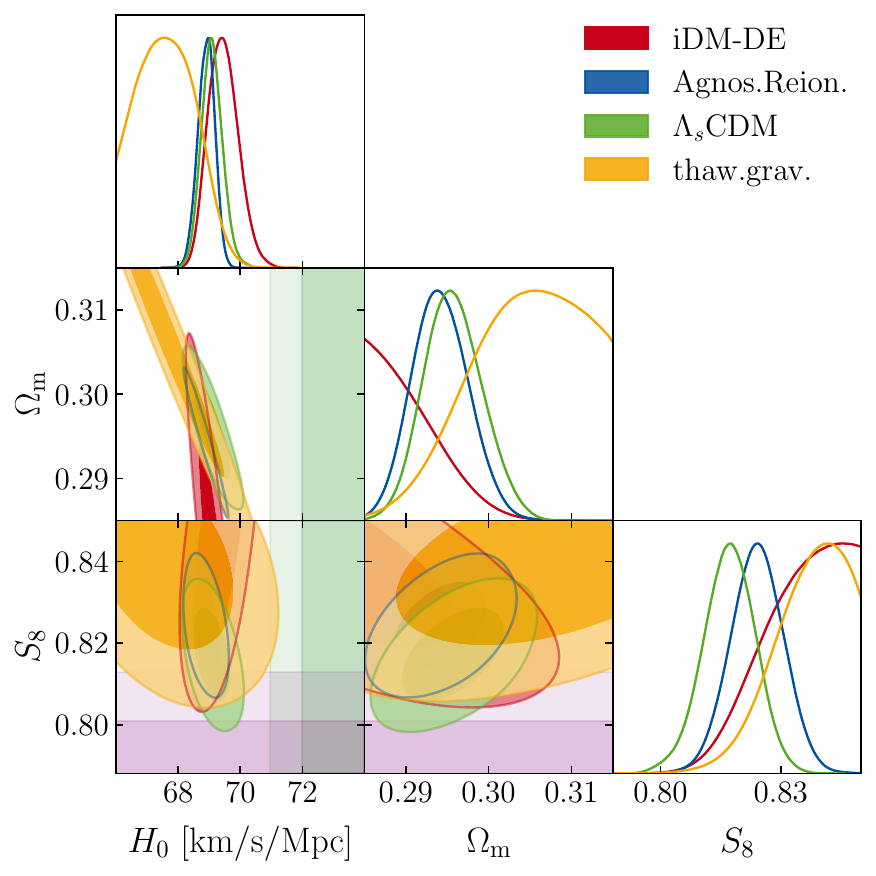}
    \caption{Same as \cref{fig:overviews}, but with the impact of an additional prior on $M_B$\,. For comparisons between different datasets for a given individual model, see \cref{app:permodel}.}
    \label{fig:overviews_BSM}
\end{figure}

\section{Parameter constraint tables\label{app:tables}}

\enlargethispage*{1em}
In \cref{tab:constr_model_BS,tab:constr_BS,tab:constr_BS_Ok,tab:constr_BS_w0wa,tab:constr_B,tab:constr} we give the parameter constraints of the individual models for various data and extension combinations. \Cref{tab:constr_BS} give CMB+BAO+SN constraints, \cref{tab:constr_B} gives CMB+BAO constraints, and \cref{tab:constr} gives CMB-only constraints on the $\Lambda$CDM parameters. Additionally, for the CMB+BAO+SN data we give model specific parameter constraints in \cref{tab:constr_model_BS} and $\Lambda$CDM parameters constraints for extensions through curvature in \cref{tab:constr_BS_Ok} and for extensions through CPL dark energy in \cref{tab:constr_BS_w0wa}.

\begin{table}[tp]
    \centering
\resizebox{\textwidth}{!}{
\begin{tabular}{c c c c c c}
\toprule
Model & Parameter & CMB+BAO+SN & +$\Omega_k$ & +$w_0, w_a$ & no ACT  \\
\midrule
 \arrayrulecolor[HTML]{e6e6e6}
\switch & $z_t$ & $6.0 \pm 2.4$ & ${5.3}^{+2.9}_{-2.5}$ & --- & ${4.8}^{+2.6}_{-1.9}$ \\ \midrule
\idedm & $\xi_\mathrm{ide}$ & $> -0.10$ & $> -0.10$ & --- & $> -0.21$ \\ 
\arrayrulecolor[HTML]{808080} \midrule \arrayrulecolor[HTML]{e6e6e6}
\multirow{3}{*}{\thawing} & $\lambda_\mathrm{TG}$ & $1.43 \pm 0.32$ & $1.35 \pm 0.31$ & --- & $ 1.21 \pm 0.36$ \\ 
& $\sigma_\mathrm{i}/M_\mathrm{pl}$  & ${-0.195}^{+0.076}_{-0.069}$ & ${-0.189}^{+0.067}_{-0.062}$ & --- & $-0.309 \pm 0.075$\\ 
& $\xi \sigma_\mathrm{i}^2/M_\mathrm{pl}^2$ &$-0.040 \pm 0.023$  & $-0.038 \pm 0.021$ &  --- & $-0.055 \pm 0.029$ \\ 
\arrayrulecolor[HTML]{808080} \midrule \arrayrulecolor[HTML]{e6e6e6}
\multirow{3}{*}{\ede} & $f_\mathrm{ede}(z_c)$ & $< 0.090$ & $< 0.088$ & $0.096 \pm 0.025$ &  $< 0.084$ \\ 
 & $\log_{10}(z_c)$ & $-3.44 \pm 0.10$ & ${-3.442}^{+0.099}_{-0.091}$ & $3.475 \pm 0.065$ &  $< -3.2$\\ 
 & $\Theta_i$ & unconstrained & unconstrained & ${1.9}^{+0.8}_{-1.1}$ &  unconstrained \\ \midrule
\multirow{3}{*}{\emg} & $\sigma_i/M_\mathrm{pl}$  & $0.34 \pm 0.14$ & $0.32 \pm 0.15$ & --- & $0.28 \pm 0.13$ \\ 
& $\alpha_\mathrm{EMG}$ & $2.54 \pm 0.34$ & $> 1.9$ & --- &  $2.64 \pm 0.37$\\ 
 & $\xi \sigma_\mathrm{i}^2/M_\mathrm{pl}^2$ & $< 0.033$ & $< 0.035$ & --- & unconstrained \\ \midrule
\multirow{2}{*}{\rnr} & $\sigma_\mathrm{i}/M_\mathrm{pl}$ & $< 0.51$ & $< 0.57$ & --- & unconstrained \\
 & $\alpha_\mathrm{EMG}$ & $< 3.1$ & $2.22 \pm 0.38$ & --- & unconstrained \\ \midrule
\multirow{4}{*}{\coldnede} & $f_\mathrm{NEDE}$  & $0.135 \pm 0.024$ & $< 0.13$ & $< 0.12$ & $< 0.15$ \\  
 &  $\log_{10}(z_c)$ & $3.455 \pm 0.044$ & $3.399 \pm 0.098$ & $> 3.0$ & $> 3.4$ \\ 
  & $3w$ & $2.55 \pm 0.26$ & $> 1.7$ & $> 1.5$ & $> 1.6$ \\ 
  & $\Omega_\phi$ & $< 0.0030$ & $< 0.0027$ & $< 0.0028$ &  $< 0.0026$ \\ 

\arrayrulecolor[HTML]{808080} \midrule \arrayrulecolor[HTML]{e6e6e6}

\multirow{4}{*}{\modrec} & $\Delta z$ & $2.9 \pm 3.4$ & $1.6 \pm 4.4$ & $0.9 \pm 3.7$ & $0.4 \pm 4.0$ \\ 
 & $A_\mathrm{gauss}$ & unconstrained & unconstrained &  unconstrained & unconstrained \\ 
 & $z_\mathrm{gauss}$ & ${1464}^{+27}_{-25}$ & ${1390}^{+99}_{-6}$ & $1467 \pm 26$ & ${1429}^{+51}_{-46}$ \\ 
 & $\sigma_\mathrm{gauss}$ & unconstrained & unconstrained & unconstrained & unconstrained \\ \midrule
\me & $m_\mathrm{e}^\mathrm{early}/m_\mathrm{e}^\mathrm{late}$ & $1.0056 \pm 0.0048$ & $1.0008 \pm 0.0072$ & ${1.0008}^{+0.0069}_{-0.0075}$ & $1.0105 \pm 0.0061$ \\ \arrayrulecolor[HTML]{808080} \midrule \arrayrulecolor[HTML]{e6e6e6}

  \neff & $N_\mathrm{eff}$ & $< 0.14$ & $< 0.14$ & $< 0.14$ & $< 0.46$ \\ 
\midrule
\sidr & $N_\mathrm{idr}$ & $< 0.11$ & $< 0.11$ & $< 0.10$ &  $< 0.42$ \\ 
\midrule
\multirow{2}{*}{\wzdr} & $N_\mathrm{wzdr}$ & $< 0.25$ & $< 0.24$  & $< 0.25$ & unconstrained \\ 
  & $\log_{10}(z_t)$ & unconstrained & unconstrained & unconstrained & unconstrained \\  
  \midrule
\multirow{4}{*}{\hotnede} & $\Delta N_\mathrm{eff}$ & $0.223 \pm 0.096$ & $< 0.15$ & $< 0.14$ & $< 0.61$ \\  
& $f_\mathrm{idm}$ & $< 0.022$ & $< 0.021$ & $> 2.4$ & $< 0.020$ \\
& $\log_{10}(z_\mathrm{stop})$ & $4.22^{+0.50}_{-0.68}$ & $> 2.4$ & $< 0.023$ & $> 2.4$ \\
\arrayrulecolor{black} \bottomrule
\end{tabular}
}
    \caption{Constraints on the model parameters of the competing models. The first column shows the model of which the parameters in column two are investigated. Column three contains the baseline results, while column four and five contain the results from \cref{ssec:curved} and \cref{ssec:w0wa}, respectively. Finally, column 6 contains the results for \cref{ssec:noACT}.}
    \label{tab:constr_model_BS}
\end{table}

\begin{table}[tp]
    \centering
\resizebox{\textwidth}{!}{
\begin{tabular}{ccccccc}
\toprule
Model & $H_0$ & $\Omega_\mathrm{m}$ & $\sum m_\nu~[\mathrm{meV}]$ & $\sigma_8$ & $n_\mathrm{s}$ & $\tau_\mathrm{reio}$ \\
\midrule \arrayrulecolor[HTML]{CCCCCC}
\lcdm & $68.39 \pm 0.28$ & $0.3009 \pm 0.0035$ & $<84$ & ${0.8217}^{+0.0058}_{-0.0055}$ & $0.9732 \pm 0.0028$ & $0.0649 \pm 0.0060$ \\
\switch & $68.49 \pm 0.28$ & $0.3008 \pm 0.0034$ & $<84$ & $0.8217 \pm 0.0052$ & $0.9727 \pm 0.0030$ & $0.0632 \pm 0.0062$ \\
\nosroll & $68.46 \pm 0.31$ & $0.2993 \pm 0.0038$ & $<120$ & $0.8276 \pm 0.0072$ & $0.9754 \pm 0.0030$ & $0.083 \pm 0.010$ \\
\idedm & $68.66 \pm 0.34$ & ${0.2901}^{+0.0088}_{-0.0093}$ & $<75$ & $0.851 \pm 0.024$ & $0.9731 \pm 0.0028$ & $0.0636 \pm 0.0058$ \\ \midrule
\thawing & $65.8 \pm 1.1$ & $0.324 \pm 0.011$ & $<87$ & $0.810 \pm 0.012$ & $0.9753 \pm 0.0042$ & $0.0638 \pm 0.0060$ \\ \midrule
\ede & $69.76 \pm 0.87$ & $0.2994 \pm 0.0036$ & $<90$ & $0.8317 \pm 0.0082$ & $0.9782 \pm 0.0053$ & $0.0622 \pm 0.0061$ \\
\emg & ${69.61}^{+0.86}_{-0.83}$ & $0.3000 \pm 0.0035$ & $<66$ & $0.8257 \pm 0.0064$ & $0.9756 \pm 0.0046$ & $0.0637 \pm 0.0061$ \\
\rnr & ${68.95}^{+0.78}_{-0.66}$ & $0.3006 \pm 0.0035$ & $<72$ & $0.8248 \pm 0.0067$ & $0.9749 \pm 0.0042$ & $0.0638 \pm 0.0062$ \\
\coldnede & ${69.95}^{+0.96}_{-0.99}$ & $0.2995 \pm 0.0037$ & $<96$ & $0.8285 \pm 0.0084$ & $0.9782 \pm 0.0057$ & $0.0634 \pm 0.0063$ \\ \midrule
\modrec & $68.54 \pm 0.43$ & $0.3010 \pm 0.0036$ & $<72$ & $0.8196 \pm 0.0053$ & $0.9624 \pm 0.0060$ & $0.0617 \pm 0.0059$ \\
\me & ${69.10}^{+0.71}_{-0.68}$ & $0.2986 \pm 0.0041$ & $<123$ & $0.8252 \pm 0.0072$ & $0.9721 \pm 0.0031$ & $0.0630 \pm 0.0061$ \\ \midrule
\neff & $68.66 \pm 0.36$ & $0.3003 \pm 0.0035$ & $<75$ & $0.8241 \pm 0.0056$ & $0.9753 \pm 0.0033$ & $0.0645 \pm 0.0059$ \\
\sidr & $68.64 \pm 0.33$ & $0.3003 \pm 0.0034$ & $<75$ & $0.8232 \pm 0.0054$ & $0.9741 \pm 0.0030$ & $ 0.0649 \pm 0.0061$ \\
\wzdr & $68.96 \pm 0.45$ & $0.2996 \pm 0.0035$ & $<81$ & $0.8251 \pm 0.0059$ & $0.9764 \pm 0.0036$ & $0.0644 \pm 0.0063$ \\
\hotnede & $68.70 \pm 0.41$ & $0.3013 \pm 0.0035$ & $<72$ & $0.8188 \pm 0.0066$ & $0.9754 \pm 0.0033$ & $0.0670 \pm 0.0069$ \\
\arrayrulecolor{black}\bottomrule
\end{tabular}
}
    \caption{Parameter constraints on $\Lambda$CDM cosmological parameters for each contender when subjected to CMB+BAO+SN data. For the constraints on model parameters, see \cref{tab:constr_model_BS} instead.}
    \label{tab:constr_BS}
\end{table}
\begin{table}[tp]
    \centering
\resizebox{\textwidth}{!}{
\begin{tabular}{cccccccc}
\toprule
Model & $H_0$ & $\Omega_\mathrm{m}$ & $\sum m_\nu~[\mathrm{meV}]$ & $\sigma_8$ & $n_\mathrm{s}$ & $\tau_\mathrm{reio}$ & $\Omega_k$ \\
\midrule \arrayrulecolor[HTML]{CCCCCC}
\lcdm & $68.56 \pm 0.31$ & $0.3013 \pm 0.0039$ & $<219$ & ${0.8226}^{+0.0061}_{-0.0066}$ & $0.9714 \pm 0.0033$ & ${0.0647}^{+0.0062}_{-0.0071}$ & $0.0012 \pm 0.0012$ \\
\switch & $68.66 \pm 0.30$ & $0.3004 \pm 0.0035$ & $<102$ & ${0.8222}^{+0.0057}_{-0.0060}$ & $0.9720 \pm 0.0031$ & ${0.0618}^{+0.0061}_{-0.0056}$ & $0.0003 \pm 0.0013$ \\
\nosroll & $68.48 \pm 0.35$ & $0.3009 \pm 0.0045$ & $<189$ & $0.8252 \pm 0.0083$ & $0.9742 \pm 0.0034$ & $0.085 \pm 0.011$ & $0.0010 \pm 0.0015$ \\
\idedm & $68.78 \pm 0.38$ & $0.2912 \pm 0.0089$ & $<108$ & $0.850 \pm 0.023$ & $0.9716 \pm 0.0033$ & $0.0628 \pm 0.0060$ & $0.0011 \pm 0.0012$ \\ \midrule
\thawing & $66.0 \pm 1.1$ & $0.323 \pm 0.010$ & $<81$ & $0.811 \pm 0.011$ & $0.9756 \pm 0.0045$ & $0.0621 \pm 0.0058$ & $0.0001 \pm 0.0010$ \\ \midrule
\ede & $69.94 \pm 0.88$ & $0.3004 \pm 0.0039$ & $<144$ & $0.8314 \pm 0.0086$ & $0.9768 \pm 0.0054$ & $0.0618 \pm 0.0062$ & $0.0012 \pm 0.0012$ \\
\emg & $69.55 \pm 0.83$ & $0.3001 \pm 0.0035$ & $<78$ & $0.8258 \pm 0.0064$ & $0.9745 \pm 0.0046$ & $0.0638 \pm 0.0062$ & $0.00048 \pm 0.00097$ \\
\rnr & ${69.36}^{+0.91}_{-0.85}$ & $0.3010 \pm 0.0038$ & $<111$ & $0.8268 \pm 0.0076$ & $0.9734 \pm 0.0049$ & $0.0632 \pm 0.0064$ & $0.0012 \pm 0.0013$ \\
\coldnede & $69.9 \pm 1.0$ & $0.3010 \pm 0.0040$ & $<120$ & $0.8249 \pm 0.0094$ & $0.9756 \pm 0.0061$ & $0.0626 \pm 0.0062$ & $0.0013 \pm 0.0012$ \\ \midrule
\modrec & $68.56 \pm 0.47$ & ${0.3026}^{+0.0042}_{-0.0039}$ & $<96$ & $0.8207 \pm 0.0059$ & $0.9618 \pm 0.0061$ & $0.0608 \pm 0.0059$ & $0.0015 \pm 0.0014$ \\
\me & $68.66 \pm 0.81$ & $0.3010 \pm 0.0048$ & $<138$ & $0.8226 \pm 0.0075$ & $0.9710 \pm 0.0030$ & ${0.0632}^{+0.0060}_{-0.0063}$ & $0.0015 \pm 0.0016$ \\ \midrule
\neff & $69.60 \pm 0.43$ & $0.2958 \pm 0.0034$ & $<96$ & $0.8278 \pm 0.0057$ & $0.9739 \pm 0.0037$ & $0.0657 \pm 0.0064$ & $0.0009 \pm 0.0011$ \\
\sidr & $68.78 \pm 0.36$ & $0.3004 \pm 0.0036$ & $<96$ & $0.8240 \pm 0.0060$ & $0.9727 \pm 0.0034$ & $0.0647 \pm 0.0062$ & $0.0009 \pm 0.0011$ \\
\wzdr & $68.99 \pm 0.44$ & $0.3003 \pm 0.0036$ & $<99$ & $0.8253 \pm 0.0063$ & $0.9751 \pm 0.0041$ & $0.0637 \pm 0.0061$ & $0.0007 \pm 0.0012$ \\
\hotnede & $68.82 \pm 0.40$ & $0.3020 \pm 0.0037$ & $<93$ & $0.8201 \pm 0.0067$ & $0.9737 \pm 0.0035$ & $0.0650 \pm 0.0062$ & $0.0013 \pm 0.0013$ \\
\arrayrulecolor{black}\bottomrule
\end{tabular}
}
    \caption{Same as \cref{tab:constr_BS}, but when additional curvature is allowed for the models.}
    \label{tab:constr_BS_Ok}
\end{table}

\begin{table}[tp]
    \centering
\resizebox{\textwidth}{!}{
\begin{tabular}{ccccccccc}
\toprule
Model & $H_0$ & $\Omega_\mathrm{m}$ & $\sum m_\nu~[\mathrm{meV}]$ & $\sigma_8$ & $n_\mathrm{s}$ & $\tau_\mathrm{reio}$ & $w_0$ & $w_a$ \\
\midrule \arrayrulecolor[HTML]{CCCCCC}
\lcdm & $67.53 \pm 0.60$ & $0.3101 \pm 0.0057$ & $<114$ & $0.8164 \pm 0.0081$ & $0.9722 \pm 0.0030$ & $0.0622 \pm 0.0062$ & $-0.846 \pm 0.054$ & $-0.52 \pm 0.20$ \\
\switch & --- & --- & --- & --- & --- & --- & --- & --- \\
\nosroll & $67.49 \pm 0.65$ & $0.3091 \pm 0.0064$ & $<183$ & $0.8180 \pm 0.0090$ & $0.9749 \pm 0.0033$ & $0.082 \pm 0.012$ & $-0.870 \pm 0.063$ & $-0.41 \pm 0.25$ \\
\idedm & --- & --- & --- & --- & --- & --- & --- & --- \\ \midrule
\thawing & --- & --- & --- & --- & --- & --- & --- & --- \\ \midrule
\ede & $68.63 \pm 0.94$ & $0.3097 \pm 0.0059$ & $<147$ & $0.8224 \pm 0.0094$ & $0.9768 \pm 0.0054$ & $0.0608 \pm 0.0062$ & $-0.849 \pm 0.056$ & $-0.50 \pm 0.21$ \\
\emg & --- & --- & --- & --- & --- & --- & --- & --- \\
\rnr & --- & --- & --- & --- & --- & --- & --- & --- \\
\coldnede & $68.7 \pm 1.0$ & $0.3101 \pm 0.0058$ & $<129$ & $0.8163 \pm 0.0098$ & $0.9754 \pm 0.0062$ & $0.0616 \pm 0.0061$ & $-0.844 \pm 0.055$ & $-0.53 \pm 0.22$ \\ \midrule
\modrec & $67.45 \pm 0.67$ & $0.3117 \pm 0.0058$ & $<129$ & ${0.8130}^{+0.0076}_{-0.0079}$ & $0.9631 \pm 0.0062$ & $0.0596 \pm 0.0055$ & $-0.839 \pm 0.052$ & $-0.57 \pm 0.21$ \\
\me & $67.59 \pm 0.84$ & $0.3104 \pm 0.0058$ & $<153$ & ${0.8162}^{+0.0085}_{-0.0081}$ & $0.9717 \pm 0.0033$ & ${0.0624}^{+0.0072}_{-0.0067}$ & $-0.842 \pm 0.059$ & $-0.55 \pm 0.26$ \\ \midrule
\neff & $67.73 \pm 0.63$ & $0.3098 \pm 0.0057$ & $<120$ & $0.8170 \pm 0.0080$ & $0.9742 \pm 0.0035$ & $0.0624 \pm 0.0062$ & $-0.847 \pm 0.054$ & $-0.50 \pm 0.21$ \\
\sidr & $67.69 \pm 0.62$ & $0.3101 \pm 0.0058$ & $<123$ & $0.8159 \pm 0.0082$ & $0.9730 \pm 0.0030$ & $0.0625 \pm 0.0061$ & $-0.844 \pm 0.055$ & $-0.52 \pm 0.21$ \\
\wzdr & $67.97 \pm 0.68$ & $0.3096 \pm 0.0054$ & $<126$ & ${0.8172}^{+0.0080}_{-0.0084}$ & $0.9755 \pm 0.0037$ & $0.0624 \pm 0.0064$ & $-0.853 \pm 0.053$ & $-0.48 \pm 0.21$ \\
\hotnede & $67.79 \pm 0.64$ & $0.3109 \pm 0.0057$ & $<105$ & $0.8135 \pm 0.0084$ & $0.9740 \pm 0.0034$ & $0.0627 \pm 0.0062$ & $-0.838 \pm 0.054$ & $-0.57 \pm 0.20$ \\
\arrayrulecolor{black}\bottomrule
\end{tabular}
}
    \caption{Same as \cref{tab:constr_BS}, but when additional CPL dark energy ($w_0,w_a$) is allowed for the models.}
    \label{tab:constr_BS_w0wa}
\end{table}

\begin{table}[tp]
    \centering
\resizebox{\textwidth}{!}{
\begin{tabular}{ccccccc}
\toprule
Model & $H_0$ & $\Omega_\mathrm{m}$ & $\sum m_\nu~[\mathrm{meV}]$ & $\sigma_8$ & $n_\mathrm{s}$ & $\tau_\mathrm{reio}$ \\
\midrule \arrayrulecolor[HTML]{CCCCCC}
\lcdm & $68.50 \pm 0.28$ & $0.2995 \pm 0.0035$ & $<81$ & $0.8225 \pm 0.0053$ & $0.9736 \pm 0.0029$ & $0.0659 \pm 0.0065$ \\
\switch & $68.60 \pm 0.30$ & $0.2993 \pm 0.0033$ & $<72$ & $0.8222 \pm 0.0052$ & $0.9728 \pm 0.0029$ & ${0.0639}^{+0.0065}_{-0.0061}$ \\
\nosroll & $68.58 \pm 0.31$ & $0.2978 \pm 0.0038$ & $<120$ & $0.8283 \pm 0.0073$ & $0.9758 \pm 0.0030$ & $0.084 \pm 0.010$ \\
\idedm & $69.50 \pm 0.73$ & $0.262 \pm 0.025$ & $<90$ & $0.937 \pm 0.084$ & $0.9731 \pm 0.0029$ & $0.0629 \pm 0.0061$ \\ \midrule
\thawing & $65.7 \pm 1.9$ & $0.325 \pm 0.018$ & $<75$ & $0.809 \pm 0.016$ & $0.9759 \pm 0.0040$ & $0.0623 \pm 0.0058$ \\ \midrule
\ede & $69.95 \pm 0.92$ & $0.2979 \pm 0.0037$ & $<81$ & $0.8324 \pm 0.0081$ & $0.9791 \pm 0.0053$ & $0.0627 \pm 0.0061$ \\
\emg & ${69.62}^{+0.96}_{-0.88}$ & $0.2990 \pm 0.0036$ & $<66$ & ${0.8254}^{+0.0069}_{-0.0065}$ & ${0.9758}^{+0.0050}_{-0.0047}$ & $0.0646 \pm 0.0062$ \\
\rnr & ${69.10}^{+0.80}_{-0.68}$ & $0.2992 \pm 0.0035$ & $<66$ & $0.8255 \pm 0.0066$ & $0.9754 \pm 0.0042$ & $0.0641 \pm 0.0060$ \\
\coldnede & $70.1 \pm 1.0$ & $0.2984 \pm 0.0037$ & $<90$ & $0.8289 \pm 0.0093$ & $0.9789 \pm 0.0061$ & $0.0645 \pm 0.0068$ \\ \midrule
\modrec & $68.63 \pm 0.44$ & ${0.3002}^{+0.0038}_{-0.0035}$ & $<66$ & $0.8207 \pm 0.0052$ & $0.9625 \pm 0.0062$ & $0.0620 \pm 0.0061$ \\
\me & $69.30 \pm 0.70$ & $0.2968 \pm 0.0042$ & $<138$ & ${0.8259}^{+0.0073}_{-0.0070}$ & $0.9722 \pm 0.0031$ & $0.0632 \pm 0.0063$ \\ \midrule
\neff & $68.82 \pm 0.38$ & $0.2986 \pm 0.0035$ & $<78$ & $0.8246 \pm 0.0057$ & $0.9757 \pm 0.0034$ & $0.0653 \pm 0.0061$ \\
\sidr & $68.78 \pm 0.35$ & $0.2987 \pm 0.0035$ & $<66$ & $0.8236 \pm 0.0053$ & $0.9744 \pm 0.0030$ & $0.0652 \pm 0.0061$ \\
\wzdr & $69.08 \pm 0.48$ & $0.2984 \pm 0.0035$ & $<75$ & $0.8261 \pm 0.0057$ & $0.9769 \pm 0.0035$ & $0.0648 \pm 0.0059$ \\
\hotnede & $68.75 \pm 0.36$ & $0.3001 \pm 0.0036$ & $<66$ & $0.8204 \pm 0.0062$ & $0.9754 \pm 0.0032$ & $0.0670 \pm 0.0064$ \\
\arrayrulecolor{black}\bottomrule
\end{tabular}
}
    \caption{Same as \cref{tab:constr_BS}, but with only CMB+BAO data.}
    \label{tab:constr_B}
\end{table}

\begin{table}[tp]
    \centering
\resizebox{\textwidth}{!}{
\begin{tabular}{ccccccc}
\toprule
Model & $H_0$ & $\Omega_\mathrm{m}$ & $\sum m_\nu~[\mathrm{meV}]$ & $\sigma_8$ & $n_\mathrm{s}$ & $\tau_\mathrm{reio}$ \\
\midrule \arrayrulecolor[HTML]{CCCCCC}
\lcdm & $67.73 \pm 0.60$ & $0.3096 \pm 0.0078$ & $<147$ & $0.8174 \pm 0.0080$ & $0.9713 \pm 0.0032$ & $0.0623 \pm 0.0063$ \\
\switch & $68.04 \pm 0.69$ & $0.3066 \pm 0.0085$ & $<156$ & $0.8176 \pm 0.0084$ & $0.9717 \pm 0.0032$ & $0.0623 \pm 0.0062$ \\
\nosroll & $67.42 \pm 1.00$ & $0.312 \pm 0.012$ & $<291$ & $0.814 \pm 0.014$ & $0.9738 \pm 0.0035$ & $0.085 \pm 0.011$ \\
\idedm & $70.2 \pm 1.6$ & ${0.229}^{+0.048}_{-0.045}$ & $<177$ & $1.11 \pm 0.22$ & $0.9723 \pm 0.0033$ & $0.0615 \pm 0.0060$ \\ \midrule
\thawing & $64.8 \pm 2.5$ & $0.339 \pm 0.027$ & $<177$ & $0.799 \pm 0.022$ & $0.9736 \pm 0.0044$ & $0.0610 \pm 0.0062$ \\ \midrule
\ede & $69.0 \pm 1.1$ & $0.309 \pm 0.011$ & $<234$ & $0.824 \pm 0.012$ & $0.9767 \pm 0.0055$ & $0.0611 \pm 0.0063$ \\
\emg & $68.9 \pm 1.1$ & $0.3094 \pm 0.0086$ & $<162$ & $0.8203 \pm 0.0096$ & ${0.9734}^{+0.0055}_{-0.0053}$ & $0.0620 \pm 0.0061$ \\
\rnr & ${68.31}^{+0.95}_{-0.91}$ & $0.3099 \pm 0.0086$ & $<174$ & ${0.8196}^{+0.0096}_{-0.0099}$ & $0.9732 \pm 0.0048$ & $0.0618 \pm 0.0062$ \\
\coldnede & $68.8 \pm 1.3$ & $0.312 \pm 0.010$ & $<177$ & $0.816 \pm 0.013$ & $0.9738 \pm 0.0065$ & $0.0620 \pm 0.0064$ \\ \midrule
\modrec & $67.1 \pm 1.1$ & $0.317 \pm 0.011$ & $<177$ & $0.811 \pm 0.010$ & $0.9634 \pm 0.0062$ & $0.0600 \pm 0.0057$ \\
\me & unconstrained & $0.346 \pm 0.026$ & $<153$ & $0.794 \pm 0.017$ & $0.9696 \pm 0.0034$ & $0.0609 \pm 0.0062$ \\ \midrule
\neff & $68.01 \pm 0.68$ & $0.3084 \pm 0.0081$ & $<150$ & ${0.8190}^{+0.0082}_{-0.0085}$ & $0.9731 \pm 0.0037$ & $0.0627 \pm 0.0062$ \\
\sidr & $68.01 \pm 0.64$ & $0.3080 \pm 0.0078$ & $<144$ & $0.8184 \pm 0.0081$ & $0.9723 \pm 0.0033$ & $0.0627 \pm 0.0062$ \\
\wzdr & $68.41 \pm 0.76$ & $0.3059 \pm 0.0083$ & $<147$ & ${0.8207}^{+0.0086}_{-0.0088}$ & $0.9748 \pm 0.0039$ & $0.0627 \pm 0.0063$ \\
\hotnede & $67.80 \pm 0.70$ & $0.3131 \pm 0.0089$ & $<159$ & $0.812 \pm 0.010$ & $0.9729 \pm 0.0036$ & $0.0633 \pm 0.0067$ \\
\arrayrulecolor{black}\bottomrule
\end{tabular}
}
    \caption{Same as \cref{tab:constr_BS}, but with only CMB data.}
    \label{tab:constr}
\end{table}

\section{Detailed results per model}\label{app:permodel}

\begin{figure}[htbp]
    \centering
    \includegraphics[width=0.99\linewidth]{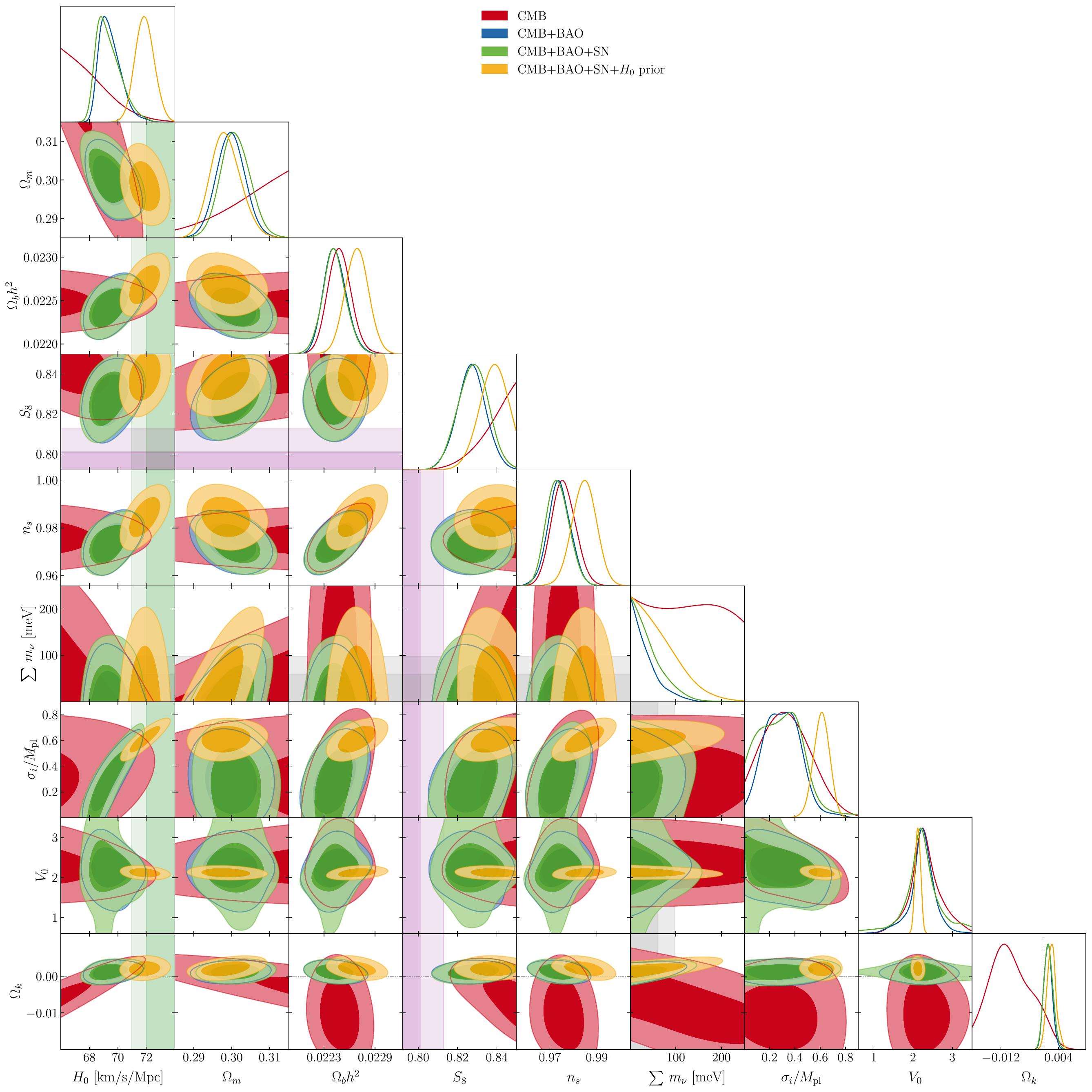}
    \caption{68\% and 95\% CL contours of the \rnr{} model, when a curved universe parameterized by $\Omega_k$ is assumed. The datasets are CMB, CMB+BAO, CMB+BAO+SN, and CMB+BAO+SN+$M_B$ prior as shown in the legend.}
    \label{app:fig:RnR_Ok}
\end{figure}

In \cref{app:fig:RnR_Ok} we show the results for the curved RnR model with the CMB, CMB+BAO, CMB+BAO+SN, and CMB+BAO+SN+$M_B$~prior data combinations. The same figures for all other cases can be found at \url{https://doi.org/10.5281/zenodo.21277319} \cite{ZENODO}, where we also include the figures for the cases without ACT data and with the $\alpha_s/\beta_s$ extension, as well as extensions through curvature and CPL dark energy. The same repository also contains the chain files for the sake of reproducibility.



\begin{figure}[tp]
    \centering
    \includegraphics[width=0.99\linewidth]{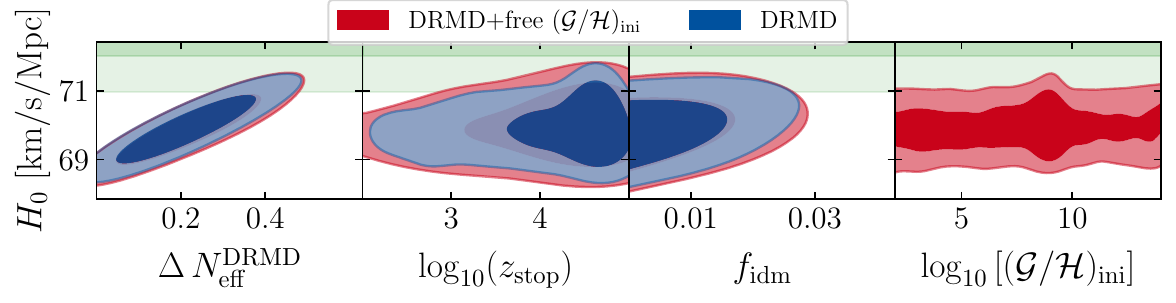}
    \caption{Impact of varying the initial value of the ratio between the interaction rate and the Hubble rate $\log_{10} (\mathcal{G}/\mathcal{H})_{\rm ini}$ in the DRMD model for the CMB+BAO+SN+$M_B$ dataset.}
    \label{fig:DRMD_free_G}
\end{figure}

In \cref{fig:DRMD_free_G}, we show that the initial value of the ratio between the interaction rate and the Hubble rate is unconstrained for the DRMD model, and has no impact on the constraints for the Hubble parameter. We conclude that we can safely fix it in the analysis.

\section{Nuisance parameter correlation}\label{app:nuisance_corr}

\begin{figure}[tp]
    \centering
    \includegraphics[width=0.65\linewidth]{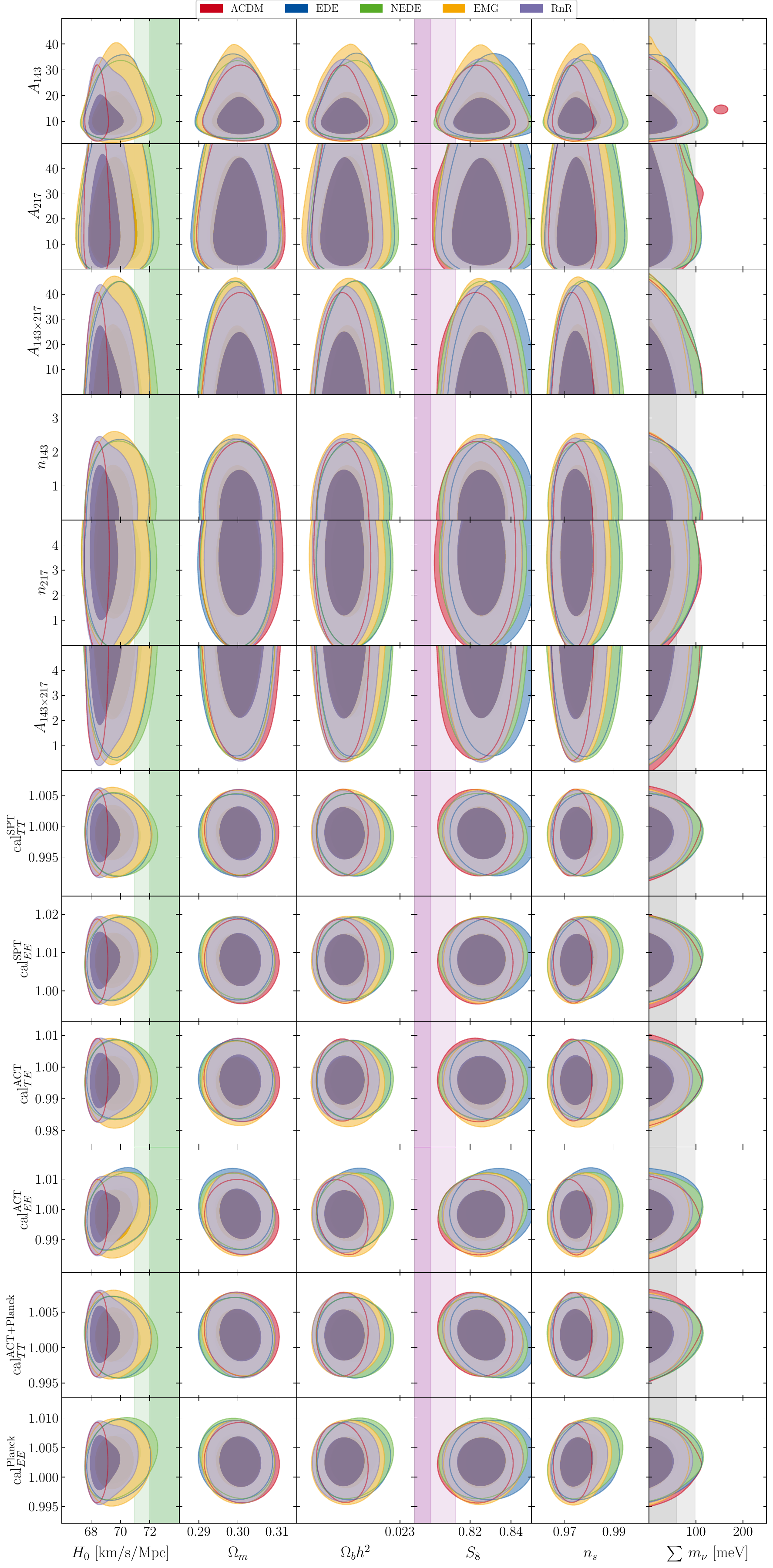}
    \caption{Correlation between cosmological parameters and nuisance parameters in the CMB+BAO+SN dataset case for models of group E.}
    \label{fig:nuisance_early}
\end{figure}

\begin{figure}[tp]
    \centering
    \includegraphics[width=0.65\linewidth]{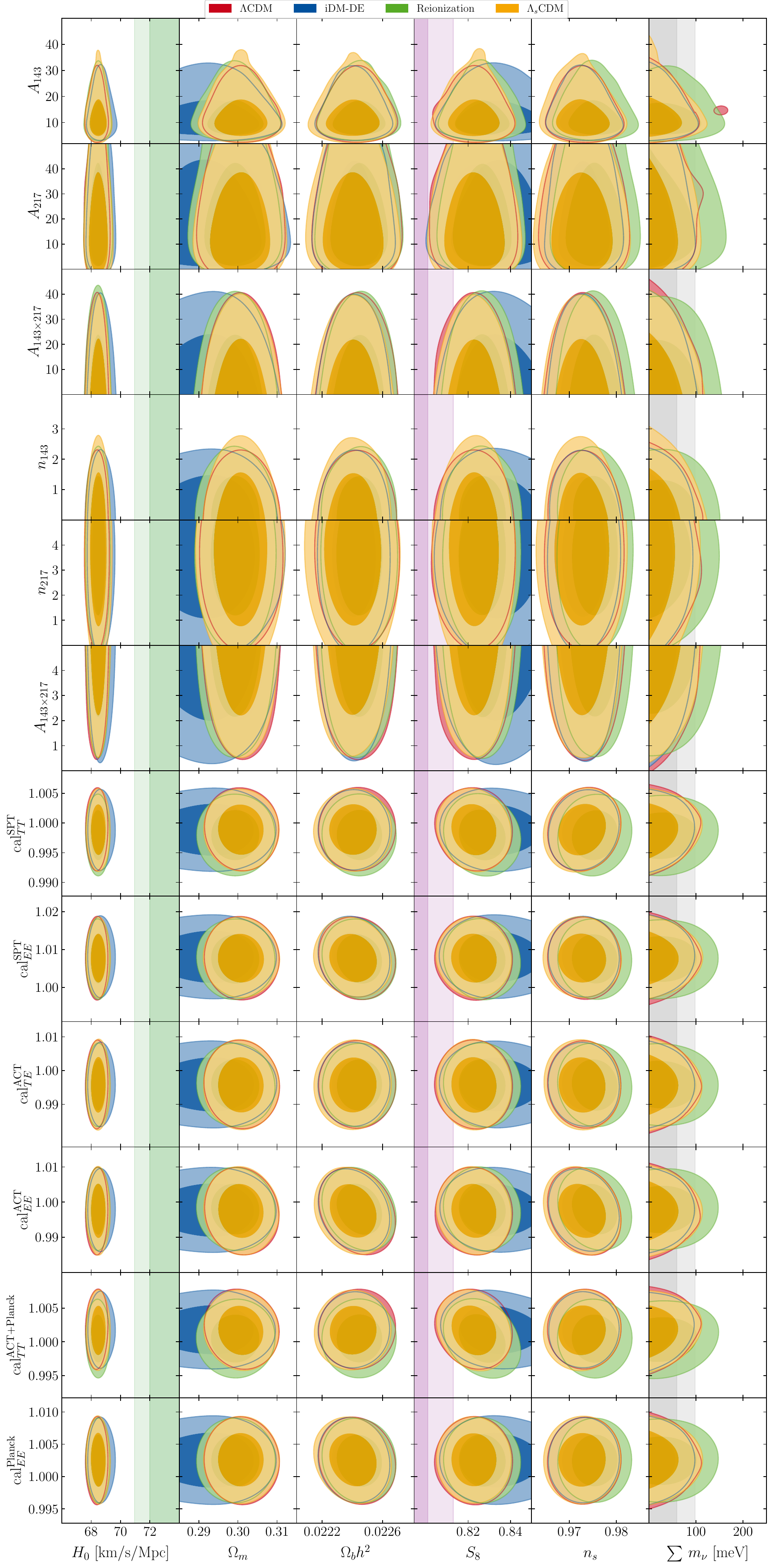}
    \caption{Correlation between cosmological parameters and nuisance parameters in the CMB+BAO+SN dataset case for models of group L.}
    \label{fig:nuisance_late}
\end{figure}

\begin{figure}[tp]
    \centering
    \includegraphics[width=0.65\linewidth]{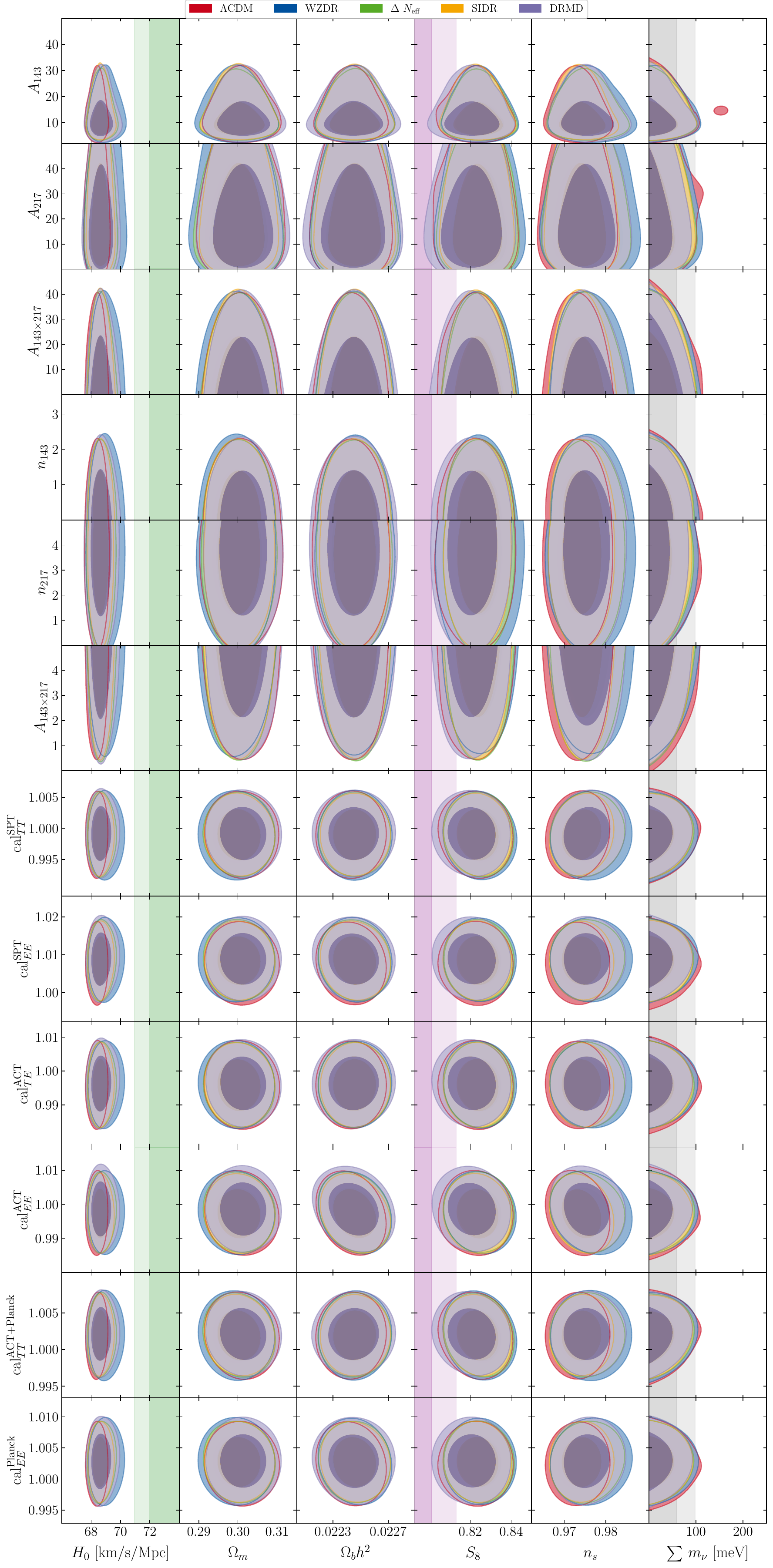}
    \caption{Correlation between cosmological parameters and nuisance parameters in the CMB+BAO+SN dataset case for models of group R.}
    \label{fig:nuisance_rad}
\end{figure}

\begin{figure}[tp]
    \centering
    \includegraphics[width=0.65\linewidth]{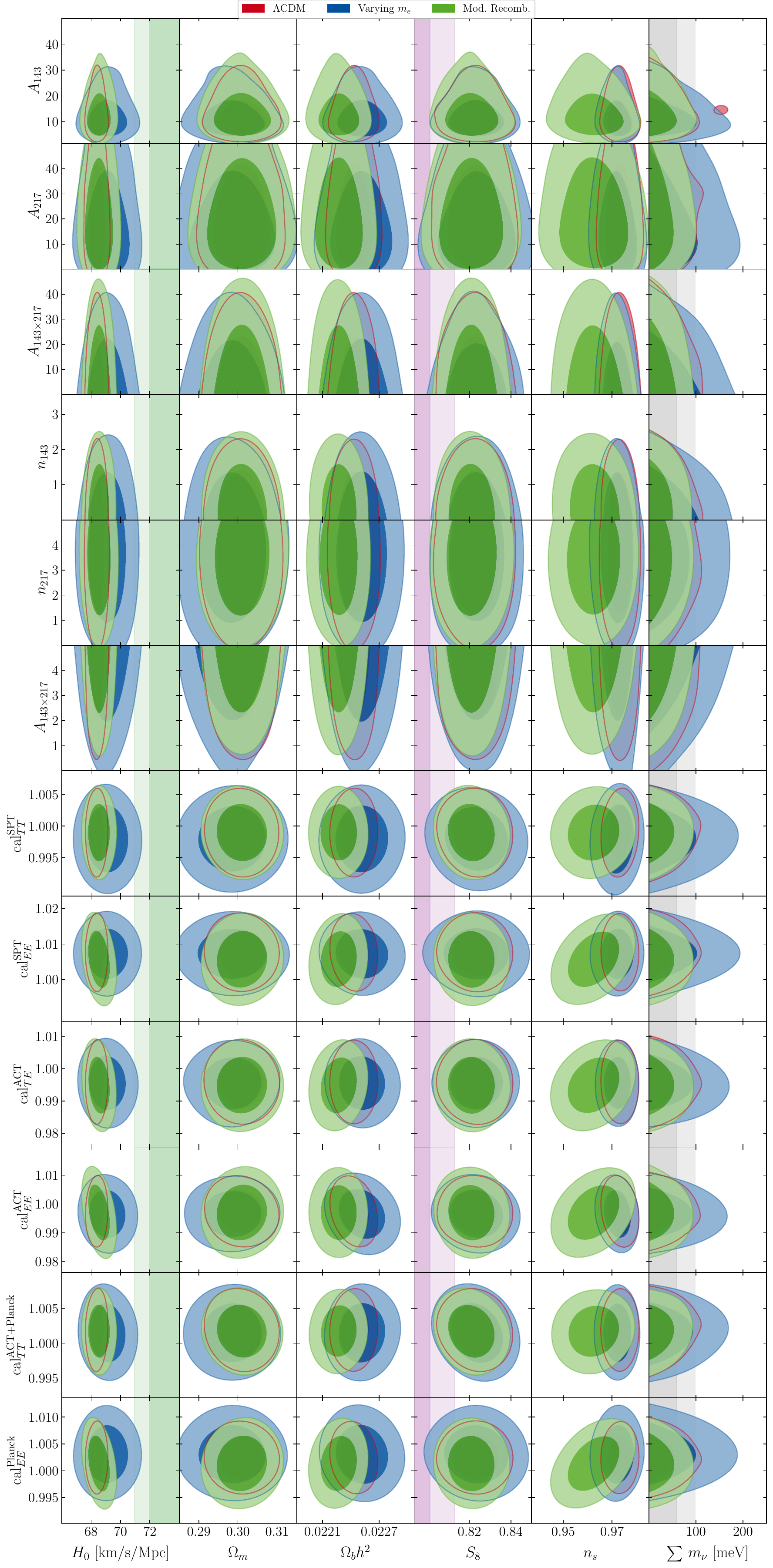}
    \caption{Correlation between cosmological parameters and nuisance parameters in the CMB+BAO+SN dataset case for models of group M.}
    \label{fig:nuisance_mod}
\end{figure}

In \cref{fig:nuisance_early,fig:nuisance_late,fig:nuisance_mod,fig:nuisance_rad} we show the 68\% and 95\% CL contours showing the correlation between the cosmological parameters and the nuisance parameters of the Planck \texttt{CamSpec} and lite ACT and SPT likelihoods discussed in \cref{ssec:data}. We see that there are no very strong correlations. The strongest one is between the EE calibration parameters and the scalar tilt $n_s$ in the 4-parameter modified recombination model. 

\bibliography{bibliography}
\end{document}